\documentclass[acmsmall]{acmart}

\usepackage{xspace}
\usepackage{graphicx}
\usepackage{xcolor}
\usepackage[normalem]{ulem} 
\usepackage{amsmath}
\usepackage{enumitem}
\usepackage{semantic} 
\usepackage{siunitx}
\usepackage{scalefnt}
\usepackage{multicol}

\usepackage{hyperref}
\usepackage{breakurl}

\usepackage{booktabs}
\usepackage{multirow}
\usepackage{colortbl}
\usepackage{makecell}
\usepackage{ragged2e}

\usepackage{caption}
\usepackage{subcaption}
\usepackage{wrapfig}
\usepackage{adjustbox}

\usepackage{pifont}
\usepackage{upquote} 
\usepackage{listings}

\usepackage{algorithm}
\usepackage{algorithmic}

\usepackage{tikz}
\usepackage{forest} 
\usetikzlibrary{arrows.meta}
\usetikzlibrary{backgrounds}
\usetikzlibrary{calc}
\usetikzlibrary{decorations.pathreplacing}
\usetikzlibrary{fit}
\usetikzlibrary{positioning}
\usetikzlibrary{shapes.geometric}

\usepackage{flushend} 
\usepackage{datetime}
\usepackage{fmtcount} 
\usepackage[framemethod=TikZ]{mdframed} 
\usepackage{pgf}

\newcommand{\DefMacro}[2]{\expandafter\newcommand\csname rmk-#1\endcsname{#2}}
\newcommand{\UseMacro}[1]{\csname rmk-#1\endcsname}
\newcommand{\eg}{e.g.}
\newcommand{\ie}{i.e.}

\newcommand{\MyPara}[1]{\vspace{2pt}\noindent\textbf{#1}.}

\newcommand{\InputWithSpace}[1]{\bgroup\def\arraystretch{1.2}\input{#1}\egroup}
\newcommand{\Code}[1]{{\ifmmode{\mathtt{#1}}\else$\mathtt{#1}$\fi}}
\newcommand{\CodeIn}[1]{\texttt{\small #1}}

\newcolumntype{R}[1]{>{\RaggedLeft\arraybackslash}p{#1}}
\newcolumntype{L}[1]{>{\RaggedRight\arraybackslash}p{#1}}

\definecolor{gray}{RGB}{211,211,211}

\let\origthelstnumber\thelstnumber
\makeatletter
\newcommand*\Suppressnumber{%
  \lst@AddToHook{OnNewLine}{%
    \let\thelstnumber\relax%
     \advance\c@lstnumber-\@ne\relax%
    }%
}
\newcommand*\Reactivatenumber[1]{%
  \setcounter{lstnumber}{\numexpr#1-1\relax}
  \lst@AddToHook{OnNewLine}{%
   \let\thelstnumber\origthelstnumber%
   \refstepcounter{lstnumber}
  }%
}
\makeatother

\makeatletter
\lst@Key{countblanklines}{true}[t]%
    {\lstKV@SetIf{#1}\lst@ifcountblanklines}

\lst@AddToHook{OnEmptyLine}{%
    \lst@ifnumberblanklines\else%
       \lst@ifcountblanklines\else%
         \advance\c@lstnumber-\@ne\relax%
       \fi%
    \fi}
\makeatother

\lstdefinelanguage{java-pretty}
{
  language=java,
  numbers=left,
  basicstyle=\footnotesize\ttfamily,
  numberstyle=\footnotesize,
  breaklines=true,
  columns=fullflexible,
  xleftmargin=16pt,
  showstringspaces=false,
}

\lstdefinelanguage{js-pretty}
{
  language=java-pretty,
  keywords={typeof, new, true, false, catch, function, return, null, switch, var, if, while, do, else, case, break, for, in, let, const, await, async, yield, export, import, from, as, default, class, extends, super, this, try, throw},
}

\lstdefinelanguage{cpp-pretty}
{
  language=C++,
  numbers=left,
  basicstyle=\footnotesize\ttfamily,
  numberstyle=\footnotesize,
  breaklines=true,
  columns=fullflexible,
  xleftmargin=16pt,
  showstringspaces=false,
}

\lstdefinelanguage{patch-pretty}
{
  morecomment=[f][\color{gray}]{diff},     
  morecomment=[f][\color{gray}]{index},    
  morecomment=[f][\color{gray}]{@@},       
  morecomment=[f][\color{gray}]{---},      
  morecomment=[f][\color{gray}]{+++},      
  morecomment=[f][\color{green!50!black}]{+}, 
  morecomment=[f][\color{red!70!black}]{-},   
  morestring=[b]",
  alsoletter={+,-,@},
  sensitive=true,
  keywords={bool}
}

\lstdefinestyle{patch-pretty}
{
  language=patch-pretty,
  numbers=left,
  basicstyle=\footnotesize\ttfamily,
  numberstyle=\footnotesize,
  breaklines=true,
  columns=fullflexible,   
  keepspaces=true,        
  xleftmargin=16pt,
  rulecolor=\color{black!20},
  backgroundcolor=\color{gray!2},
  showstringspaces=false,
  tabsize=4,
}

\DefMacro{empirical-hotspot-collected}{111}
\DefMacro{empirical-hotspot-studied}{87}
\DefMacro{empirical-hotspot-art-tc}{38}
\DefMacro{empirical-hotspot-art-bm}{24}
\DefMacro{empirical-hotspot-art-app}{3}
\DefMacro{empirical-hotspot-art-other}{22}
\DefMacro{empirical-hotspot-sym-pd}{10}
\DefMacro{empirical-hotspot-sym-pr}{28}
\DefMacro{empirical-hotspot-sym-log}{19}
\DefMacro{empirical-hotspot-sym-crash}{8}
\DefMacro{empirical-hotspot-sym-nc}{10}
\DefMacro{empirical-hotspot-sym-other}{12}
\DefMacro{empirical-hotspot-rc-spec}{14}
\DefMacro{empirical-hotspot-rc-lc}{3}
\DefMacro{empirical-hotspot-rc-opt}{36}
\DefMacro{empirical-hotspot-rc-codegen}{21}
\DefMacro{empirical-hotspot-rc-runtime}{12}
\DefMacro{empirical-hotspot-rc-other}{1}
\DefMacro{empirical-graal-collected}{39}
\DefMacro{empirical-graal-studied}{18}
\DefMacro{empirical-graal-art-tc}{9}
\DefMacro{empirical-graal-art-bm}{2}
\DefMacro{empirical-graal-art-app}{6}
\DefMacro{empirical-graal-art-other}{1}
\DefMacro{empirical-graal-sym-pd}{6}
\DefMacro{empirical-graal-sym-pr}{3}
\DefMacro{empirical-graal-sym-log}{2}
\DefMacro{empirical-graal-sym-crash}{1}
\DefMacro{empirical-graal-sym-nc}{4}
\DefMacro{empirical-graal-sym-other}{2}
\DefMacro{empirical-graal-rc-spec}{2}
\DefMacro{empirical-graal-rc-lc}{1}
\DefMacro{empirical-graal-rc-opt}{8}
\DefMacro{empirical-graal-rc-codegen}{6}
\DefMacro{empirical-graal-rc-runtime}{1}
\DefMacro{empirical-graal-rc-other}{0}
\DefMacro{empirical-v8-collected}{75}
\DefMacro{empirical-v8-studied}{51}
\DefMacro{empirical-v8-art-tc}{36}
\DefMacro{empirical-v8-art-bm}{9}
\DefMacro{empirical-v8-art-app}{5}
\DefMacro{empirical-v8-art-other}{1}
\DefMacro{empirical-v8-sym-pd}{18}
\DefMacro{empirical-v8-sym-pr}{12}
\DefMacro{empirical-v8-sym-log}{12}
\DefMacro{empirical-v8-sym-crash}{4}
\DefMacro{empirical-v8-sym-nc}{1}
\DefMacro{empirical-v8-sym-other}{4}
\DefMacro{empirical-v8-rc-spec}{22}
\DefMacro{empirical-v8-rc-lc}{2}
\DefMacro{empirical-v8-rc-opt}{17}
\DefMacro{empirical-v8-rc-codegen}{6}
\DefMacro{empirical-v8-rc-runtime}{2}
\DefMacro{empirical-v8-rc-other}{2}
\DefMacro{empirical-sm-collected}{47}
\DefMacro{empirical-sm-studied}{35}
\DefMacro{empirical-sm-art-tc}{10}
\DefMacro{empirical-sm-art-bm}{12}
\DefMacro{empirical-sm-art-app}{12}
\DefMacro{empirical-sm-art-other}{1}
\DefMacro{empirical-sm-sym-pd}{13}
\DefMacro{empirical-sm-sym-pr}{17}
\DefMacro{empirical-sm-sym-log}{2}
\DefMacro{empirical-sm-sym-crash}{1}
\DefMacro{empirical-sm-sym-nc}{1}
\DefMacro{empirical-sm-sym-other}{1}
\DefMacro{empirical-sm-rc-spec}{17}
\DefMacro{empirical-sm-rc-lc}{1}
\DefMacro{empirical-sm-rc-opt}{6}
\DefMacro{empirical-sm-rc-codegen}{5}
\DefMacro{empirical-sm-rc-runtime}{2}
\DefMacro{empirical-sm-rc-other}{4}
\DefMacro{empirical-total-collected}{272}
\DefMacro{empirical-total-studied}{191}
\DefMacro{empirical-total-art-tc}{93}
\DefMacro{empirical-total-art-bm}{47}
\DefMacro{empirical-total-art-app}{26}
\DefMacro{empirical-total-art-other}{25}
\DefMacro{empirical-total-sym-pd}{47}
\DefMacro{empirical-total-sym-pr}{60}
\DefMacro{empirical-total-sym-log}{35}
\DefMacro{empirical-total-sym-crash}{14}
\DefMacro{empirical-total-sym-nc}{16}
\DefMacro{empirical-total-sym-other}{19}
\DefMacro{empirical-total-rc-spec}{55}
\DefMacro{empirical-total-rc-lc}{7}
\DefMacro{empirical-total-rc-opt}{67}
\DefMacro{empirical-total-rc-codegen}{38}
\DefMacro{empirical-total-rc-runtime}{17}
\DefMacro{empirical-total-rc-other}{7}
\DefMacro{empirical-total-art-tc-percentage}{48.69\%}
\DefMacro{empirical-total-art-bm-percentage}{24.61\%}
\DefMacro{empirical-total-art-app-percentage}{13.61\%}
\DefMacro{empirical-total-art-other-percentage}{13.09\%}

\DefMacro{empirical-total-sym-pd-percentage}{24.61\%}
\DefMacro{empirical-total-sym-pr-percentage}{31.41\%}
\DefMacro{empirical-total-sym-log-percentage}{18.32\%}
\DefMacro{empirical-total-sym-crash-percentage}{7.33\%}
\DefMacro{empirical-total-sym-nc-percentage}{8.38\%}
\DefMacro{empirical-total-sym-other-percentage}{9.95\%}

\DefMacro{empirical-total-rc-spec-percentage}{28.80\%}
\DefMacro{empirical-total-rc-lc-percentage}{3.66\%}
\DefMacro{empirical-total-rc-opt-percentage}{35.08\%}
\DefMacro{empirical-total-rc-codegen-percentage}{19.90\%}
\DefMacro{empirical-total-rc-runtime-percentage}{8.90\%}
\DefMacro{empirical-total-rc-other-percentage}{3.66\%}

\DefMacro{empirical-hotspot-irrelevant}{11}
\DefMacro{empirical-hotspot-enhancement}{8}
\DefMacro{empirical-hotspot-duplicate}{5}
\DefMacro{empirical-hotspot-notanissue}{0}

\DefMacro{empirical-graal-irrelevant}{2}
\DefMacro{empirical-graal-enhancement}{2}
\DefMacro{empirical-graal-duplicate}{1}
\DefMacro{empirical-graal-notanissue}{16}

\DefMacro{empirical-v8-irrelevant}{4}
\DefMacro{empirical-v8-enhancement}{8}
\DefMacro{empirical-v8-duplicate}{0}
\DefMacro{empirical-v8-notanissue}{12}

\DefMacro{empirical-sm-irrelevant}{1}
\DefMacro{empirical-sm-enhancement}{4}
\DefMacro{empirical-sm-duplicate}{0}
\DefMacro{empirical-sm-notanissue}{7}

\DefMacro{empirical-total-irrelevant}{18}
\DefMacro{empirical-total-enhancement}{22}
\DefMacro{empirical-total-duplicate}{6}
\DefMacro{empirical-total-notanissue}{35}
\DefMacro{table-tdt-codec-tmpl}{1157}
\DefMacro{table-tdt-codec-gen}{6355}
\DefMacro{table-tdt-codec-time}{2167.55}
\DefMacro{table-tdt-codec-lp0-dup}{3}
\DefMacro{table-tdt-codec-lp0-fp}{75}
\DefMacro{table-tdt-codec-lp0-tp}{2}
\DefMacro{table-tdt-codec-lp0-fp-filtered}{5}
\DefMacro{table-tdt-codec-lp0-dup-filtered}{2}
\DefMacro{table-tdt-codec-lp0-th0}{1824}
\DefMacro{table-tdt-codec-lp0-th1}{3504}
\DefMacro{table-tdt-codec-lp0-th2}{756}
\DefMacro{table-tdt-codec-lp0-th3}{191}
\DefMacro{table-tdt-codec-lp0-th4}{80}
\DefMacro{table-tdt-codec-lp1-dup}{16}
\DefMacro{table-tdt-codec-lp1-fp}{73}
\DefMacro{table-tdt-codec-lp1-tp}{2}
\DefMacro{table-tdt-codec-lp1-fp-filtered}{18}
\DefMacro{table-tdt-codec-lp1-dup-filtered}{2}
\DefMacro{table-tdt-codec-lp1-th0}{1804}
\DefMacro{table-tdt-codec-lp1-th1}{3480}
\DefMacro{table-tdt-codec-lp1-th2}{808}
\DefMacro{table-tdt-codec-lp1-th3}{172}
\DefMacro{table-tdt-codec-lp1-th4}{91}
\DefMacro{table-tdt-codec-lp2-dup}{11}
\DefMacro{table-tdt-codec-lp2-fp}{259}
\DefMacro{table-tdt-codec-lp2-tp}{2}
\DefMacro{table-tdt-codec-lp2-fp-filtered}{14}
\DefMacro{table-tdt-codec-lp2-dup-filtered}{3}
\DefMacro{table-tdt-codec-lp2-th0}{1417}
\DefMacro{table-tdt-codec-lp2-th1}{2989}
\DefMacro{table-tdt-codec-lp2-th2}{1209}
\DefMacro{table-tdt-codec-lp2-th3}{467}
\DefMacro{table-tdt-codec-lp2-th4}{273}
\DefMacro{table-tdt-codec-lp3-dup}{11}
\DefMacro{table-tdt-codec-lp3-fp}{259}
\DefMacro{table-tdt-codec-lp3-tp}{2}
\DefMacro{table-tdt-codec-lp3-fp-filtered}{13}
\DefMacro{table-tdt-codec-lp3-dup-filtered}{2}
\DefMacro{table-tdt-codec-lp3-th0}{1403}
\DefMacro{table-tdt-codec-lp3-th1}{3035}
\DefMacro{table-tdt-codec-lp3-th2}{1204}
\DefMacro{table-tdt-codec-lp3-th3}{441}
\DefMacro{table-tdt-codec-lp3-th4}{272}
\DefMacro{table-tdt-codec-rp0-dup}{0}
\DefMacro{table-tdt-codec-rp0-fp}{0}
\DefMacro{table-tdt-codec-rp0-tp}{0}
\DefMacro{table-tdt-codec-rp0-fp-filtered}{0}
\DefMacro{table-tdt-codec-rp0-dup-filtered}{0}
\DefMacro{table-tdt-codec-rp0-th0}{4926}
\DefMacro{table-tdt-codec-rp0-th1}{1220}
\DefMacro{table-tdt-codec-rp0-th2}{201}
\DefMacro{table-tdt-codec-rp0-th3}{8}
\DefMacro{table-tdt-codec-rp0-th4}{0}
\DefMacro{table-tdt-codec-rp1-dup}{0}
\DefMacro{table-tdt-codec-rp1-fp}{0}
\DefMacro{table-tdt-codec-rp1-tp}{0}
\DefMacro{table-tdt-codec-rp1-fp-filtered}{0}
\DefMacro{table-tdt-codec-rp1-dup-filtered}{0}
\DefMacro{table-tdt-codec-rp1-th0}{5202}
\DefMacro{table-tdt-codec-rp1-th1}{971}
\DefMacro{table-tdt-codec-rp1-th2}{175}
\DefMacro{table-tdt-codec-rp1-th3}{7}
\DefMacro{table-tdt-codec-rp1-th4}{0}
\DefMacro{table-tdt-compress-tmpl}{792}
\DefMacro{table-tdt-compress-gen}{6582}
\DefMacro{table-tdt-compress-time}{2276.14}
\DefMacro{table-tdt-compress-lp0-dup}{20}
\DefMacro{table-tdt-compress-lp0-fp}{289}
\DefMacro{table-tdt-compress-lp0-tp}{1}
\DefMacro{table-tdt-compress-lp0-fp-filtered}{22}
\DefMacro{table-tdt-compress-lp0-dup-filtered}{2}
\DefMacro{table-tdt-compress-lp0-th0}{860}
\DefMacro{table-tdt-compress-lp0-th1}{3102}
\DefMacro{table-tdt-compress-lp0-th2}{1745}
\DefMacro{table-tdt-compress-lp0-th3}{564}
\DefMacro{table-tdt-compress-lp0-th4}{311}
\DefMacro{table-tdt-compress-lp1-dup}{8}
\DefMacro{table-tdt-compress-lp1-fp}{355}
\DefMacro{table-tdt-compress-lp1-tp}{1}
\DefMacro{table-tdt-compress-lp1-fp-filtered}{9}
\DefMacro{table-tdt-compress-lp1-dup-filtered}{1}
\DefMacro{table-tdt-compress-lp1-th0}{848}
\DefMacro{table-tdt-compress-lp1-th1}{2950}
\DefMacro{table-tdt-compress-lp1-th2}{1809}
\DefMacro{table-tdt-compress-lp1-th3}{611}
\DefMacro{table-tdt-compress-lp1-th4}{364}
\DefMacro{table-tdt-compress-lp2-dup}{12}
\DefMacro{table-tdt-compress-lp2-fp}{822}
\DefMacro{table-tdt-compress-lp2-tp}{1}
\DefMacro{table-tdt-compress-lp2-fp-filtered}{13}
\DefMacro{table-tdt-compress-lp2-dup-filtered}{1}
\DefMacro{table-tdt-compress-lp2-th0}{586}
\DefMacro{table-tdt-compress-lp2-th1}{2000}
\DefMacro{table-tdt-compress-lp2-th2}{1975}
\DefMacro{table-tdt-compress-lp2-th3}{1186}
\DefMacro{table-tdt-compress-lp2-th4}{835}
\DefMacro{table-tdt-compress-lp3-dup}{19}
\DefMacro{table-tdt-compress-lp3-fp}{840}
\DefMacro{table-tdt-compress-lp3-tp}{1}
\DefMacro{table-tdt-compress-lp3-fp-filtered}{21}
\DefMacro{table-tdt-compress-lp3-dup-filtered}{2}
\DefMacro{table-tdt-compress-lp3-th0}{560}
\DefMacro{table-tdt-compress-lp3-th1}{2128}
\DefMacro{table-tdt-compress-lp3-th2}{1870}
\DefMacro{table-tdt-compress-lp3-th3}{1163}
\DefMacro{table-tdt-compress-lp3-th4}{861}
\DefMacro{table-tdt-compress-rp0-dup}{0}
\DefMacro{table-tdt-compress-rp0-fp}{1}
\DefMacro{table-tdt-compress-rp0-tp}{0}
\DefMacro{table-tdt-compress-rp0-fp-filtered}{0}
\DefMacro{table-tdt-compress-rp0-dup-filtered}{0}
\DefMacro{table-tdt-compress-rp0-th0}{4578}
\DefMacro{table-tdt-compress-rp0-th1}{1451}
\DefMacro{table-tdt-compress-rp0-th2}{519}
\DefMacro{table-tdt-compress-rp0-th3}{33}
\DefMacro{table-tdt-compress-rp0-th4}{1}
\DefMacro{table-tdt-compress-rp1-dup}{0}
\DefMacro{table-tdt-compress-rp1-fp}{0}
\DefMacro{table-tdt-compress-rp1-tp}{0}
\DefMacro{table-tdt-compress-rp1-fp-filtered}{0}
\DefMacro{table-tdt-compress-rp1-dup-filtered}{0}
\DefMacro{table-tdt-compress-rp1-th0}{5249}
\DefMacro{table-tdt-compress-rp1-th1}{1067}
\DefMacro{table-tdt-compress-rp1-th2}{253}
\DefMacro{table-tdt-compress-rp1-th3}{13}
\DefMacro{table-tdt-compress-rp1-th4}{0}
\DefMacro{table-tdt-lang-tmpl}{1472}
\DefMacro{table-tdt-lang-gen}{12368}
\DefMacro{table-tdt-lang-time}{3180.85}
\DefMacro{table-tdt-lang-lp0-dup}{0}
\DefMacro{table-tdt-lang-lp0-fp}{87}
\DefMacro{table-tdt-lang-lp0-tp}{1}
\DefMacro{table-tdt-lang-lp0-fp-filtered}{1}
\DefMacro{table-tdt-lang-lp0-dup-filtered}{1}
\DefMacro{table-tdt-lang-lp0-th0}{2319}
\DefMacro{table-tdt-lang-lp0-th1}{6412}
\DefMacro{table-tdt-lang-lp0-th2}{3141}
\DefMacro{table-tdt-lang-lp0-th3}{408}
\DefMacro{table-tdt-lang-lp0-th4}{88}
\DefMacro{table-tdt-lang-lp1-dup}{0}
\DefMacro{table-tdt-lang-lp1-fp}{88}
\DefMacro{table-tdt-lang-lp1-tp}{0}
\DefMacro{table-tdt-lang-lp1-fp-filtered}{1}
\DefMacro{table-tdt-lang-lp1-dup-filtered}{1}
\DefMacro{table-tdt-lang-lp1-th0}{2317}
\DefMacro{table-tdt-lang-lp1-th1}{6495}
\DefMacro{table-tdt-lang-lp1-th2}{3325}
\DefMacro{table-tdt-lang-lp1-th3}{142}
\DefMacro{table-tdt-lang-lp1-th4}{89}
\DefMacro{table-tdt-lang-lp2-dup}{0}
\DefMacro{table-tdt-lang-lp2-fp}{646}
\DefMacro{table-tdt-lang-lp2-tp}{0}
\DefMacro{table-tdt-lang-lp2-fp-filtered}{0}
\DefMacro{table-tdt-lang-lp2-dup-filtered}{0}
\DefMacro{table-tdt-lang-lp2-th0}{1619}
\DefMacro{table-tdt-lang-lp2-th1}{4096}
\DefMacro{table-tdt-lang-lp2-th2}{4213}
\DefMacro{table-tdt-lang-lp2-th3}{1794}
\DefMacro{table-tdt-lang-lp2-th4}{646}
\DefMacro{table-tdt-lang-lp3-dup}{0}
\DefMacro{table-tdt-lang-lp3-fp}{748}
\DefMacro{table-tdt-lang-lp3-tp}{1}
\DefMacro{table-tdt-lang-lp3-fp-filtered}{1}
\DefMacro{table-tdt-lang-lp3-dup-filtered}{1}
\DefMacro{table-tdt-lang-lp3-th0}{1634}
\DefMacro{table-tdt-lang-lp3-th1}{3937}
\DefMacro{table-tdt-lang-lp3-th2}{4325}
\DefMacro{table-tdt-lang-lp3-th3}{1723}
\DefMacro{table-tdt-lang-lp3-th4}{749}
\DefMacro{table-tdt-lang-rp0-dup}{10}
\DefMacro{table-tdt-lang-rp0-fp}{2}
\DefMacro{table-tdt-lang-rp0-tp}{2}
\DefMacro{table-tdt-lang-rp0-fp-filtered}{16}
\DefMacro{table-tdt-lang-rp0-dup-filtered}{6}
\DefMacro{table-tdt-lang-rp0-th0}{9357}
\DefMacro{table-tdt-lang-rp0-th1}{2450}
\DefMacro{table-tdt-lang-rp0-th2}{541}
\DefMacro{table-tdt-lang-rp0-th3}{2}
\DefMacro{table-tdt-lang-rp0-th4}{18}
\DefMacro{table-tdt-lang-rp1-dup}{0}
\DefMacro{table-tdt-lang-rp1-fp}{0}
\DefMacro{table-tdt-lang-rp1-tp}{0}
\DefMacro{table-tdt-lang-rp1-fp-filtered}{0}
\DefMacro{table-tdt-lang-rp1-dup-filtered}{0}
\DefMacro{table-tdt-lang-rp1-th0}{10141}
\DefMacro{table-tdt-lang-rp1-th1}{1974}
\DefMacro{table-tdt-lang-rp1-th2}{253}
\DefMacro{table-tdt-lang-rp1-th3}{0}
\DefMacro{table-tdt-lang-rp1-th4}{0}
\DefMacro{table-tdt-math-tmpl}{656}
\DefMacro{table-tdt-math-gen}{5387}
\DefMacro{table-tdt-math-time}{1901.12}
\DefMacro{table-tdt-math-lp0-dup}{36}
\DefMacro{table-tdt-math-lp0-fp}{113}
\DefMacro{table-tdt-math-lp0-tp}{3}
\DefMacro{table-tdt-math-lp0-fp-filtered}{42}
\DefMacro{table-tdt-math-lp0-dup-filtered}{6}
\DefMacro{table-tdt-math-lp0-th0}{600}
\DefMacro{table-tdt-math-lp0-th1}{3150}
\DefMacro{table-tdt-math-lp0-th2}{1320}
\DefMacro{table-tdt-math-lp0-th3}{162}
\DefMacro{table-tdt-math-lp0-th4}{155}
\DefMacro{table-tdt-math-lp1-dup}{21}
\DefMacro{table-tdt-math-lp1-fp}{85}
\DefMacro{table-tdt-math-lp1-tp}{3}
\DefMacro{table-tdt-math-lp1-fp-filtered}{27}
\DefMacro{table-tdt-math-lp1-dup-filtered}{6}
\DefMacro{table-tdt-math-lp1-th0}{550}
\DefMacro{table-tdt-math-lp1-th1}{3169}
\DefMacro{table-tdt-math-lp1-th2}{1390}
\DefMacro{table-tdt-math-lp1-th3}{166}
\DefMacro{table-tdt-math-lp1-th4}{112}
\DefMacro{table-tdt-math-lp2-dup}{14}
\DefMacro{table-tdt-math-lp2-fp}{158}
\DefMacro{table-tdt-math-lp2-tp}{2}
\DefMacro{table-tdt-math-lp2-fp-filtered}{16}
\DefMacro{table-tdt-math-lp2-dup-filtered}{2}
\DefMacro{table-tdt-math-lp2-th0}{456}
\DefMacro{table-tdt-math-lp2-th1}{2447}
\DefMacro{table-tdt-math-lp2-th2}{1700}
\DefMacro{table-tdt-math-lp2-th3}{610}
\DefMacro{table-tdt-math-lp2-th4}{174}
\DefMacro{table-tdt-math-lp3-dup}{12}
\DefMacro{table-tdt-math-lp3-fp}{180}
\DefMacro{table-tdt-math-lp3-tp}{2}
\DefMacro{table-tdt-math-lp3-fp-filtered}{17}
\DefMacro{table-tdt-math-lp3-dup-filtered}{5}
\DefMacro{table-tdt-math-lp3-th0}{455}
\DefMacro{table-tdt-math-lp3-th1}{2398}
\DefMacro{table-tdt-math-lp3-th2}{1723}
\DefMacro{table-tdt-math-lp3-th3}{614}
\DefMacro{table-tdt-math-lp3-th4}{197}
\DefMacro{table-tdt-math-rp0-dup}{2}
\DefMacro{table-tdt-math-rp0-fp}{1}
\DefMacro{table-tdt-math-rp0-tp}{1}
\DefMacro{table-tdt-math-rp0-fp-filtered}{5}
\DefMacro{table-tdt-math-rp0-dup-filtered}{3}
\DefMacro{table-tdt-math-rp0-th0}{3994}
\DefMacro{table-tdt-math-rp0-th1}{1063}
\DefMacro{table-tdt-math-rp0-th2}{303}
\DefMacro{table-tdt-math-rp0-th3}{21}
\DefMacro{table-tdt-math-rp0-th4}{6}
\DefMacro{table-tdt-math-rp1-dup}{0}
\DefMacro{table-tdt-math-rp1-fp}{1}
\DefMacro{table-tdt-math-rp1-tp}{0}
\DefMacro{table-tdt-math-rp1-fp-filtered}{0}
\DefMacro{table-tdt-math-rp1-dup-filtered}{0}
\DefMacro{table-tdt-math-rp1-th0}{4346}
\DefMacro{table-tdt-math-rp1-th1}{834}
\DefMacro{table-tdt-math-rp1-th2}{197}
\DefMacro{table-tdt-math-rp1-th3}{9}
\DefMacro{table-tdt-math-rp1-th4}{1}
\DefMacro{table-tdt-statistics-tmpl}{1062}
\DefMacro{table-tdt-statistics-gen}{8567}
\DefMacro{table-tdt-statistics-time}{2296.09}
\DefMacro{table-tdt-statistics-lp0-dup}{0}
\DefMacro{table-tdt-statistics-lp0-fp}{0}
\DefMacro{table-tdt-statistics-lp0-tp}{0}
\DefMacro{table-tdt-statistics-lp0-fp-filtered}{0}
\DefMacro{table-tdt-statistics-lp0-dup-filtered}{0}
\DefMacro{table-tdt-statistics-lp0-th0}{657}
\DefMacro{table-tdt-statistics-lp0-th1}{5928}
\DefMacro{table-tdt-statistics-lp0-th2}{1959}
\DefMacro{table-tdt-statistics-lp0-th3}{23}
\DefMacro{table-tdt-statistics-lp0-th4}{0}
\DefMacro{table-tdt-statistics-lp1-dup}{0}
\DefMacro{table-tdt-statistics-lp1-fp}{0}
\DefMacro{table-tdt-statistics-lp1-tp}{0}
\DefMacro{table-tdt-statistics-lp1-fp-filtered}{0}
\DefMacro{table-tdt-statistics-lp1-dup-filtered}{0}
\DefMacro{table-tdt-statistics-lp1-th0}{630}
\DefMacro{table-tdt-statistics-lp1-th1}{6100}
\DefMacro{table-tdt-statistics-lp1-th2}{1813}
\DefMacro{table-tdt-statistics-lp1-th3}{24}
\DefMacro{table-tdt-statistics-lp1-th4}{0}
\DefMacro{table-tdt-statistics-lp2-dup}{0}
\DefMacro{table-tdt-statistics-lp2-fp}{0}
\DefMacro{table-tdt-statistics-lp2-tp}{0}
\DefMacro{table-tdt-statistics-lp2-fp-filtered}{0}
\DefMacro{table-tdt-statistics-lp2-dup-filtered}{0}
\DefMacro{table-tdt-statistics-lp2-th0}{477}
\DefMacro{table-tdt-statistics-lp2-th1}{4291}
\DefMacro{table-tdt-statistics-lp2-th2}{3202}
\DefMacro{table-tdt-statistics-lp2-th3}{597}
\DefMacro{table-tdt-statistics-lp2-th4}{0}
\DefMacro{table-tdt-statistics-lp3-dup}{0}
\DefMacro{table-tdt-statistics-lp3-fp}{0}
\DefMacro{table-tdt-statistics-lp3-tp}{0}
\DefMacro{table-tdt-statistics-lp3-fp-filtered}{0}
\DefMacro{table-tdt-statistics-lp3-dup-filtered}{0}
\DefMacro{table-tdt-statistics-lp3-th0}{473}
\DefMacro{table-tdt-statistics-lp3-th1}{4278}
\DefMacro{table-tdt-statistics-lp3-th2}{3277}
\DefMacro{table-tdt-statistics-lp3-th3}{539}
\DefMacro{table-tdt-statistics-lp3-th4}{0}
\DefMacro{table-tdt-statistics-rp0-dup}{0}
\DefMacro{table-tdt-statistics-rp0-fp}{0}
\DefMacro{table-tdt-statistics-rp0-tp}{0}
\DefMacro{table-tdt-statistics-rp0-fp-filtered}{0}
\DefMacro{table-tdt-statistics-rp0-dup-filtered}{0}
\DefMacro{table-tdt-statistics-rp0-th0}{6191}
\DefMacro{table-tdt-statistics-rp0-th1}{1787}
\DefMacro{table-tdt-statistics-rp0-th2}{566}
\DefMacro{table-tdt-statistics-rp0-th3}{23}
\DefMacro{table-tdt-statistics-rp0-th4}{0}
\DefMacro{table-tdt-statistics-rp1-dup}{0}
\DefMacro{table-tdt-statistics-rp1-fp}{0}
\DefMacro{table-tdt-statistics-rp1-tp}{0}
\DefMacro{table-tdt-statistics-rp1-fp-filtered}{0}
\DefMacro{table-tdt-statistics-rp1-dup-filtered}{0}
\DefMacro{table-tdt-statistics-rp1-th0}{6781}
\DefMacro{table-tdt-statistics-rp1-th1}{1452}
\DefMacro{table-tdt-statistics-rp1-th2}{324}
\DefMacro{table-tdt-statistics-rp1-th3}{10}
\DefMacro{table-tdt-statistics-rp1-th4}{0}
\DefMacro{table-tdt-text-tmpl}{1212}
\DefMacro{table-tdt-text-gen}{10592}
\DefMacro{table-tdt-text-time}{2330.62}
\DefMacro{table-tdt-text-lp0-dup}{0}
\DefMacro{table-tdt-text-lp0-fp}{65}
\DefMacro{table-tdt-text-lp0-tp}{0}
\DefMacro{table-tdt-text-lp0-fp-filtered}{0}
\DefMacro{table-tdt-text-lp0-dup-filtered}{0}
\DefMacro{table-tdt-text-lp0-th0}{3957}
\DefMacro{table-tdt-text-lp0-th1}{3870}
\DefMacro{table-tdt-text-lp0-th2}{2141}
\DefMacro{table-tdt-text-lp0-th3}{559}
\DefMacro{table-tdt-text-lp0-th4}{65}
\DefMacro{table-tdt-text-lp1-dup}{0}
\DefMacro{table-tdt-text-lp1-fp}{57}
\DefMacro{table-tdt-text-lp1-tp}{0}
\DefMacro{table-tdt-text-lp1-fp-filtered}{0}
\DefMacro{table-tdt-text-lp1-dup-filtered}{0}
\DefMacro{table-tdt-text-lp1-th0}{3842}
\DefMacro{table-tdt-text-lp1-th1}{3978}
\DefMacro{table-tdt-text-lp1-th2}{2004}
\DefMacro{table-tdt-text-lp1-th3}{711}
\DefMacro{table-tdt-text-lp1-th4}{57}
\DefMacro{table-tdt-text-lp2-dup}{0}
\DefMacro{table-tdt-text-lp2-fp}{1221}
\DefMacro{table-tdt-text-lp2-tp}{0}
\DefMacro{table-tdt-text-lp2-fp-filtered}{1}
\DefMacro{table-tdt-text-lp2-dup-filtered}{1}
\DefMacro{table-tdt-text-lp2-th0}{3107}
\DefMacro{table-tdt-text-lp2-th1}{3260}
\DefMacro{table-tdt-text-lp2-th2}{1972}
\DefMacro{table-tdt-text-lp2-th3}{1031}
\DefMacro{table-tdt-text-lp2-th4}{1222}
\DefMacro{table-tdt-text-lp3-dup}{0}
\DefMacro{table-tdt-text-lp3-fp}{1096}
\DefMacro{table-tdt-text-lp3-tp}{0}
\DefMacro{table-tdt-text-lp3-fp-filtered}{0}
\DefMacro{table-tdt-text-lp3-dup-filtered}{0}
\DefMacro{table-tdt-text-lp3-th0}{3132}
\DefMacro{table-tdt-text-lp3-th1}{3203}
\DefMacro{table-tdt-text-lp3-th2}{1936}
\DefMacro{table-tdt-text-lp3-th3}{1225}
\DefMacro{table-tdt-text-lp3-th4}{1096}
\DefMacro{table-tdt-text-rp0-dup}{12}
\DefMacro{table-tdt-text-rp0-fp}{25}
\DefMacro{table-tdt-text-rp0-tp}{1}
\DefMacro{table-tdt-text-rp0-fp-filtered}{28}
\DefMacro{table-tdt-text-rp0-dup-filtered}{16}
\DefMacro{table-tdt-text-rp0-th0}{8430}
\DefMacro{table-tdt-text-rp0-th1}{1662}
\DefMacro{table-tdt-text-rp0-th2}{427}
\DefMacro{table-tdt-text-rp0-th3}{20}
\DefMacro{table-tdt-text-rp0-th4}{53}
\DefMacro{table-tdt-text-rp1-dup}{0}
\DefMacro{table-tdt-text-rp1-fp}{0}
\DefMacro{table-tdt-text-rp1-tp}{0}
\DefMacro{table-tdt-text-rp1-fp-filtered}{0}
\DefMacro{table-tdt-text-rp1-dup-filtered}{0}
\DefMacro{table-tdt-text-rp1-th0}{8723}
\DefMacro{table-tdt-text-rp1-th1}{1621}
\DefMacro{table-tdt-text-rp1-th2}{248}
\DefMacro{table-tdt-text-rp1-th3}{0}
\DefMacro{table-tdt-text-rp1-th4}{0}
\DefMacro{table-tdt-vectorz-tmpl}{785}
\DefMacro{table-tdt-vectorz-gen}{6255}
\DefMacro{table-tdt-vectorz-time}{2123.96}
\DefMacro{table-tdt-vectorz-lp0-dup}{18}
\DefMacro{table-tdt-vectorz-lp0-fp}{67}
\DefMacro{table-tdt-vectorz-lp0-tp}{2}
\DefMacro{table-tdt-vectorz-lp0-fp-filtered}{20}
\DefMacro{table-tdt-vectorz-lp0-dup-filtered}{2}
\DefMacro{table-tdt-vectorz-lp0-th0}{1462}
\DefMacro{table-tdt-vectorz-lp0-th1}{3229}
\DefMacro{table-tdt-vectorz-lp0-th2}{1208}
\DefMacro{table-tdt-vectorz-lp0-th3}{269}
\DefMacro{table-tdt-vectorz-lp0-th4}{87}
\DefMacro{table-tdt-vectorz-lp1-dup}{11}
\DefMacro{table-tdt-vectorz-lp1-fp}{73}
\DefMacro{table-tdt-vectorz-lp1-tp}{2}
\DefMacro{table-tdt-vectorz-lp1-fp-filtered}{13}
\DefMacro{table-tdt-vectorz-lp1-dup-filtered}{2}
\DefMacro{table-tdt-vectorz-lp1-th0}{1393}
\DefMacro{table-tdt-vectorz-lp1-th1}{3344}
\DefMacro{table-tdt-vectorz-lp1-th2}{1203}
\DefMacro{table-tdt-vectorz-lp1-th3}{229}
\DefMacro{table-tdt-vectorz-lp1-th4}{86}
\DefMacro{table-tdt-vectorz-lp2-dup}{9}
\DefMacro{table-tdt-vectorz-lp2-fp}{411}
\DefMacro{table-tdt-vectorz-lp2-tp}{2}
\DefMacro{table-tdt-vectorz-lp2-fp-filtered}{11}
\DefMacro{table-tdt-vectorz-lp2-dup-filtered}{2}
\DefMacro{table-tdt-vectorz-lp2-th0}{1108}
\DefMacro{table-tdt-vectorz-lp2-th1}{2653}
\DefMacro{table-tdt-vectorz-lp2-th2}{1300}
\DefMacro{table-tdt-vectorz-lp2-th3}{772}
\DefMacro{table-tdt-vectorz-lp2-th4}{422}
\DefMacro{table-tdt-vectorz-lp3-dup}{14}
\DefMacro{table-tdt-vectorz-lp3-fp}{416}
\DefMacro{table-tdt-vectorz-lp3-tp}{2}
\DefMacro{table-tdt-vectorz-lp3-fp-filtered}{16}
\DefMacro{table-tdt-vectorz-lp3-dup-filtered}{2}
\DefMacro{table-tdt-vectorz-lp3-th0}{1102}
\DefMacro{table-tdt-vectorz-lp3-th1}{2652}
\DefMacro{table-tdt-vectorz-lp3-th2}{1327}
\DefMacro{table-tdt-vectorz-lp3-th3}{742}
\DefMacro{table-tdt-vectorz-lp3-th4}{432}
\DefMacro{table-tdt-vectorz-rp0-dup}{0}
\DefMacro{table-tdt-vectorz-rp0-fp}{0}
\DefMacro{table-tdt-vectorz-rp0-tp}{0}
\DefMacro{table-tdt-vectorz-rp0-fp-filtered}{0}
\DefMacro{table-tdt-vectorz-rp0-dup-filtered}{0}
\DefMacro{table-tdt-vectorz-rp0-th0}{4786}
\DefMacro{table-tdt-vectorz-rp0-th1}{1149}
\DefMacro{table-tdt-vectorz-rp0-th2}{307}
\DefMacro{table-tdt-vectorz-rp0-th3}{13}
\DefMacro{table-tdt-vectorz-rp0-th4}{0}
\DefMacro{table-tdt-vectorz-rp1-dup}{0}
\DefMacro{table-tdt-vectorz-rp1-fp}{0}
\DefMacro{table-tdt-vectorz-rp1-tp}{0}
\DefMacro{table-tdt-vectorz-rp1-fp-filtered}{0}
\DefMacro{table-tdt-vectorz-rp1-dup-filtered}{0}
\DefMacro{table-tdt-vectorz-rp1-th0}{5123}
\DefMacro{table-tdt-vectorz-rp1-th1}{945}
\DefMacro{table-tdt-vectorz-rp1-th2}{174}
\DefMacro{table-tdt-vectorz-rp1-th3}{13}
\DefMacro{table-tdt-vectorz-rp1-th4}{0}
\DefMacro{table-tdt-zxing-tmpl}{1182}
\DefMacro{table-tdt-zxing-gen}{10326}
\DefMacro{table-tdt-zxing-time}{3208.59}
\DefMacro{table-tdt-zxing-lp0-dup}{85}
\DefMacro{table-tdt-zxing-lp0-fp}{215}
\DefMacro{table-tdt-zxing-lp0-tp}{1}
\DefMacro{table-tdt-zxing-lp0-fp-filtered}{86}
\DefMacro{table-tdt-zxing-lp0-dup-filtered}{1}
\DefMacro{table-tdt-zxing-lp0-th0}{2316}
\DefMacro{table-tdt-zxing-lp0-th1}{4861}
\DefMacro{table-tdt-zxing-lp0-th2}{2278}
\DefMacro{table-tdt-zxing-lp0-th3}{570}
\DefMacro{table-tdt-zxing-lp0-th4}{301}
\DefMacro{table-tdt-zxing-lp1-dup}{115}
\DefMacro{table-tdt-zxing-lp1-fp}{197}
\DefMacro{table-tdt-zxing-lp1-tp}{1}
\DefMacro{table-tdt-zxing-lp1-fp-filtered}{116}
\DefMacro{table-tdt-zxing-lp1-dup-filtered}{1}
\DefMacro{table-tdt-zxing-lp1-th0}{2489}
\DefMacro{table-tdt-zxing-lp1-th1}{4727}
\DefMacro{table-tdt-zxing-lp1-th2}{2247}
\DefMacro{table-tdt-zxing-lp1-th3}{550}
\DefMacro{table-tdt-zxing-lp1-th4}{313}
\DefMacro{table-tdt-zxing-lp2-dup}{137}
\DefMacro{table-tdt-zxing-lp2-fp}{752}
\DefMacro{table-tdt-zxing-lp2-tp}{1}
\DefMacro{table-tdt-zxing-lp2-fp-filtered}{138}
\DefMacro{table-tdt-zxing-lp2-dup-filtered}{1}
\DefMacro{table-tdt-zxing-lp2-th0}{1520}
\DefMacro{table-tdt-zxing-lp2-th1}{3559}
\DefMacro{table-tdt-zxing-lp2-th2}{2743}
\DefMacro{table-tdt-zxing-lp2-th3}{1614}
\DefMacro{table-tdt-zxing-lp2-th4}{890}
\DefMacro{table-tdt-zxing-lp3-dup}{131}
\DefMacro{table-tdt-zxing-lp3-fp}{912}
\DefMacro{table-tdt-zxing-lp3-tp}{1}
\DefMacro{table-tdt-zxing-lp3-fp-filtered}{132}
\DefMacro{table-tdt-zxing-lp3-dup-filtered}{1}
\DefMacro{table-tdt-zxing-lp3-th0}{1512}
\DefMacro{table-tdt-zxing-lp3-th1}{3639}
\DefMacro{table-tdt-zxing-lp3-th2}{2673}
\DefMacro{table-tdt-zxing-lp3-th3}{1458}
\DefMacro{table-tdt-zxing-lp3-th4}{1044}
\DefMacro{table-tdt-zxing-rp0-dup}{0}
\DefMacro{table-tdt-zxing-rp0-fp}{0}
\DefMacro{table-tdt-zxing-rp0-tp}{0}
\DefMacro{table-tdt-zxing-rp0-fp-filtered}{0}
\DefMacro{table-tdt-zxing-rp0-dup-filtered}{0}
\DefMacro{table-tdt-zxing-rp0-th0}{7936}
\DefMacro{table-tdt-zxing-rp0-th1}{1866}
\DefMacro{table-tdt-zxing-rp0-th2}{508}
\DefMacro{table-tdt-zxing-rp0-th3}{16}
\DefMacro{table-tdt-zxing-rp0-th4}{0}
\DefMacro{table-tdt-zxing-rp1-dup}{0}
\DefMacro{table-tdt-zxing-rp1-fp}{0}
\DefMacro{table-tdt-zxing-rp1-tp}{0}
\DefMacro{table-tdt-zxing-rp1-fp-filtered}{0}
\DefMacro{table-tdt-zxing-rp1-dup-filtered}{0}
\DefMacro{table-tdt-zxing-rp1-th0}{8271}
\DefMacro{table-tdt-zxing-rp1-th1}{1642}
\DefMacro{table-tdt-zxing-rp1-th2}{327}
\DefMacro{table-tdt-zxing-rp1-th3}{86}
\DefMacro{table-tdt-zxing-rp1-th4}{0}

\DefMacro{table-tdt-topk500_100_50-codec-tmpl}{1157}
\DefMacro{table-tdt-topk500_100_50-codec-gen}{6355}
\DefMacro{table-tdt-topk500_100_50-codec-time}{264.09}
\DefMacro{table-tdt-topk500_100_50-codec-lp0-dup}{3}
\DefMacro{table-tdt-topk500_100_50-codec-lp0-fp}{27}
\DefMacro{table-tdt-topk500_100_50-codec-lp0-tp}{3}
\DefMacro{table-tdt-topk500_100_50-codec-lp0-th0}{1824}
\DefMacro{table-tdt-topk500_100_50-codec-lp0-th1}{4353}
\DefMacro{table-tdt-topk500_100_50-codec-lp0-th2}{110}
\DefMacro{table-tdt-topk500_100_50-codec-lp0-th3}{35}
\DefMacro{table-tdt-topk500_100_50-codec-lp0-th4}{33}
\DefMacro{table-tdt-topk500_100_50-codec-lp1-dup}{1}
\DefMacro{table-tdt-topk500_100_50-codec-lp1-fp}{31}
\DefMacro{table-tdt-topk500_100_50-codec-lp1-tp}{1}
\DefMacro{table-tdt-topk500_100_50-codec-lp1-th0}{1804}
\DefMacro{table-tdt-topk500_100_50-codec-lp1-th1}{4383}
\DefMacro{table-tdt-topk500_100_50-codec-lp1-th2}{108}
\DefMacro{table-tdt-topk500_100_50-codec-lp1-th3}{27}
\DefMacro{table-tdt-topk500_100_50-codec-lp1-th4}{33}
\DefMacro{table-tdt-topk500_100_50-codec-lp2-dup}{0}
\DefMacro{table-tdt-topk500_100_50-codec-lp2-fp}{30}
\DefMacro{table-tdt-topk500_100_50-codec-lp2-tp}{1}
\DefMacro{table-tdt-topk500_100_50-codec-lp2-th0}{1417}
\DefMacro{table-tdt-topk500_100_50-codec-lp2-th1}{4697}
\DefMacro{table-tdt-topk500_100_50-codec-lp2-th2}{177}
\DefMacro{table-tdt-topk500_100_50-codec-lp2-th3}{33}
\DefMacro{table-tdt-topk500_100_50-codec-lp2-th4}{31}
\DefMacro{table-tdt-topk500_100_50-codec-lp3-dup}{1}
\DefMacro{table-tdt-topk500_100_50-codec-lp3-fp}{31}
\DefMacro{table-tdt-topk500_100_50-codec-lp3-tp}{2}
\DefMacro{table-tdt-topk500_100_50-codec-lp3-th0}{1403}
\DefMacro{table-tdt-topk500_100_50-codec-lp3-th1}{4723}
\DefMacro{table-tdt-topk500_100_50-codec-lp3-th2}{160}
\DefMacro{table-tdt-topk500_100_50-codec-lp3-th3}{35}
\DefMacro{table-tdt-topk500_100_50-codec-lp3-th4}{34}
\DefMacro{table-tdt-topk500_100_50-codec-rp0-dup}{0}
\DefMacro{table-tdt-topk500_100_50-codec-rp0-fp}{0}
\DefMacro{table-tdt-topk500_100_50-codec-rp0-tp}{0}
\DefMacro{table-tdt-topk500_100_50-codec-rp0-th0}{4926}
\DefMacro{table-tdt-topk500_100_50-codec-rp0-th1}{1361}
\DefMacro{table-tdt-topk500_100_50-codec-rp0-th2}{67}
\DefMacro{table-tdt-topk500_100_50-codec-rp0-th3}{1}
\DefMacro{table-tdt-topk500_100_50-codec-rp0-th4}{0}
\DefMacro{table-tdt-topk500_100_50-codec-rp1-dup}{0}
\DefMacro{table-tdt-topk500_100_50-codec-rp1-fp}{0}
\DefMacro{table-tdt-topk500_100_50-codec-rp1-tp}{0}
\DefMacro{table-tdt-topk500_100_50-codec-rp1-th0}{5202}
\DefMacro{table-tdt-topk500_100_50-codec-rp1-th1}{1097}
\DefMacro{table-tdt-topk500_100_50-codec-rp1-th2}{53}
\DefMacro{table-tdt-topk500_100_50-codec-rp1-th3}{3}
\DefMacro{table-tdt-topk500_100_50-codec-rp1-th4}{0}
\DefMacro{table-tdt-topk500_100_50-compress-tmpl}{792}
\DefMacro{table-tdt-topk500_100_50-compress-gen}{6582}
\DefMacro{table-tdt-topk500_100_50-compress-time}{187.46}
\DefMacro{table-tdt-topk500_100_50-compress-lp0-dup}{0}
\DefMacro{table-tdt-topk500_100_50-compress-lp0-fp}{36}
\DefMacro{table-tdt-topk500_100_50-compress-lp0-tp}{0}
\DefMacro{table-tdt-topk500_100_50-compress-lp0-th0}{860}
\DefMacro{table-tdt-topk500_100_50-compress-lp0-th1}{5543}
\DefMacro{table-tdt-topk500_100_50-compress-lp0-th2}{113}
\DefMacro{table-tdt-topk500_100_50-compress-lp0-th3}{30}
\DefMacro{table-tdt-topk500_100_50-compress-lp0-th4}{36}
\DefMacro{table-tdt-topk500_100_50-compress-lp1-dup}{0}
\DefMacro{table-tdt-topk500_100_50-compress-lp1-fp}{47}
\DefMacro{table-tdt-topk500_100_50-compress-lp1-tp}{0}
\DefMacro{table-tdt-topk500_100_50-compress-lp1-th0}{848}
\DefMacro{table-tdt-topk500_100_50-compress-lp1-th1}{5533}
\DefMacro{table-tdt-topk500_100_50-compress-lp1-th2}{127}
\DefMacro{table-tdt-topk500_100_50-compress-lp1-th3}{27}
\DefMacro{table-tdt-topk500_100_50-compress-lp1-th4}{47}
\DefMacro{table-tdt-topk500_100_50-compress-lp2-dup}{0}
\DefMacro{table-tdt-topk500_100_50-compress-lp2-fp}{40}
\DefMacro{table-tdt-topk500_100_50-compress-lp2-tp}{0}
\DefMacro{table-tdt-topk500_100_50-compress-lp2-th0}{586}
\DefMacro{table-tdt-topk500_100_50-compress-lp2-th1}{5715}
\DefMacro{table-tdt-topk500_100_50-compress-lp2-th2}{201}
\DefMacro{table-tdt-topk500_100_50-compress-lp2-th3}{40}
\DefMacro{table-tdt-topk500_100_50-compress-lp2-th4}{40}
\DefMacro{table-tdt-topk500_100_50-compress-lp3-dup}{0}
\DefMacro{table-tdt-topk500_100_50-compress-lp3-fp}{37}
\DefMacro{table-tdt-topk500_100_50-compress-lp3-tp}{0}
\DefMacro{table-tdt-topk500_100_50-compress-lp3-th0}{560}
\DefMacro{table-tdt-topk500_100_50-compress-lp3-th1}{5751}
\DefMacro{table-tdt-topk500_100_50-compress-lp3-th2}{191}
\DefMacro{table-tdt-topk500_100_50-compress-lp3-th3}{43}
\DefMacro{table-tdt-topk500_100_50-compress-lp3-th4}{37}
\DefMacro{table-tdt-topk500_100_50-compress-rp0-dup}{0}
\DefMacro{table-tdt-topk500_100_50-compress-rp0-fp}{0}
\DefMacro{table-tdt-topk500_100_50-compress-rp0-tp}{0}
\DefMacro{table-tdt-topk500_100_50-compress-rp0-th0}{4578}
\DefMacro{table-tdt-topk500_100_50-compress-rp0-th1}{1850}
\DefMacro{table-tdt-topk500_100_50-compress-rp0-th2}{142}
\DefMacro{table-tdt-topk500_100_50-compress-rp0-th3}{12}
\DefMacro{table-tdt-topk500_100_50-compress-rp0-th4}{0}
\DefMacro{table-tdt-topk500_100_50-compress-rp1-dup}{0}
\DefMacro{table-tdt-topk500_100_50-compress-rp1-fp}{0}
\DefMacro{table-tdt-topk500_100_50-compress-rp1-tp}{0}
\DefMacro{table-tdt-topk500_100_50-compress-rp1-th0}{5249}
\DefMacro{table-tdt-topk500_100_50-compress-rp1-th1}{1218}
\DefMacro{table-tdt-topk500_100_50-compress-rp1-th2}{112}
\DefMacro{table-tdt-topk500_100_50-compress-rp1-th3}{3}
\DefMacro{table-tdt-topk500_100_50-compress-rp1-th4}{0}
\DefMacro{table-tdt-topk500_100_50-math-tmpl}{656}
\DefMacro{table-tdt-topk500_100_50-math-gen}{5387}
\DefMacro{table-tdt-topk500_100_50-math-time}{212.78}
\DefMacro{table-tdt-topk500_100_50-math-lp0-dup}{24}
\DefMacro{table-tdt-topk500_100_50-math-lp0-fp}{20}
\DefMacro{table-tdt-topk500_100_50-math-lp0-tp}{5}
\DefMacro{table-tdt-topk500_100_50-math-lp0-th0}{600}
\DefMacro{table-tdt-topk500_100_50-math-lp0-th1}{4568}
\DefMacro{table-tdt-topk500_100_50-math-lp0-th2}{122}
\DefMacro{table-tdt-topk500_100_50-math-lp0-th3}{48}
\DefMacro{table-tdt-topk500_100_50-math-lp0-th4}{49}
\DefMacro{table-tdt-topk500_100_50-math-lp1-dup}{9}
\DefMacro{table-tdt-topk500_100_50-math-lp1-fp}{31}
\DefMacro{table-tdt-topk500_100_50-math-lp1-tp}{5}
\DefMacro{table-tdt-topk500_100_50-math-lp1-th0}{550}
\DefMacro{table-tdt-topk500_100_50-math-lp1-th1}{4620}
\DefMacro{table-tdt-topk500_100_50-math-lp1-th2}{153}
\DefMacro{table-tdt-topk500_100_50-math-lp1-th3}{19}
\DefMacro{table-tdt-topk500_100_50-math-lp1-th4}{45}
\DefMacro{table-tdt-topk500_100_50-math-lp2-dup}{3}
\DefMacro{table-tdt-topk500_100_50-math-lp2-fp}{21}
\DefMacro{table-tdt-topk500_100_50-math-lp2-tp}{3}
\DefMacro{table-tdt-topk500_100_50-math-lp2-th0}{456}
\DefMacro{table-tdt-topk500_100_50-math-lp2-th1}{4661}
\DefMacro{table-tdt-topk500_100_50-math-lp2-th2}{208}
\DefMacro{table-tdt-topk500_100_50-math-lp2-th3}{35}
\DefMacro{table-tdt-topk500_100_50-math-lp2-th4}{27}
\DefMacro{table-tdt-topk500_100_50-math-lp3-dup}{2}
\DefMacro{table-tdt-topk500_100_50-math-lp3-fp}{24}
\DefMacro{table-tdt-topk500_100_50-math-lp3-tp}{2}
\DefMacro{table-tdt-topk500_100_50-math-lp3-th0}{455}
\DefMacro{table-tdt-topk500_100_50-math-lp3-th1}{4654}
\DefMacro{table-tdt-topk500_100_50-math-lp3-th2}{215}
\DefMacro{table-tdt-topk500_100_50-math-lp3-th3}{35}
\DefMacro{table-tdt-topk500_100_50-math-lp3-th4}{28}
\DefMacro{table-tdt-topk500_100_50-math-rp0-dup}{1}
\DefMacro{table-tdt-topk500_100_50-math-rp0-fp}{2}
\DefMacro{table-tdt-topk500_100_50-math-rp0-tp}{3}
\DefMacro{table-tdt-topk500_100_50-math-rp0-th0}{3994}
\DefMacro{table-tdt-topk500_100_50-math-rp0-th1}{1268}
\DefMacro{table-tdt-topk500_100_50-math-rp0-th2}{116}
\DefMacro{table-tdt-topk500_100_50-math-rp0-th3}{3}
\DefMacro{table-tdt-topk500_100_50-math-rp0-th4}{6}
\DefMacro{table-tdt-topk500_100_50-math-rp1-dup}{0}
\DefMacro{table-tdt-topk500_100_50-math-rp1-fp}{0}
\DefMacro{table-tdt-topk500_100_50-math-rp1-tp}{0}
\DefMacro{table-tdt-topk500_100_50-math-rp1-th0}{4346}
\DefMacro{table-tdt-topk500_100_50-math-rp1-th1}{947}
\DefMacro{table-tdt-topk500_100_50-math-rp1-th2}{92}
\DefMacro{table-tdt-topk500_100_50-math-rp1-th3}{2}
\DefMacro{table-tdt-topk500_100_50-math-rp1-th4}{0}
\DefMacro{table-tdt-topk500_100_50-statistics-tmpl}{1062}
\DefMacro{table-tdt-topk500_100_50-statistics-gen}{8567}
\DefMacro{table-tdt-topk500_100_50-statistics-time}{133.10}
\DefMacro{table-tdt-topk500_100_50-statistics-lp0-dup}{0}
\DefMacro{table-tdt-topk500_100_50-statistics-lp0-fp}{0}
\DefMacro{table-tdt-topk500_100_50-statistics-lp0-tp}{0}
\DefMacro{table-tdt-topk500_100_50-statistics-lp0-th0}{657}
\DefMacro{table-tdt-topk500_100_50-statistics-lp0-th1}{7704}
\DefMacro{table-tdt-topk500_100_50-statistics-lp0-th2}{204}
\DefMacro{table-tdt-topk500_100_50-statistics-lp0-th3}{2}
\DefMacro{table-tdt-topk500_100_50-statistics-lp0-th4}{0}
\DefMacro{table-tdt-topk500_100_50-statistics-lp1-dup}{0}
\DefMacro{table-tdt-topk500_100_50-statistics-lp1-fp}{0}
\DefMacro{table-tdt-topk500_100_50-statistics-lp1-tp}{0}
\DefMacro{table-tdt-topk500_100_50-statistics-lp1-th0}{630}
\DefMacro{table-tdt-topk500_100_50-statistics-lp1-th1}{7724}
\DefMacro{table-tdt-topk500_100_50-statistics-lp1-th2}{210}
\DefMacro{table-tdt-topk500_100_50-statistics-lp1-th3}{3}
\DefMacro{table-tdt-topk500_100_50-statistics-lp1-th4}{0}
\DefMacro{table-tdt-topk500_100_50-statistics-lp2-dup}{0}
\DefMacro{table-tdt-topk500_100_50-statistics-lp2-fp}{0}
\DefMacro{table-tdt-topk500_100_50-statistics-lp2-tp}{0}
\DefMacro{table-tdt-topk500_100_50-statistics-lp2-th0}{477}
\DefMacro{table-tdt-topk500_100_50-statistics-lp2-th1}{7827}
\DefMacro{table-tdt-topk500_100_50-statistics-lp2-th2}{240}
\DefMacro{table-tdt-topk500_100_50-statistics-lp2-th3}{23}
\DefMacro{table-tdt-topk500_100_50-statistics-lp2-th4}{0}
\DefMacro{table-tdt-topk500_100_50-statistics-lp3-dup}{0}
\DefMacro{table-tdt-topk500_100_50-statistics-lp3-fp}{0}
\DefMacro{table-tdt-topk500_100_50-statistics-lp3-tp}{0}
\DefMacro{table-tdt-topk500_100_50-statistics-lp3-th0}{473}
\DefMacro{table-tdt-topk500_100_50-statistics-lp3-th1}{7806}
\DefMacro{table-tdt-topk500_100_50-statistics-lp3-th2}{276}
\DefMacro{table-tdt-topk500_100_50-statistics-lp3-th3}{12}
\DefMacro{table-tdt-topk500_100_50-statistics-lp3-th4}{0}
\DefMacro{table-tdt-topk500_100_50-statistics-rp0-dup}{0}
\DefMacro{table-tdt-topk500_100_50-statistics-rp0-fp}{0}
\DefMacro{table-tdt-topk500_100_50-statistics-rp0-tp}{0}
\DefMacro{table-tdt-topk500_100_50-statistics-rp0-th0}{6191}
\DefMacro{table-tdt-topk500_100_50-statistics-rp0-th1}{2245}
\DefMacro{table-tdt-topk500_100_50-statistics-rp0-th2}{128}
\DefMacro{table-tdt-topk500_100_50-statistics-rp0-th3}{3}
\DefMacro{table-tdt-topk500_100_50-statistics-rp0-th4}{0}
\DefMacro{table-tdt-topk500_100_50-statistics-rp1-dup}{0}
\DefMacro{table-tdt-topk500_100_50-statistics-rp1-fp}{0}
\DefMacro{table-tdt-topk500_100_50-statistics-rp1-tp}{0}
\DefMacro{table-tdt-topk500_100_50-statistics-rp1-th0}{6781}
\DefMacro{table-tdt-topk500_100_50-statistics-rp1-th1}{1684}
\DefMacro{table-tdt-topk500_100_50-statistics-rp1-th2}{97}
\DefMacro{table-tdt-topk500_100_50-statistics-rp1-th3}{5}
\DefMacro{table-tdt-topk500_100_50-statistics-rp1-th4}{0}
\DefMacro{table-tdt-topk500_100_50-text-tmpl}{1212}
\DefMacro{table-tdt-topk500_100_50-text-gen}{10592}
\DefMacro{table-tdt-topk500_100_50-text-time}{171.57}
\DefMacro{table-tdt-topk500_100_50-text-lp0-dup}{0}
\DefMacro{table-tdt-topk500_100_50-text-lp0-fp}{0}
\DefMacro{table-tdt-topk500_100_50-text-lp0-tp}{0}
\DefMacro{table-tdt-topk500_100_50-text-lp0-th0}{3957}
\DefMacro{table-tdt-topk500_100_50-text-lp0-th1}{6509}
\DefMacro{table-tdt-topk500_100_50-text-lp0-th2}{124}
\DefMacro{table-tdt-topk500_100_50-text-lp0-th3}{2}
\DefMacro{table-tdt-topk500_100_50-text-lp0-th4}{0}
\DefMacro{table-tdt-topk500_100_50-text-lp1-dup}{0}
\DefMacro{table-tdt-topk500_100_50-text-lp1-fp}{1}
\DefMacro{table-tdt-topk500_100_50-text-lp1-tp}{0}
\DefMacro{table-tdt-topk500_100_50-text-lp1-th0}{3842}
\DefMacro{table-tdt-topk500_100_50-text-lp1-th1}{6584}
\DefMacro{table-tdt-topk500_100_50-text-lp1-th2}{142}
\DefMacro{table-tdt-topk500_100_50-text-lp1-th3}{23}
\DefMacro{table-tdt-topk500_100_50-text-lp1-th4}{1}
\DefMacro{table-tdt-topk500_100_50-text-lp2-dup}{0}
\DefMacro{table-tdt-topk500_100_50-text-lp2-fp}{7}
\DefMacro{table-tdt-topk500_100_50-text-lp2-tp}{0}
\DefMacro{table-tdt-topk500_100_50-text-lp2-th0}{3107}
\DefMacro{table-tdt-topk500_100_50-text-lp2-th1}{7252}
\DefMacro{table-tdt-topk500_100_50-text-lp2-th2}{185}
\DefMacro{table-tdt-topk500_100_50-text-lp2-th3}{41}
\DefMacro{table-tdt-topk500_100_50-text-lp2-th4}{7}
\DefMacro{table-tdt-topk500_100_50-text-lp3-dup}{0}
\DefMacro{table-tdt-topk500_100_50-text-lp3-fp}{9}
\DefMacro{table-tdt-topk500_100_50-text-lp3-tp}{0}
\DefMacro{table-tdt-topk500_100_50-text-lp3-th0}{3132}
\DefMacro{table-tdt-topk500_100_50-text-lp3-th1}{7230}
\DefMacro{table-tdt-topk500_100_50-text-lp3-th2}{171}
\DefMacro{table-tdt-topk500_100_50-text-lp3-th3}{50}
\DefMacro{table-tdt-topk500_100_50-text-lp3-th4}{9}
\DefMacro{table-tdt-topk500_100_50-text-rp0-dup}{0}
\DefMacro{table-tdt-topk500_100_50-text-rp0-fp}{18}
\DefMacro{table-tdt-topk500_100_50-text-rp0-tp}{4}
\DefMacro{table-tdt-topk500_100_50-text-rp0-th0}{8430}
\DefMacro{table-tdt-topk500_100_50-text-rp0-th1}{2101}
\DefMacro{table-tdt-topk500_100_50-text-rp0-th2}{37}
\DefMacro{table-tdt-topk500_100_50-text-rp0-th3}{2}
\DefMacro{table-tdt-topk500_100_50-text-rp0-th4}{22}
\DefMacro{table-tdt-topk500_100_50-text-rp1-dup}{0}
\DefMacro{table-tdt-topk500_100_50-text-rp1-fp}{0}
\DefMacro{table-tdt-topk500_100_50-text-rp1-tp}{0}
\DefMacro{table-tdt-topk500_100_50-text-rp1-th0}{8723}
\DefMacro{table-tdt-topk500_100_50-text-rp1-th1}{1856}
\DefMacro{table-tdt-topk500_100_50-text-rp1-th2}{13}
\DefMacro{table-tdt-topk500_100_50-text-rp1-th3}{0}
\DefMacro{table-tdt-topk500_100_50-text-rp1-th4}{0}
\DefMacro{table-tdt-topk500_100_50-vectorz-tmpl}{785}
\DefMacro{table-tdt-topk500_100_50-vectorz-gen}{6255}
\DefMacro{table-tdt-topk500_100_50-vectorz-time}{226.77}
\DefMacro{table-tdt-topk500_100_50-vectorz-lp0-dup}{0}
\DefMacro{table-tdt-topk500_100_50-vectorz-lp0-fp}{24}
\DefMacro{table-tdt-topk500_100_50-vectorz-lp0-tp}{0}
\DefMacro{table-tdt-topk500_100_50-vectorz-lp0-th0}{1462}
\DefMacro{table-tdt-topk500_100_50-vectorz-lp0-th1}{4595}
\DefMacro{table-tdt-topk500_100_50-vectorz-lp0-th2}{161}
\DefMacro{table-tdt-topk500_100_50-vectorz-lp0-th3}{13}
\DefMacro{table-tdt-topk500_100_50-vectorz-lp0-th4}{24}
\DefMacro{table-tdt-topk500_100_50-vectorz-lp1-dup}{0}
\DefMacro{table-tdt-topk500_100_50-vectorz-lp1-fp}{24}
\DefMacro{table-tdt-topk500_100_50-vectorz-lp1-tp}{0}
\DefMacro{table-tdt-topk500_100_50-vectorz-lp1-th0}{1393}
\DefMacro{table-tdt-topk500_100_50-vectorz-lp1-th1}{4664}
\DefMacro{table-tdt-topk500_100_50-vectorz-lp1-th2}{158}
\DefMacro{table-tdt-topk500_100_50-vectorz-lp1-th3}{16}
\DefMacro{table-tdt-topk500_100_50-vectorz-lp1-th4}{24}
\DefMacro{table-tdt-topk500_100_50-vectorz-lp2-dup}{0}
\DefMacro{table-tdt-topk500_100_50-vectorz-lp2-fp}{31}
\DefMacro{table-tdt-topk500_100_50-vectorz-lp2-tp}{0}
\DefMacro{table-tdt-topk500_100_50-vectorz-lp2-th0}{1108}
\DefMacro{table-tdt-topk500_100_50-vectorz-lp2-th1}{4910}
\DefMacro{table-tdt-topk500_100_50-vectorz-lp2-th2}{161}
\DefMacro{table-tdt-topk500_100_50-vectorz-lp2-th3}{45}
\DefMacro{table-tdt-topk500_100_50-vectorz-lp2-th4}{31}
\DefMacro{table-tdt-topk500_100_50-vectorz-lp3-dup}{0}
\DefMacro{table-tdt-topk500_100_50-vectorz-lp3-fp}{28}
\DefMacro{table-tdt-topk500_100_50-vectorz-lp3-tp}{0}
\DefMacro{table-tdt-topk500_100_50-vectorz-lp3-th0}{1102}
\DefMacro{table-tdt-topk500_100_50-vectorz-lp3-th1}{4904}
\DefMacro{table-tdt-topk500_100_50-vectorz-lp3-th2}{174}
\DefMacro{table-tdt-topk500_100_50-vectorz-lp3-th3}{47}
\DefMacro{table-tdt-topk500_100_50-vectorz-lp3-th4}{28}
\DefMacro{table-tdt-topk500_100_50-vectorz-rp0-dup}{0}
\DefMacro{table-tdt-topk500_100_50-vectorz-rp0-fp}{0}
\DefMacro{table-tdt-topk500_100_50-vectorz-rp0-tp}{0}
\DefMacro{table-tdt-topk500_100_50-vectorz-rp0-th0}{4786}
\DefMacro{table-tdt-topk500_100_50-vectorz-rp0-th1}{1371}
\DefMacro{table-tdt-topk500_100_50-vectorz-rp0-th2}{96}
\DefMacro{table-tdt-topk500_100_50-vectorz-rp0-th3}{2}
\DefMacro{table-tdt-topk500_100_50-vectorz-rp0-th4}{0}
\DefMacro{table-tdt-topk500_100_50-vectorz-rp1-dup}{0}
\DefMacro{table-tdt-topk500_100_50-vectorz-rp1-fp}{0}
\DefMacro{table-tdt-topk500_100_50-vectorz-rp1-tp}{0}
\DefMacro{table-tdt-topk500_100_50-vectorz-rp1-th0}{5123}
\DefMacro{table-tdt-topk500_100_50-vectorz-rp1-th1}{1060}
\DefMacro{table-tdt-topk500_100_50-vectorz-rp1-th2}{71}
\DefMacro{table-tdt-topk500_100_50-vectorz-rp1-th3}{1}
\DefMacro{table-tdt-topk500_100_50-vectorz-rp1-th4}{0}
\DefMacro{table-tdt-topk500_100_50-zxing-tmpl}{1182}
\DefMacro{table-tdt-topk500_100_50-zxing-gen}{10326}
\DefMacro{table-tdt-topk500_100_50-zxing-time}{113.78}
\DefMacro{table-tdt-topk500_100_50-zxing-lp0-dup}{0}
\DefMacro{table-tdt-topk500_100_50-zxing-lp0-fp}{17}
\DefMacro{table-tdt-topk500_100_50-zxing-lp0-tp}{0}
\DefMacro{table-tdt-topk500_100_50-zxing-lp0-th0}{2316}
\DefMacro{table-tdt-topk500_100_50-zxing-lp0-th1}{7589}
\DefMacro{table-tdt-topk500_100_50-zxing-lp0-th2}{383}
\DefMacro{table-tdt-topk500_100_50-zxing-lp0-th3}{21}
\DefMacro{table-tdt-topk500_100_50-zxing-lp0-th4}{17}
\DefMacro{table-tdt-topk500_100_50-zxing-lp1-dup}{0}
\DefMacro{table-tdt-topk500_100_50-zxing-lp1-fp}{12}
\DefMacro{table-tdt-topk500_100_50-zxing-lp1-tp}{0}
\DefMacro{table-tdt-topk500_100_50-zxing-lp1-th0}{2489}
\DefMacro{table-tdt-topk500_100_50-zxing-lp1-th1}{7431}
\DefMacro{table-tdt-topk500_100_50-zxing-lp1-th2}{381}
\DefMacro{table-tdt-topk500_100_50-zxing-lp1-th3}{13}
\DefMacro{table-tdt-topk500_100_50-zxing-lp1-th4}{12}
\DefMacro{table-tdt-topk500_100_50-zxing-lp2-dup}{0}
\DefMacro{table-tdt-topk500_100_50-zxing-lp2-fp}{24}
\DefMacro{table-tdt-topk500_100_50-zxing-lp2-tp}{1}
\DefMacro{table-tdt-topk500_100_50-zxing-lp2-th0}{1520}
\DefMacro{table-tdt-topk500_100_50-zxing-lp2-th1}{8364}
\DefMacro{table-tdt-topk500_100_50-zxing-lp2-th2}{349}
\DefMacro{table-tdt-topk500_100_50-zxing-lp2-th3}{68}
\DefMacro{table-tdt-topk500_100_50-zxing-lp2-th4}{25}
\DefMacro{table-tdt-topk500_100_50-zxing-lp3-dup}{0}
\DefMacro{table-tdt-topk500_100_50-zxing-lp3-fp}{18}
\DefMacro{table-tdt-topk500_100_50-zxing-lp3-tp}{0}
\DefMacro{table-tdt-topk500_100_50-zxing-lp3-th0}{1512}
\DefMacro{table-tdt-topk500_100_50-zxing-lp3-th1}{8356}
\DefMacro{table-tdt-topk500_100_50-zxing-lp3-th2}{365}
\DefMacro{table-tdt-topk500_100_50-zxing-lp3-th3}{75}
\DefMacro{table-tdt-topk500_100_50-zxing-lp3-th4}{18}
\DefMacro{table-tdt-topk500_100_50-zxing-rp0-dup}{0}
\DefMacro{table-tdt-topk500_100_50-zxing-rp0-fp}{0}
\DefMacro{table-tdt-topk500_100_50-zxing-rp0-tp}{0}
\DefMacro{table-tdt-topk500_100_50-zxing-rp0-th0}{7936}
\DefMacro{table-tdt-topk500_100_50-zxing-rp0-th1}{2374}
\DefMacro{table-tdt-topk500_100_50-zxing-rp0-th2}{16}
\DefMacro{table-tdt-topk500_100_50-zxing-rp0-th3}{0}
\DefMacro{table-tdt-topk500_100_50-zxing-rp0-th4}{0}
\DefMacro{table-tdt-topk500_100_50-zxing-rp1-dup}{0}
\DefMacro{table-tdt-topk500_100_50-zxing-rp1-fp}{0}
\DefMacro{table-tdt-topk500_100_50-zxing-rp1-tp}{0}
\DefMacro{table-tdt-topk500_100_50-zxing-rp1-th0}{8271}
\DefMacro{table-tdt-topk500_100_50-zxing-rp1-th1}{2051}
\DefMacro{table-tdt-topk500_100_50-zxing-rp1-th2}{4}
\DefMacro{table-tdt-topk500_100_50-zxing-rp1-th3}{0}
\DefMacro{table-tdt-topk500_100_50-zxing-rp1-th4}{0}

\DefMacro{table-tdt-topk500_100_50-lang-time}{172.14}
\DefMacro{table-tdt-total-time}{19484.92}
\DefMacro{table-tdt-topk500_100_50-total-time}{1481.69}
\DefMacro{table-tdt-total-time-reduction}{92.40}
\DefMacro{table-single-tier-codec-time}{5038.53}
\DefMacro{table-single-tier-codec-topk500-time}{570.48}
\DefMacro{table-single-tier-compress-time}{2762.75}
\DefMacro{table-single-tier-compress-topk500-time}{339.52}
\DefMacro{table-single-tier-lang-time}{3692.30}
\DefMacro{table-single-tier-lang-topk500-time}{318.02}
\DefMacro{table-single-tier-math-time}{3423.68}
\DefMacro{table-single-tier-math-topk500-time}{378.22}
\DefMacro{table-single-tier-statistics-time}{4221.85}
\DefMacro{table-single-tier-statistics-topk500-time}{179.25}
\DefMacro{table-single-tier-text-time}{4055.27}
\DefMacro{table-single-tier-text-topk500-time}{392.75}
\DefMacro{table-single-tier-vectorz-time}{3568.98}
\DefMacro{table-single-tier-vectorz-topk500-time}{438.52}
\DefMacro{table-single-tier-zxing-time}{3624.03}
\DefMacro{table-single-tier-zxing-topk500-time}{216.17}

\newcommand{\NumOfBugs}{11\xspace}
\newcommand{\NumOfFixed}{6\xspace}
\newcommand{\NumOfFound}{12\xspace}

\newcommand{\JDKBug}[1]{\href{https://bugs.openjdk.org/browse/#1}{#1}\xspace}
\newcommand{\VBug}[1]{\href{https://issues.chromium.org/issues/#1}{V8-#1}\xspace}
\newcommand{\GraalBug}[1]{\href{https://github.com/oracle/graal/issues/#1}{Graal-#1}\xspace}
\newcommand{\MozillaBug}[1]{\href{https://bugzilla.mozilla.org/show_bug.cgi?id=#1}{Mozilla-#1}\xspace}

\newcommand{\JIT}{JIT\xspace}

\newcommand{\JVM}{JVM\xspace}
\newcommand{\AOT}{AOT\xspace}
\newcommand{\JDK}{JDK\xspace}

\newcommand{\Chrome}{Chrome\xspace}
\newcommand{\Firefox}{Firefox\xspace}
\newcommand{\GitHub}{GitHub\xspace}
\newcommand{\SpringFramework}{Spring Framework\xspace}
\newcommand{\Java}{Java\xspace}
\newcommand{\JS}{JavaScript\xspace}

\newcommand{\HotSpot}{HotSpot\xspace}
\newcommand{\Graal}{Graal\xspace}
\newcommand{\GC}{GC\xspace}
\newcommand{\VEight}{V8\xspace}
\newcommand{\SpiderMonkey}{SpiderMonkey\xspace}
\newcommand{\COne}{C1\xspace}
\newcommand{\CTwo}{C2\xspace}
\newcommand{\Maglev}{Maglev\xspace}
\newcommand{\Sparkplug}{Sparkplug\xspace}
\newcommand{\TurboFan}{TurboFan\xspace}

\newcommand{\Select}{\ding{51}}

\newcommand{\StFixed}{Fixed\xspace}
\newcommand{\StConfirmed}{Confirmed\xspace}
\newcommand{\StDup}{Duplicate\xspace}
\newcommand{\RcOpt}{Optimization\xspace}
\newcommand{\RcSpec}{Speculation\xspace}
\newcommand{\RcCodegen}{Codegen\xspace}
\newcommand{\RcUnknown}{Unknown\xspace}

\newcommand{\Tool}{\textsc{Jittery}\xspace}
\newcommand{\Lejit}{LeJit\xspace}
\newcommand{\Artemis}{Artemis\xspace}
\newcommand{\JavaFuzzer}{Java* Fuzzer\xspace}
\newcommand{\functional}{functional\xspace}

\newcommand{\level}{layer\xspace} 
\newcommand{\Level}{Layer\xspace}
\newcommand{\levels}{layers\xspace} 
\newcommand{\Levels}{Layers\xspace} 
\newcommand{\leveled}{layered\xspace} 
\newcommand{\Leveled}{Layered\xspace} 
\newcommand{\xname}{\leveled differential performance testing\xspace}
\newcommand{\xName}{\Leveled Differential Performance Testing\xspace}

\definecolor{rqgray}{gray}{0.85}
\newmdenv[
  skipabove=10pt,
  skipbelow=10pt,
  leftmargin=1pt,
  rightmargin=1pt,
  innertopmargin=8pt,
  innerbottommargin=8pt,
  innerleftmargin=10pt,
  innerrightmargin=10pt,
  backgroundcolor=rqgray,
  linecolor=rqgray,
  roundcorner=10pt,
]{rqanswer}
\newcommand{\answerrqbox}[2]{%
  \begin{rqanswer}
    \textbf{Answer to #1.} #2
  \end{rqanswer}
}

\newcommand{\timediff}[2]{%
  \pgfmathparse{(#1-#2)/#1*100}%
  \pgfmathprintnumber[
    fixed,
    precision=2,
    zerofill
  ]{\pgfmathresult}%
}

\newcommand{\timediffr}[2]{%
  \pgfmathparse{(#2-#1)/#1*100}%
  \pgfmathprintnumber[
    fixed,
    precision=2,
    zerofill
  ]{\pgfmathresult}%
}

\setcopyright{cc}
\setcctype{by}
\acmDOI{10.1145/3798245}
\acmYear{2026}
\acmJournal{PACMPL}
\acmVolume{10}
\acmNumber{OOPSLA1}
\acmArticle{137}
\acmMonth{4}
\received{2025-10-10}
\received[accepted]{2026-02-17}

\begin{document}

\title{Understanding and Finding JIT Compiler Performance Bugs}

\author{Zijian Yi}
\orcid{0009-0000-5616-642X}
\affiliation{%
  \institution{The University of Texas at Austin}
  \city{}
  \country{USA}
}
\email{zijianyi@utexas.edu}

\author{Cheng Ding}
\orcid{0009-0006-0260-8484}
\affiliation{%
  \institution{The University of Texas at Austin}
  \city{}
  \country{USA}
}
\email{cheng.ding@utexas.edu}

\author{August Shi}
\orcid{0000-0001-8239-3124}
\affiliation{%
  \institution{The University of Texas at Austin}
  \city{}
  \country{USA}
}
\email{august@utexas.edu}

\author{Milos Gligoric}
\orcid{0000-0002-5894-7649}
\affiliation{%
  \institution{The University of Texas at Austin}
  \city{}
  \country{USA}
}
\email{gligoric@utexas.edu}


\begin{abstract}
Just-in-time (\JIT) compilers are key components for many popular
programming languages with managed runtimes (\eg, \Java and \JS).
\JIT compilers perform optimizations and generate native code at
runtime based on dynamic profiling data, to improve the execution
performance of the running application.
Like other software systems, \JIT compilers might have software bugs,
and prior work has developed a number of automated techniques for
detecting functional bugs (\ie, generated native code does not
semantically match that of the original code).
However, no prior work has targeted \JIT compiler performance bugs,
which can cause significant performance degradation while an
application is running.
These performance bugs are challenging to detect due to the complexity
and dynamic nature of \JIT compilers.
In this paper, we present the first work on demystifying \JIT
performance bugs.
First, we perform an empirical study across four popular \JIT
compilers for \Java and \JS.
Our manual analysis of \UseMacro{empirical-total-studied} bug reports
uncovers common triggers of performance bugs, patterns in which these
bugs manifest, and their root causes.
Second, informed by these insights, we propose \xname, a lightweight
technique to automatically detect \JIT compiler performance bugs, and
implement it in a tool called \Tool.
We incorporate practical optimizations into \Tool such as test
prioritization, which reduces testing time by
\UseMacro{table-tdt-total-time-reduction}\% without compromising
bug-detection capability, and automatic filtering of false-positives
and duplicates, which substantially reduces manual inspection effort.
Using \Tool, we discovered \NumOfFound previously unknown performance
bugs in the Oracle \HotSpot and \Graal \JIT compilers, with \NumOfBugs
confirmed and \NumOfFixed fixed by developers.

\end{abstract}


\maketitle

\section{Introduction}
\label{sec:intro}

Modern optimizing compilers are responsible for generating both
\emph{correct} and \emph{efficient} code~\cite{alfred2007compilers,
  cooper2022engineering}.
An optimizing compiler takes a program as an input ($p$) and produces
a semantically equivalent program in another form (e.g., intermediate
representation or native code), \ie, $compiler(p) = p'$.

Just-in-time (\JIT) compilation is a technique used by many
programming language implementations to improve performance
\emph{during} program execution.
It originates from Smalltalk~\cite{deutsch1984efficient}, and has been
widely adopted in many popular programming languages like
Java~\cite{cramer1997compiling}, JavaScript~\cite{turbofan}, and
Python~\cite{PyPy, Python_JIT}.
Unlike ahead-of-time (\AOT) compilation, \JIT compilation happens
during the execution of a program.  An input to a \JIT compiler is an
intermediate representation of (or part of) a program ($a$) and the
output is commonly native code, \ie, $jit(a) = o$.
\JIT compilers collect profiling
data while the program is running and use this data to guide the
compilation process.  They selectively compile code that is frequently
executed and worth optimizing for future execution,
so that the performance gain from \JIT compilation outweighs
the compilation overhead~\cite{kulkarni2011jit}.  Speculative
optimizations (\ie, optimizations that are only applicable under
certain assumptions) may also be performed based on the profiling
data~\cite{chambers1989customization, arnold2005survey,
  zheng2017empirical}.  

Like many software systems, a \JIT compiler can itself have unintended
software bugs. These bugs can be broadly categorized into \functional
bugs or performance bugs. A \functional bug in the \JIT compiler is
when the compiler does not produce code that matches the semantics of
the original code, \ie, we say that a \JIT compiler has a \functional
bug if $semantics(a) \neq semantics(jit(a))$.  Finding \functional bugs
in \JIT compilers has been a popular topic in recent years with many
notable results~\cite{javafuzzer, chen2016coverage, zhao2022history,
  jia2023detecting, li2023validating, ZangETAL22JAttack,
  ZangETAL24LeJit, gross2023fuzzilli, wang2023fuzzjit,
  wachterdumpling, polito2022interpreter}.
On the other hand, we say that a \JIT compiler has a \emph{performance
bug} when it delivers unexpected performance behavior on certain input
programs.  We distinguish two cases.
First, the \JIT compiler itself takes an excessive amount of time to
perform its task, \ie, $time(jit(a)) > threshold$, where the threshold
denotes an upper bound on the expected compilation time as defined by
users or developers.
We refer to this kind of performance bug as \textit{long compilation}
in this paper (also referred to as ``compile time hog'' in some
literature).
Second, the \JIT compiler produces code that takes too long to
execute, \ie, $time(run(a)) < time(run(o))$. Namely, we observe an
unexpected performance behavior of the \JIT compiled program compared
to an unoptimized version of that code.
We refer to this as \textit{high-order performance bugs}, where the root
causes of performance degradation come not from the program $a$ itself
but from the compiler that transforms the program.
Since \JIT compilation happens at runtime, both \textit{long
  compilation} and \textit{high-order performance bugs} can lead to
similar effects.  There is little prior work on understanding and
detecting performance bugs in \JIT compilers.

Existing research on performance bugs~\cite{jin2012understanding,
  zaman2012qualitative, selakovic2016performance, han2016empirical,
  azad2023empirical} mostly focuses on performance bugs in application
  code that can be fixed by modifying the application source itself.
In contrast, performance bugs in \JIT compilers cannot be easily,
sometimes not even possibly, resolved at the application level.  While
certain workarounds at the source-code level can make specific code
patterns more amenable to \JIT optimizations~\cite{gong2015jitprof},
it remains desirable to improve \JIT compilers themselves to make them
more flexible and robust, thereby reducing the burden on developers.
There has also been prior work on identifying \textit{long
  compilation} bugs in C~\cite{le2014compiler} and
  Markdown~\cite{li2021understanding} compilers and on reducing
  compilation time in C++~\cite{AlAwarETAL25Yalla, dietrich2017chash},
  as well as studies on \textit{high-order performance bugs} that
  primarily target missed optimizations in \AOT
  compilers~\cite{barany2018finding, theodoridis2022finding,
  liu2023exploring, gao2024shoot, theodoridis2024refined}.  One study
  that explores \JIT compilers~\cite{pevcimuth2023diagnosing} largely
  treats them as \AOT compilers, focusing solely on
  optimization-related issues.  However, \JIT compilers exhibit
  several unique aspects like profiling, speculation, and interactions
  with other runtime components that are not present in \AOT compilers.
  These aspects can also introduce performance bugs, yet they remain
  unexplored by prior work.

To bridge this gap in understanding \JIT compiler performance bugs, we
present the first in-depth empirical study of \JIT compiler
performance bugs from real-world \Java and \JS issue trackers.  Our
analysis of \UseMacro{empirical-total-studied} bug reports reveals
common triggers, patterns in which these bugs manifest, and their root
causes.
Specifically, we find that (1)~nearly half of the bugs can be exposed
by small, focused micro-benchmarks rather than full benchmark suites;
(2)~they are most commonly detected through comparative signals such
as performance regressions or differences across equivalent
executions; and (3)~beyond traditional optimization and code
generation issues, \JIT-specific components like speculation and
runtime interaction are also significant sources of bugs.
These characteristics provide insights on how to design
better testing and development methods to detect and combat
performance bugs in \JIT compilers.

Based on these insights, we propose \Tool, a tool that implements
\emph{\xname} to automatically detect \JIT compiler performance bugs.
\Tool generates a large number of small random programs, executes each
under two differential \JIT configurations (\eg, different compiler
tiers or compiler versions), and flags programs whose execution times
diverge significantly as potential bug candidates.
Since rigorous benchmarking of every generated program is
prohibitively expensive, we organize detection into a series of \levels
with increasing iteration counts: early \levels cheaply discard programs
that show no performance anomaly, while later \levels apply more
thorough measurement only to the surviving candidates.  This lets us
test many programs quickly (efficiency) while still obtaining
reliable measurements for promising candidates (accuracy).
\Tool further leverages runtime information from earlier \levels to
prioritize candidates for subsequent \levels.
Finally, it uses heuristics to remove potential false positives and
duplicates, which greatly reduce manual inspection effort.
\Tool leads to the discovery of \NumOfFound previously unknown
performance bugs in the Oracle \HotSpot and \Graal \JIT compilers,
with \NumOfBugs confirmed and \NumOfFixed already fixed by the
developers.

\vspace{5pt}
\noindent
The main contributions of this paper include the following:
\begin{itemize}[topsep=2pt,itemsep=3pt,partopsep=0ex,parsep=0ex,leftmargin=*]
    \item[$\star$] \textbf{Empirical study}. The first in-depth
      empirical study of real-world \JIT compiler performance bugs. We
      analyze \UseMacro{empirical-total-studied} \JIT compiler
      performance bugs in 4 widely used \JIT compilers and summarize
      their characteristics. Based on our empirical study, we provide
      insights on challenges and potential solutions for detecting and
      avoiding JIT compiler performance bugs.
    \item[$\star$] \textbf{Dataset}.  We make a dataset of \JIT
      compiler performance bugs, which we built during our empirical
      study, publicly available. We expect that this dataset can be
      used to help future work to develop techniques for detecting
      and preventing \JIT compiler performance bugs.
    \item[$\star$] \textbf{Approach}.  We developed \Tool, a
      lightweight tool based on \xname to automatically detect \JIT
      compiler performance bugs with practical optimizations to
      further improve efficiency and reduce manual effort.
    \item[$\star$] \textbf{Results}.  \Tool led to the
      discovery of \NumOfFound previously unknown performance bugs in
      the Oracle \HotSpot and \Graal \JIT compilers, including some
      involving aspects unique to \JIT compilers like speculation.
      \NumOfBugs of these bugs have been confirmed, with some already
      fixed.
\end{itemize}

\noindent
Our dataset and scripts are publicly available at
\url{https://github.com/EngineeringSoftware/jittery}.


\section{Background}
\label{sec:background}

This section provides background on just-in-time (\JIT) compilers,
especially focusing on unique features of \JIT compilers that
differentiate them from ahead-of-time (\AOT) compilers.
We discuss a brief overview of \JIT compilers (Section~\ref{sec:bg-jit}),
speculative optimization (Section~\ref{sec:bg-specopt}), and tiered
compilation (Section~\ref{sec:bg-tiered}).

\subsection{Overview of \JIT Compilers}
\label{sec:bg-jit}

\begin{figure}[t]
    \centering
    \includegraphics[width=0.95\columnwidth]{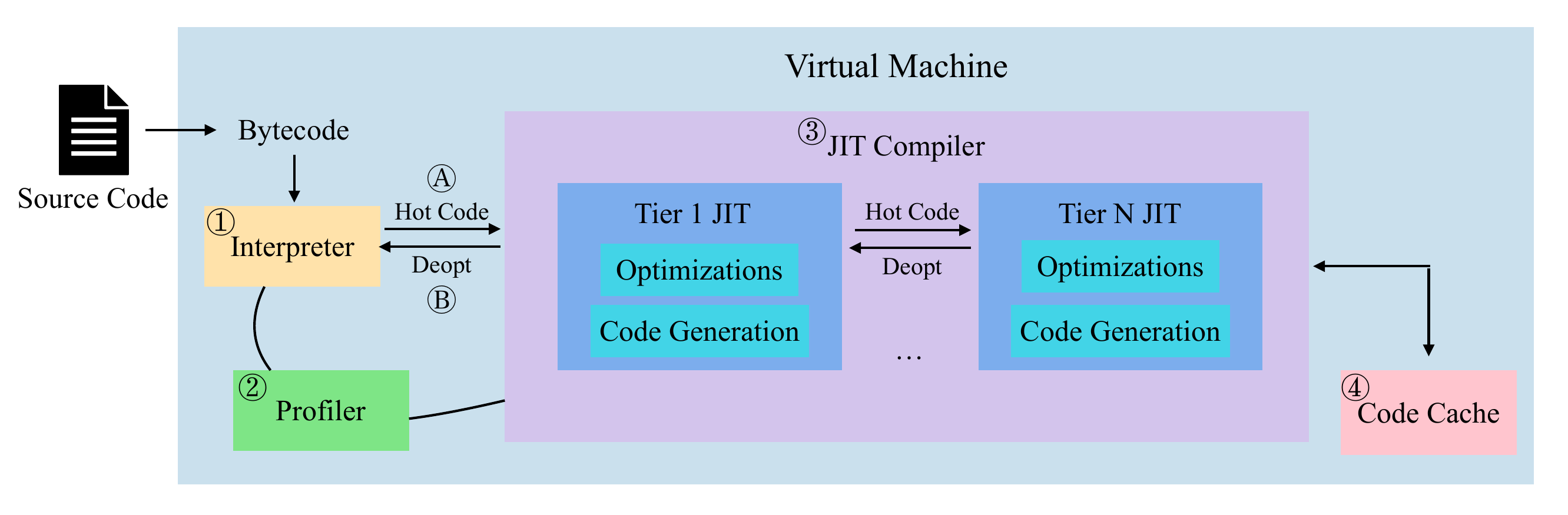}
    \vspace{-12pt}
    \caption{High-level workflow of running a program through \JIT
      compilation, which happens while the program is running.
      Commonly, \JIT compilers
      optimize code snippets that are frequently
      executed.\label{fig:bg-jit}}
\end{figure}

Figure~\ref{fig:bg-jit} shows the overview of the \JIT compilation
process for general-purpose programming languages like \Java and
\JS.
Program execution begins in the interpreter (\textcircled{1}). While
the interpreter runs the program, the profiler (\textcircled{2})
collects profiling data, \eg, counters recording the number of times a
method is invoked or a loop is entered.
Once profiling thresholds are reached, the \JIT compiler
(\textcircled{3}) is triggered to compile any ``hot code''
(\textcircled{A}), i.e., code regions that have been frequently
executed and would likely continue to be frequently executed in the
future. The \JIT compiler then performs optimization and code
generation for this hot code, following steps similar to those in \AOT
compilers, but often incorporating speculative optimizations guided by
the collected profiling data (Section~\ref{sec:bg-specopt}).
Furthermore, modern runtimes often contain more than one \JIT compiler
with different trade-offs (Section~\ref{sec:bg-tiered}).  Because
speculations may occasionally fail to hold, execution must sometimes
revert to the interpreter or a lower tier \JIT compiler through a
process known as deoptimization
(\textcircled{B})~\cite{zheng2017empirical}. Finally, since native
code is generated dynamically at runtime, the \JIT compiler must
carefully manage the memory region used to store compiled code,
commonly referred to as the code cache, ensuring that other parts of
the code invoke and use the correct compiled code.

There are two major types of decisions a \JIT compiler needs to make
based on the collected profiling data: 1)~what code to compile, and
2)~what speculative optimizations to apply.  In practice, both
decisions are typically guided by hand-crafted heuristics derived from
empirical experience and human expertise~\cite{kulkarni2011jit,
  chambers1989customization}.

Let the time of compiling a given code region be denoted as $C$, time
to execute that piece of code in the interpreter mode as $T_i$, time
to execute the compiled code as $T_c$, and the expected number of
future executions as $N$. For compilation to be profitable, the
following condition should hold:
\begin{equation}
    (T_i - T_c) \times N > C
\end{equation}

\subsection{Speculative Optimization}
\label{sec:bg-specopt}

Speculative optimization is a technique where \JIT compilers make
educated guesses about the behavior of a program based on runtime
profiling data.
Unlike optimizations in \AOT compilers, whose correctness is ensured
by static analysis, speculative optimizations may not always hold true
and they are guarded by runtime checks. When a speculation fails, the
execution falls back to a safe state through \emph{deoptimization}.

\begin{figure}[t]
    \subcaptionbox{
        \label{figure:bg-specopt-source-java}
    }
    {
\begin{lstlisting}[language=java-pretty]
if (obj == null) {
    throw new NullPointerException();
}
val = obj.f;
\end{lstlisting}
    }
    \hspace{1.5cm}
    \subcaptionbox{
        \label{figure:bg-specopt-gen-java}
    }
    {
\begin{lstlisting}[language=java-pretty]
val = obj.f;

// If 'obj' is null
// A signal is handled by JVM
\end{lstlisting}
    }
    \vspace{-10pt}
    \caption{Example of a null-check elimination speculative
      optimization in \HotSpot: (a)~a common null check code pattern
      in Java source code; (b)~the pseudocode generated by \HotSpot
      with null-check eliminated.  The optimization removes the null
      check and uses an implicit guard when speculation
      fails.\label{fig:bg-specopt-java}}
\end{figure}

Figure~\ref{fig:bg-specopt-java} illustrates an example of speculative
optimization for null check elimination in the
\HotSpot~\cite{kawahito2000effective}.
The source code in Figure~\ref{figure:bg-specopt-source-java} shows a
common code pattern in \Java, where field access is preceded by an
explicit null check. In most programs, it is rare for the null check
to fail, which can be confirmed based on the profiling data collected
after executing that part of the code many times. The \JIT compiler
then speculates that the check will always succeed (without formally
proving it), generating code similar to the pseudocode shown in
Figure~\ref{figure:bg-specopt-gen-java}.  It will directly access the
field without any additional checks.  If a null pointer dereference is
ever encountered, the operating system will send a signal to execution
runtime, whose handler performs deoptimization and resumes execution
in a safe state. This form of speculative optimization can improve
performance by eliminating redundant null checks, while also reducing
compilation time and code size by avoiding the need to generate code
for exception creation, throwing, and handling.

\begin{figure}[t]
    \subcaptionbox{
        \label{figure:bg-specopt-js-source}
    }
    {
\begin{lstlisting}[language=js-pretty]
function add(a, b) {
  return a + b;
}
for (let i = 0; i < 100000; i++) {
  add(i, i + 1);
}
console.log(add("Hello ", "world"));
\end{lstlisting}
    }
    \hspace{1.5cm}
    \subcaptionbox{
        \label{figure:bg-specopt-js-gen}
    }
    {
\begin{lstlisting}[language=js-pretty]
function add(a, b) {
  if (!is_smi(a) || !is_smi(b)) {
    deoptimize();
  }
  return raw_numeric_add(a, b);
}
\end{lstlisting}
    }
    \vspace{-12pt}
    \caption{Example of type specialization for add operator
      speculative optimization in \VEight: (a)~add operator is
      overloaded for many different types in JavaScript source code;
      (b)~the pseudocode generated by \VEight with type specialization
      based on profiling data. The optimization specializes the add
      operator for small integer addition and uses an explicit guard
      to ensure the assumption holds.\label{fig:bg-specopt-js}}
\end{figure}

Speculative optimization is more extensive in dynamically typed
languages like \JS.  Figure~\ref{fig:bg-specopt-js} shows an example
of speculative optimization in \VEight~\cite{v8_spec}. The source code
in Figure~\ref{figure:bg-specopt-js-source} shows a simple
\CodeIn{add} function.  Since the `+' operator is overloaded for
multiple types, a general code generation strategy would be to
generate code that checks the types of the operands at runtime.
However, if the profiling data shows that the function is frequently
called with small integers (\CodeIn{smi}), \VEight can speculate that
the operands are always of type \CodeIn{smi} and specialize code that
only provides fast path for the case as shown in
Figure~\ref{figure:bg-specopt-js-gen}. When this speculation fails,
\eg, if a call is made with a non-integer operand, the generated code
triggers a deoptimization event. During deoptimization, the runtime
system reconstructs the corresponding interpreter or baseline stack
frame from metadata embedded in the optimized code, discards the
invalid optimized frame, and transfers control back to the interpreter
or a less-optimized version of the code. These steps ensure that
execution can safely continue without violating the program's
semantics.

\subsection{Tiered Compilation}
\label{sec:bg-tiered}

Compilation at runtime takes time and resources, so trade-offs have to
be made between the compilation overhead and the efficiency of the
generated code. For example, for simple functions, several basic
optimizations are enough to generate efficient code. However, for
complex and frequently invoked functions, more sophisticated
optimizations are needed to gain better performance.

\begin{figure}[t]
  \centering
\tikzstyle{every node} = [font=\scriptsize] 
\tikzstyle{textnode} = [rectangle, inner sep=2pt, text centered, text depth=.25ex]
\tikzstyle{textnodeBiggerSize} = [rectangle, inner sep=2pt, text centered, text depth=.25ex, font=\footnotesize]
\tikzstyle{labelnode} = [rectangle, inner sep=1pt, text centered, text depth=.25ex] 
\tikzstyle{levelnode} = [circle, draw, thick, minimum size=5mm, font=\footnotesize]
\tikzstyle{conditionnode} = [regular polygon, regular polygon sides=6, shape border rotate=30, draw, minimum size=8mm, thick]
\tikzstyle{point} = [circle, inner sep=0pt, minimum size=1pt, fill=white] 
\tikzstyle{ra} = [->, shorten >=1pt, >=stealth, semithick]

\definecolor{backcolor1}{HTML}{F0EBE0} 
\definecolor{backcolor2}{HTML}{D6E2C8} 
\definecolor{backcolor3}{HTML}{C8D4E8} 

\begin{tikzpicture}[scale=1.0, node distance=12mm]

  \node[levelnode] (level0) at (0,0) { L0 };
  \node[levelnode] (level1) [right=of level0] { L1 };
  \node[levelnode] (level2) [right=of level1] { L2 };
  \node[levelnode] (level3) [right=of level2] { L3 };
  \node[levelnode] (level4) [right=of level3] { L4 };

  \path let \p1 = (level0.center), \p2 = (level1.north) in node[conditionnode] (condition) at (\x1 + 8mm, \y2 + 11mm) { };

  \draw [ra] (level0.north) |- (condition);
  \draw [ra] (condition.corner 5) -| (level2.north);
  \draw [ra] (condition.corner 6) -| (level3.north);
  \draw [ra] (level2.east) -- (level3.west);
  \draw [ra] (level3.east) -- (level4.west);

  \node[labelnode] (c2congested) [above=2.5mm of level2, fill=backcolor2] {C2 congested};
  \node[labelnode] (normal) at (c2congested -| level3) [fill=backcolor2] {Normal};

  \node[textnode] (noprofiling) [below=4mm of level1] { No Profiling };
  \node[textnode] (limitedprofiling) [below=4mm of level2] { Limited Profiling };
  \node[textnode] (fullprofiling) [below=4mm of level3] { Full Profiling };
  \node[textnodeBiggerSize] (c1) [below=0.5mm of limitedprofiling] { C1 Compiler };
  \node[textnodeBiggerSize] (interpreter) at (level0 |- c1) { Interpreter };
  \node[textnodeBiggerSize] (c2) at (level4 |- c1) { C2 Compiler };

  \draw [ra] (level3.south) -- ++(0,-2mm) -| (level1.south);

  \path let \p1 = (level2.center), \p2 = (level1.south) in
    node[labelnode, fill=backcolor2] (trivial) at (\x1, \y2 - 2mm) {Trivial};

  \path let \p1 = (interpreter.south), \p2 = (interpreter.center) in node [point] (anchor3) at (\x2 - 8mm, \y1) {};
  \path let \p1 = (interpreter.south), \p2 = (interpreter.center) in node [point] (anchor4) at (\x2 + 8mm, \y1) {};
  \path let \p1 = (c2.south), \p2 = (c2.center) in node [point] (anchor5) at (\x2 - 8mm, \y1) {};
  \path let \p1 = (c2.south), \p2 = (c2.center) in node [point] (anchor6) at (\x2 + 8mm, \y1) {};
  \path let \p1 = (anchor4.center), \p2 = (condition.north) in node [point] (anchor1) at (\x1, \y2 + 2mm) {};
  \path let \p1 = (anchor5.center), \p2 = (condition.north) in node [point] (anchor2) at (\x1, \y2 + 2mm) {};

  \begin{scope}[on background layer]
    \fill [fill=backcolor1,rounded corners] (anchor3) rectangle (anchor1);
    \fill [fill=backcolor2,rounded corners] (anchor4) rectangle (anchor2);
    \fill [fill=backcolor3,rounded corners] (anchor6) rectangle (anchor2);
  \end{scope}
\end{tikzpicture}
  \vspace{-5pt}
  \caption{Common transitions among execution levels (L0--L4) in
    \HotSpot tiered compilation. Execution starts in the interpreter
    (L0), then typically moves to L2 or L3 based on queue/load
    heuristics; methods with sufficient profiling in L3 are promoted
    to L4, while trivial methods may end at L1.\label{fig:bg-tier}}
\end{figure}
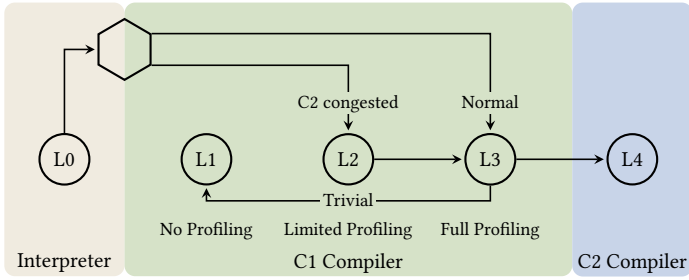

To balance these trade-offs, modern \JIT compilers often rely on
\emph{tiered compilation}.  The code under execution transitions
between different levels of execution based on the profiling data and
the runtime environment, and the \JIT chooses the appropriate
compilation and profiling strategies for different execution levels.
For example, in \HotSpot, there are 5 execution
levels~\cite{HsTieredPolicy} and two \JIT compilers: \COne and
\CTwo~\cite{HsGlossary}.  \COne (a.k.a., client compiler) is a low
overhead compiler that generates code quickly but with fewer
optimizations.  \CTwo (a.k.a., server compiler) is a high overhead
compiler that generates highly optimized code.
Figure~\ref{fig:bg-tier} shows the transitions between different
execution levels. Execution starts in the \emph{interpreter} (L0).
When a method becomes hot, policy heuristics first decide between L2
and L3, with \CTwo queue pressure as a key signal: under high \CTwo
load, L2 is preferred; otherwise L3 is chosen to collect richer
profiling.  The \COne queue length also adjusts thresholds to avoid
overloading the compiler. Methods running at L2 can later move to L3
when \CTwo pressure recedes. Once enough profiling is collected at L3,
methods are usually promoted to L4 (\CTwo) for peak performance. For
trivial methods, the policy may choose L1 (optimized \COne code
without profiling) instead of promoting to L4.  \SpiderMonkey has two
tiers of \JIT compilers, namely Baseline Compiler and
WarpMonkey~\cite{SM-overview}.  \VEight recently also introduced two
new \JIT compilers featuring faster compilation time named
\Maglev~\cite{maglev} and \Sparkplug~\cite{sparkplug}, besides the
existing \JIT compiler: \TurboFan~\cite{turbofan}.

Consider speculative optimization and its potential benefits. Let \(p
\in [0,1]\) denote the probability that a speculation holds at
runtime. Let \(G > 0\) represent the reduction in execution time
(e.g., in milliseconds or cycles) achieved when the speculation is
correct, and let \(D > 0\) denote the additional execution-time
overhead incurred due to deoptimization when the speculation
fails. Under these definitions, the speculation is expected to be
beneficial when the expected time saved outweighs the expected
penalty:
\begin{equation}
    p \times G > (1 - p) \times D.
\end{equation}


\section{Study Methodology}

In this section, we describe our empirical study methodology.  We
describe how we collect bug reports of interest
(Section~\ref{sec-method:collection}), the way we analyze these
reports (Section~\ref{sec-method:analysis}), the research questions we
aim to answer (Section~\ref{sec-method:rq}), and potential threats to
validity (Section~\ref{sec-method:limitations}) of our study.

\subsection{Bug Collection}
\label{sec-method:collection}

We choose four subjects for our study: \HotSpot, \Graal, \VEight, and
\SpiderMonkey. The former two are \Java \JIT engines used in Oracle
JDK, \VEight is the \JS engine in the \Chrome browser, and
\SpiderMonkey is the \JS engine in the \Firefox browser. They all
follow the basic architecture as described in
Section~\ref{sec:bg-jit}.
These subjects are widely-used and represent two kinds of programming
languages where \JIT compilers are commonly used, namely
statically-typed Object-Oriented (\eg, \Java) and dynamically-typed
(\eg, \JS) languages.  These subjects are representative of the
state-of-the-art \JIT compilers.

All of the subjects are open-sourced and have open issue tracking
systems.  Besides \Graal, which uses \GitHub issue
tracker~\cite{graaltracker}, the other three all have their own
dedicated issue tracking systems~\cite{JBSWebpage, bugzilla,
  v8tracker}. We systematically collected issues in the issue tracking
systems based on the following criteria:

\begin{itemize}[topsep=2pt,itemsep=3pt,partopsep=0ex,parsep=0ex,leftmargin=*]
    \item \textbf{Component}: We only consider performance issues that
      are related to \JIT compilers.  Other components like garbage
      collectors may also cause performance issues but they are out of
      scope of our study.  Note that if \JIT compilers cause
      performance issues in other components, we still consider them.
    \item \textbf{Type}: We only consider issues that are labeled as
      `bug' or `defect'.  These issues usually represent problems that
      manifest in practice and are reported by users.  There are also
      issues labeled as `enhancement' or `feature' which are not
      considered in our study.  Besides, we only consider issues that
      are marked as `fixed' by the developers as they are of higher
      priority and contain more details that help us understand the
      issues.
    \item \textbf{Date}: To narrow down the scope of our study, we
      only consider issues that are created after 2015 and fixed
      before 2025.  The time range is chosen to ensure that the issues
      are recent so that they reflect the current state of the \JIT
      design and fixes are merged into the codebase.
\end{itemize}

\noindent
To be precise about our steps, we give an example.
The JDK Bug System~\cite{JBSWebpage} provides a query language for
filtering issues, and we use the following query to search for issues
of interest in \HotSpot:

\begin{lstlisting}[language=sql, basicstyle=\ttfamily\footnotesize]
    labels = performance AND type = Bug AND Resolution = fixed AND
      created >= 2015-01-01 AND resolved <= 2025-01-01 AND 
      component = hotspot AND subcomponent in (compiler, jit)
      ORDER BY created ASC
\end{lstlisting}

Other issue tracking systems provide similar filtering or query
machinery to search for issues conforming with our criteria.  We found
in \Graal~\cite{graaltracker} many performance issues that are not
explicitly labeled as `performance', and thus we use a set of keywords
(\ie, `perf', `performance', `slow', `slower', `opt', `optimization',
`optimisation', `latency', `missed', `missing', `high', `higher') to
search for potential issues of interest.

\begin{table}[t!]
\centering
\caption{Subjects used in our empirical study. \label{table-empirical:subjects}}
\vspace{-10pt}
\resizebox{\textwidth}{!}{%
\begin{tabular}{llrrrrrr}
\toprule
\multirow{2}{*}{\centering \textbf{Subject}} & \multirow{2}{*}{\centering \textbf{Language}} & 

\multirow{2}{*}{\centering \textbf{Collected}} & 

\multicolumn{4}{c}{\centering \textbf{Filtered}} & \multirow{2}{*}{\centering \textbf{Studied}} \\

\cmidrule(lr){4-7}
& & & Irrelevant & Enhancement & Duplicate & Not an Issue & \\

\midrule
\HotSpot & \Java & \UseMacro{empirical-hotspot-collected} & \UseMacro{empirical-hotspot-irrelevant} & \UseMacro{empirical-hotspot-enhancement} & \UseMacro{empirical-hotspot-duplicate} & \UseMacro{empirical-hotspot-notanissue} & \UseMacro{empirical-hotspot-studied} \\
\Graal & \Java & \UseMacro{empirical-graal-collected} & \UseMacro{empirical-graal-irrelevant} & \UseMacro{empirical-graal-enhancement} & \UseMacro{empirical-graal-duplicate} & \UseMacro{empirical-graal-notanissue} & \UseMacro{empirical-graal-studied} \\
\VEight & \JS & \UseMacro{empirical-v8-collected} & \UseMacro{empirical-v8-irrelevant} & \UseMacro{empirical-v8-enhancement} & \UseMacro{empirical-v8-duplicate} & \UseMacro{empirical-v8-notanissue} & \UseMacro{empirical-v8-studied} \\
\SpiderMonkey & \JS & \UseMacro{empirical-sm-collected} & \UseMacro{empirical-sm-irrelevant} & \UseMacro{empirical-sm-enhancement} & \UseMacro{empirical-sm-duplicate} & \UseMacro{empirical-sm-notanissue} & \UseMacro{empirical-sm-studied} \\
\midrule
\textbf{Total} & - & \UseMacro{empirical-total-collected} & \UseMacro{empirical-total-irrelevant} & \UseMacro{empirical-total-enhancement} & \UseMacro{empirical-total-duplicate} & \UseMacro{empirical-total-notanissue} & \UseMacro{empirical-total-studied} \\
\bottomrule
\end{tabular}
}
\end{table}

\subsection{Bug Filtering and Analysis}
\label{sec-method:analysis}

After the initial automated search, we collected a total of
\UseMacro{empirical-total-collected} issues. We then performed a
manual review to further refine the dataset. During this process, we
removed issues that were mis-labeled (\emph{Irrelevant}), too vague or
proposal-oriented (\emph{Enhancement}), duplicates of other reports
(\emph{Duplicate}), or closed without a fix (\emph{Not an issue}).
Such cases were treated as feature requests rather than genuine
performance bugs revealed in practice. After this manual filtering,
\UseMacro{empirical-total-studied} issues remained for our study.

Table~\ref{table-empirical:subjects} shows the final number of
performance bugs we collected and studied for each subject.
The first column lists the studied subjects, followed by the
programming language, the initial number of collected bugs, the number
of bugs filtered out by category, and the final count of studied bugs.
The last row reports the total across all subjects.

For each of the remaining issues, we conducted an in-depth manual
inspection of all related information provided in each bug report,
including the bug description, developer discussions, attached files,
and associated fix commits or pull requests.

\subsection{Research Questions}
\label{sec-method:rq}

The goal of our study is to gain a deeper understanding of performance
bugs in \JIT compilers in practice and to provide actionable insights
that can help developers detect and prevent such bugs more
effectively.
To this end, we aim to answer the following research questions:

\begin{itemize}[topsep=3pt,itemsep=5pt,partopsep=0ex,parsep=0ex,leftmargin=*]
    \item \textbf{RQ1:} What input artifacts are used to trigger performance bugs?
    \item \textbf{RQ2:} What are common symptoms of performance bugs?
    \item \textbf{RQ3:} What are common root causes of performance bugs?
\end{itemize}

\noindent Answering RQ1 helps us understand how to design effective
test inputs for detecting performance bugs. Addressing RQ2 reveals the
observable symptoms that can guide the construction of reliable test
oracles. Finally, exploring RQ3 sheds light on which components of
\JIT compilers are more error-prone and how similar issues can be
mitigated in future development.

\subsection{Threats to Validity}
\label{sec-method:limitations}

Like other empirical studies, our work is subject to both external and
internal threats.

External threats concern the representativeness and generalizability
of our subjects and studied issues. We selected the most widely used
\JIT compilers across two representative programming languages and
collected issues reported over the past ten years to ensure recency
and relevance. We believe these criteria make our dataset broadly
representative of real-world \JIT compiler behavior.

Internal threats primarily stem from manual issue analysis. To
mitigate bias, we defined clear classification criteria with
illustrative examples and considered all available contextual
information.
Each issue was independently analyzed by at least two authors, with
disagreements resolved through discussion. We also make our dataset
publicly available to facilitate independent verification.  These
measures help reduce internal threats and enhance the reliability of
our findings.


\section{Case Study}
\label{sec:case}

In this section, we present motivating examples of real-world \JIT
compiler performance bugs to illustrate their unique characteristics.
Section~\ref{sec:characteristics} then provides a more systematic
analysis of these characteristics.

\begin{figure}[t]
    \subcaptionbox[t]{
        test case
        \label{figure:case-jdk8280320-tc}
    }
    {
\begin{lstlisting}[language=java-pretty]
public static void test(long n) {
  while (n > 0) {
    for (int i = 0; i < size; i++) {
      // Can be vectorized
      oBuf[i] = (short)(iBuf[i] >> 3);
    }
    n--;
  }}
\end{lstlisting}
    }
    \hspace{0.5cm}
    \subcaptionbox[t]{
        fix
        \label{figure:case-jdk8280320-fix}
    }
    {
\begin{lstlisting}[style=patch-pretty]
bool ciMethodData::load_data() {
   // ...
+  if (_invocation_counter == 0
+       && mdo->backedge_count() > 0) {
+      _invocation_counter = 1;
+  }
   _state = ... // ...
}
\end{lstlisting}
    }
    \vspace{-10pt}
    \caption{Case study of \JDKBug{JDK-8280320}: loop optimizations
      are missing during On-Stack Replacement (OSR) compilation.
      (a)~A test case that triggers OSR compilation but is not properly optimized;
      (b)~A fix that avoids skewing counter data during OSR
      compilation.  The bug is caused by incorrectly set profiling
      data, which prevents the vectorization optimization in
      \HotSpot.\label{fig:case-jdk8280320}}
\end{figure}

\MyPara{Case 1: Inaccurate profiling data}
\textbf{\JDKBug{JDK-8280320}}\footnote{All the bug labels are
clickable and redirect to related pages.}
(Figure~\ref{fig:case-jdk8280320}). This bug shows a missed loop
optimization during On-Stack Replacement (OSR)
compilation~\cite{d2018stack} in \HotSpot.
It was discovered while benchmarking AVX-related tests on x86 server
machines: the release build exhibited approximately 35$\times$ lower
throughput than the fastdebug build, because the compiled code fell
back to scalar instructions instead of the expected AVX instructions.
Figure~\ref{figure:case-jdk8280320-tc} shows a reduced test case.
The issue arises when misleading profiling data causes OSR
compilations to incorrectly skip loop optimizations. Specifically, the
invocation counter remains at zero even when the loop's backedge
counter indicates active execution, leading the compiler to
underestimate loop hotness and disable key optimization passes. The
fix, shown in Figure~\ref{figure:case-jdk8280320-fix}, correctly
initializes the profiling counter to ensure accurate scaling of
execution counts and restores the expected optimization behavior.

\textit{Observation 1}: Inaccurate profiling data in \JIT
compilers can cause missed optimizations even when the optimization
pipeline itself is correctly implemented.

\MyPara{Case 2: Long compilation}
\textbf{\GraalBug{2548}}. This bug occurs in the \SpringFramework when
the \Graal \JIT compiler attempts to compile the \CodeIn{BCrypt::key}
method. Due to a defect in the compiler's optimization logic, the
compilation process enters an infinite loop, causing the compilation
thread to run indefinitely with full CPU utilization. In contrast to
\AOT compilers--where such a problem would surface during the build
process--this issue only manifests at runtime, making it significantly
harder to detect and diagnose without inspecting the \JIT compiler's
internal logs. The problem was observed in the GraalVM EE 20.1.0
release, and at the time of discovery, no publicly available fix was
provided.

\textit{Observation 2}: Similar types of performance bugs can be
substantially harder to detect in \JIT compilers than in \AOT
compilers, because they manifest dynamically during program execution
and may cause severe runtime performance degradation.

\begin{figure}[t]
    \subcaptionbox{
        test case
        \label{figure:case-v8-42210048-tc}
    }
    {
\begin{lstlisting}[language=js-pretty]
var baseNoExtends = (function() {
    const A = class A {};
    return function baseNoExtends() { return new A;} })();
var baseExtendsObject = (function() {
    const A = class A extends Object {};
    return function baseExtendsObject() { return new A;} })();
var baseExtendsViaFactory = (function() {
    const A = (BaseClass => class A extends BaseClass {})(Object);
    return function baseExtendsViaFactory() { return new A;} })();
var TESTS = [baseNoExtends, baseExtendsObject, baseExtendsViaFactory];
function test(fn) { ... }
for (var j = 0; j < TESTS.length; ++j) { test(TESTS[j]); }
\end{lstlisting}
    }
    \hspace{0.5cm}
    \subcaptionbox{
        fix
        \label{figure:case-v8-42210048-fix}
    }
    {   
        \hspace{4pt}
\begin{lstlisting}[style=patch-pretty]
+ if (*function == function->native_context()->object_function()) {
+     if (arity == 0) {
+         NodeProperties::ChangeOp(node, javascript()->Create());
+         return Changed(node);
+     }
+     HeapObjectMatcher mnew_target(new_target);
+     if (mnew_target.HasValue() && *mnew_target.Value() != *function) {
+         for (int i = arity; i > 0; --i) node->RemoveInput(i);
+         NodeProperties::ChangeOp(node, javascript()->Create());
+         return Changed(node);
+     }
+ }
\end{lstlisting}
    }
    \vspace{-10pt}
    \caption{Case Study of \VBug{42210048}: performance degradation in
      JavaScript Object constructor subclassing. (a)~A test case that
      evaluates the performance impact of three alternative class
      initialization strategies, revealing observable performance
      differences. (b)~A fix that adds support to the JSCallReducer to
      recognize JSConstruct nodes where the target is the Object
      constructor, and reduce them to JSCreate nodes. The bug is
      caused by the lack of optimization for Object constructor calls
      in subclassing scenarios.\label{fig:case-v8-42210048}}
\end{figure}

\MyPara{Case 3: Performance difference}
\textbf{\VBug{42210048}} (Figure~\ref{fig:case-v8-42210048}).  This
bug exposes a severe performance cliff in V8's TurboFan compiler when
subclassing the \CodeIn{Object} constructor in \JS. Specifically,
instance creation becomes 3$\times$--4$\times$ slower when a class explicitly
extends \CodeIn{Object} compared to a class without an
\CodeIn{extends} clause. The issue was identified using a
microbenchmark that compares three variants: a base class without
inheritance, a class directly extending \CodeIn{Object}, and a class
extending \CodeIn{Object} through a factory function
(Figure~\ref{figure:case-v8-42210048-tc}). The slowdown is caused by
inefficient handling of class inheritance during object construction
when the superclass is \CodeIn{Object}. The fix enhances
\CodeIn{JSCallReducer} to detect \CodeIn{JSConstruct} nodes targeting
the \CodeIn{Object} constructor and to optimize them into more
efficient \CodeIn{JSCreate} nodes under specific conditions, as shown
in Figure~\ref{figure:case-v8-42210048-fix}.

\textit{Observation 3}: Semantically equivalent/similar programs can
exhibit huge performance differences, and such differences may
indicate performance bugs in \JIT compilers. Checking for such differences
can act as a useful
test oracle for detecting performance bugs.

\MyPara{Case 4: Speculation and recompilation loop}
\textbf{\JDKBug{JDK-8257594}}. This issue demonstrates a performance
bug observed through compiler log outputs, where the \JIT compiler
repeatedly emits \CodeIn{uncommontrap} events, indicating it is stuck
in a cycle of deoptimization and recompilation. In this scenario, the
compiler repeatedly invalidates optimized code when a speculative
assumption fails (e.g., a failed \CodeIn{checkcast}), falls back to
interpreter execution, and then recompiles the same method again under
the same incorrect assumptions. This process--known as a recompilation
loop--continues until the recompilation limit is reached, wasting CPU
cycles and preventing stable optimized code generation. The root cause
is an overly aggressive speculative decision in handling invalid type
checks, leading to persistent deoptimizations. Such loops can severely
impact application throughput and overall runtime efficiency.

\textit{Observation 4}: \JIT compilers can become trapped
in self-reinforcing recompilation loops caused by flawed speculative
assumptions, consuming significant resources without the ability to
converge to a stable optimized state.

\MyPara{Case 5: Interaction with garbage collector}
\textbf{\VBug{42209849}}. This issue reveals a performance regression
of up to 10$\times$ in V8 version 6.1, observed through benchmarks of
a router library.
V8 maintained a per-context weak list of all optimized JavaScript
functions, used primarily for code unlinking during deoptimization.
Although individual entries were weak references (i.e., they did not
prevent the referenced functions from being collected), the garbage
collector still had to traverse the \emph{entire} list on every
scavenge (minor \GC) cycle to update pointers and remove dead entries.
In applications that created many closures---such as router libraries
that generate closures for each registered route---this list grew
large, making the per-scavenge traversal cost increasingly expensive
and dominating overall \GC time.
The fix removed the weak list entirely and replaced the eager code
unlinking with a lazy scheme: a deoptimization mark is now checked in
the prologue of each code object upon its next activation, eliminating
the need for the costly per-\GC-cycle list traversal.

\textit{Observation 5}: The intricate interaction between \JIT
compilers and runtime subsystems such as garbage collection can
magnify minor design oversights into severe performance regressions.


\section{Characteristics of Bugs}
\label{sec:characteristics}

In this section, we study the characteristics of various aspects of
real-world \JIT compiler performance bugs to answer our research
questions (Section~\ref{sec-method:rq}).  Specifically, we study input
artifacts used to trigger performance bugs (Section~\ref{rq1}),
symptoms of performance bugs (Section~\ref{rq2}), and root causes of
performance bugs (Section~\ref{rq3}).
Table~\ref{table-empirical:char} shows the summary characteristics.

\subsection{RQ1: Input Artifacts}
\label{rq1}

\JIT compilers, which are parts of virtual machines, take programs as
input and optimize them.
To study what kind of input artifacts are used in real-world
performance bug reports, we analyzed the provided materials in bug
reports that we selected in Section~\ref{sec-method:analysis}.

\vspace{5pt}
\noindent
We identify the following common input artifacts provided in bug
reports (the top part of Table~\ref{table-empirical:char}):

\begin{itemize}[topsep=2pt,itemsep=3pt,partopsep=0ex,parsep=0ex,leftmargin=*]
    \item \textbf{Micro-benchmarks}
      (\UseMacro{empirical-total-art-tc-percentage}): Small,
      self-contained programs designed to expose specific performance
      issues in \JIT compilers. These tests simplify diagnosis.
    \item \textbf{Benchmark suites}
      (\UseMacro{empirical-total-art-bm-percentage}): Standardized
      performance benchmarks such as
      DaCapo~\cite{blackburn2006dacapo}
      and Renaissance~\cite{prokopec2019renaissance} for \JVM, as well
      as Octane~\cite{octane} and JetStream~\cite{jetstream} for \JS. These
      suites are often integrated into CI/CD pipelines to
      automatically detect performance regressions~\cite{ChromeSpeed}.
    \item \textbf{Applications}
      (\UseMacro{empirical-total-art-app-percentage}): A performance
      issue is detected in real-world application programs and then
      further investigated to originate in the \JIT compiler.
    \item \textbf{Other}
      (\UseMacro{empirical-total-art-other-percentage}): Reports that
      do not specify the reproducing artifact but directly identify
      performance problems in the \JIT compiler, often through code
      inspection or developer self-reporting.
\end{itemize}

\begin{table}[t]
    \centering
    \footnotesize
    \caption{Summary of characteristics of studied performance bugs in
      \JIT compilers. \label{table-empirical:char}}
    \vspace{-10pt}
    \begin{tabular}{|l|rrrr|r|}
    \hline
     & \rotatebox{90}{\textbf{\HotSpot}} & \rotatebox{90}{\textbf{\Graal}} & \rotatebox{90}{\textbf{\VEight}} &
     \rotatebox{90}{\textbf{SM}} & \rotatebox{90}{\textbf{Total}} \\
    \hline
    Number of bugs studied & \UseMacro{empirical-hotspot-studied} &
    \UseMacro{empirical-graal-studied} &
    \UseMacro{empirical-v8-studied} & \UseMacro{empirical-sm-studied}
    & \UseMacro{empirical-total-studied} \\
    \hline
    \rowcolor[gray]{0.85} \multicolumn{6}{|c|}{\textbf{Artifacts Used
    to Trigger Performance Bugs}} \\
    \hline
    \multirow{2}{*}{\makecell[l]{\textbf{Micro-benchmarks}: A simple test
    case or micro-benchmark \\ \textit{Example: \JDKBug{JDK-8300256},
    \GraalBug{6564}}}} &
    \multirow{2}{*}{\UseMacro{empirical-hotspot-art-tc}} &
    \multirow{2}{*}{\UseMacro{empirical-graal-art-tc}} &
    \multirow{2}{*}{\UseMacro{empirical-v8-art-tc}} &
    \multirow{2}{*}{\UseMacro{empirical-sm-art-tc}} &
    \multirow{2}{*}{\UseMacro{empirical-total-art-tc}} \\
    & & & & & \\
    \hline
    \multirow{2}{*}{\makecell[l]{\textbf{Benchmark Suites}: Standardized
    benchmark suites \\ \textit{Example:
    \VBug{40782714}, \JDKBug{JDK-8333583}}}} &
    \multirow{2}{*}{\UseMacro{empirical-hotspot-art-bm}} &
    \multirow{2}{*}{\UseMacro{empirical-graal-art-bm}} &
    \multirow{2}{*}{\UseMacro{empirical-v8-art-bm}} &
    \multirow{2}{*}{\UseMacro{empirical-sm-art-bm}} &
    \multirow{2}{*}{\UseMacro{empirical-total-art-bm}} \\
    & & & & & \\
    \hline
    \multirow{2}{*}{\makecell[l]{\textbf{Applications}: Real-world
    application performance issues that are tracked down to \JIT \\
    \textit{Example: \MozillaBug{1131523}, \GraalBug{2548}}}} &
    \multirow{2}{*}{\UseMacro{empirical-hotspot-art-app}} &
    \multirow{2}{*}{\UseMacro{empirical-graal-art-app}} &
    \multirow{2}{*}{\UseMacro{empirical-v8-art-app}} &
    \multirow{2}{*}{\UseMacro{empirical-sm-art-app}} &
    \multirow{2}{*}{\UseMacro{empirical-total-art-app}} \\
    & & & & & \\
    \hline
    \multirow{2}{*}{\makecell[l]{\textbf{Other}: All the bugs not
    included in the above categories \\ \textit{Example:
    \VBug{42208696}, \MozillaBug{1331350}}}} &
    \multirow{2}{*}{\UseMacro{empirical-hotspot-art-other}} &
    \multirow{2}{*}{\UseMacro{empirical-graal-art-other}} &
    \multirow{2}{*}{\UseMacro{empirical-v8-art-other}} &
    \multirow{2}{*}{\UseMacro{empirical-sm-art-other}} &
    \multirow{2}{*}{\UseMacro{empirical-total-art-other}} \\
    & & & & & \\
    \hline
    \rowcolor[gray]{0.85} \multicolumn{6}{|c|}{\textbf{Symptoms of
    Performance Bugs}} \\
    \hline
    \multirow{2}{*}{\makecell[l]{\textbf{Performance Regression}:
    Performance degradation with new versions \\
    \textit{Example: \JDKBug{JDK-8274983}, \VBug{41480684}}}} &
    \multirow{2}{*}{\UseMacro{empirical-hotspot-sym-pr}} &
    \multirow{2}{*}{\UseMacro{empirical-graal-sym-pr}} &
    \multirow{2}{*}{\UseMacro{empirical-v8-sym-pr}} &
    \multirow{2}{*}{\UseMacro{empirical-sm-sym-pr}} &
    \multirow{2}{*}{\UseMacro{empirical-total-sym-pr}} \\
    & & & & & \\
    \hline
    \multirow{2}{*}{\makecell[l]{\textbf{Performance Difference}: 
    Similar settings deliver different performance\\
    \textit{Example: \VBug{42208216}, \GraalBug{831}}}} &
    \multirow{2}{*}{\UseMacro{empirical-hotspot-sym-pd}} &
    \multirow{2}{*}{\UseMacro{empirical-graal-sym-pd}} &
    \multirow{2}{*}{\UseMacro{empirical-v8-sym-pd}} &
    \multirow{2}{*}{\UseMacro{empirical-sm-sym-pd}} &
    \multirow{2}{*}{\UseMacro{empirical-total-sym-pd}} \\
    & & & & & \\
    \hline
    \multirow{2}{*}{\makecell[l]{\textbf{Log}: Abnormal output
    observed from logs of \JIT compilers or other tools \\
    \textit{Example: \JDKBug{JDK-8280473}, \MozillaBug{1526315}}}} &
    \multirow{2}{*}{\UseMacro{empirical-hotspot-sym-log}} &
    \multirow{2}{*}{\UseMacro{empirical-graal-sym-log}} &
    \multirow{2}{*}{\UseMacro{empirical-v8-sym-log}} &
    \multirow{2}{*}{\UseMacro{empirical-sm-sym-log}} &
    \multirow{2}{*}{\UseMacro{empirical-total-sym-log}} \\
    & & & & & \\
    \hline
    \multirow{2}{*}{\makecell[l]{\textbf{Native Code}: Suboptimal
    native code generated by \JIT compilers \\ \textit{Example:
    \JDKBug{JDK-8176513}, \GraalBug{2010}}}} &
    \multirow{2}{*}{\UseMacro{empirical-hotspot-sym-nc}} &
    \multirow{2}{*}{\UseMacro{empirical-graal-sym-nc}} &
    \multirow{2}{*}{\UseMacro{empirical-v8-sym-nc}} &
    \multirow{2}{*}{\UseMacro{empirical-sm-sym-nc}} &
    \multirow{2}{*}{\UseMacro{empirical-total-sym-nc}} \\
    & & & & & \\
    \hline
    \multirow{2}{*}{\makecell[l]{\textbf{Crash/Hang}: Program crashes
    or hangs \\ \textit{Example: \JDKBug{JDK-8256368}, 
    \VBug{40832269}}}} &
    \multirow{2}{*}{\UseMacro{empirical-hotspot-sym-crash}} &
    \multirow{2}{*}{\UseMacro{empirical-graal-sym-crash}} &
    \multirow{2}{*}{\UseMacro{empirical-v8-sym-crash}} &
    \multirow{2}{*}{\UseMacro{empirical-sm-sym-crash}} &
    \multirow{2}{*}{\UseMacro{empirical-total-sym-crash}} \\
    & & & & & \\
    \hline
    \multirow{2}{*}{\makecell[l]{\textbf{Other}: All the bugs not
    included in the above categories \\ \textit{Example:
    \JDKBug{JDK-8253923}, \VBug{42207398}}}} &
    \multirow{2}{*}{\UseMacro{empirical-hotspot-sym-other}} &
    \multirow{2}{*}{\UseMacro{empirical-graal-sym-other}} &
    \multirow{2}{*}{\UseMacro{empirical-v8-sym-other}} &
    \multirow{2}{*}{\UseMacro{empirical-sm-sym-other}} &
    \multirow{2}{*}{\UseMacro{empirical-total-sym-other}} \\
    & & & & & \\
    \hline
    \rowcolor[gray]{0.85} \multicolumn{6}{|c|}{\textbf{Root Causes of
    Performance Bugs}} \\
    \hline
    \multirow{2}{*}{\makecell[l]{\textbf{Optimization}: Missed
    optimization opportunities or inappropriate optimizations \\
    \textit{Example: \JDKBug{JDK-8279888}, \GraalBug{701}}}} &
    \multirow{2}{*}{\UseMacro{empirical-hotspot-rc-opt}} &
    \multirow{2}{*}{\UseMacro{empirical-graal-rc-opt}} &
    \multirow{2}{*}{\UseMacro{empirical-v8-rc-opt}} &
    \multirow{2}{*}{\UseMacro{empirical-sm-rc-opt}} &
    \multirow{2}{*}{\UseMacro{empirical-total-rc-opt}} \\
    & & & & & \\
    \hline
    \multirow{2}{*}{\makecell[l]{\textbf{Speculation}: Suboptimal
    speculative decisions \\ \textit{Example: \VBug{42209544},
    \MozillaBug{1644425}}}} &
    \multirow{2}{*}{\UseMacro{empirical-hotspot-rc-spec}} &
    \multirow{2}{*}{\UseMacro{empirical-graal-rc-spec}} &
    \multirow{2}{*}{\UseMacro{empirical-v8-rc-spec}} &
    \multirow{2}{*}{\UseMacro{empirical-sm-rc-spec}} &
    \multirow{2}{*}{\UseMacro{empirical-total-rc-spec}} \\
    & & & & & \\
    \hline
    \multirow{2}{*}{\makecell[l]{\textbf{Code Generation}: Suboptimal
    native code sequence \\ \textit{Example:
    \JDKBug{JDK-8172434}, \MozillaBug{1503116}}}} &
    \multirow{2}{*}{\UseMacro{empirical-hotspot-rc-codegen}} &
    \multirow{2}{*}{\UseMacro{empirical-graal-rc-codegen}} &
    \multirow{2}{*}{\UseMacro{empirical-v8-rc-codegen}} &
    \multirow{2}{*}{\UseMacro{empirical-sm-rc-codegen}} &
    \multirow{2}{*}{\UseMacro{empirical-total-rc-codegen}} \\
    & & & & & \\
    \hline
    \multirow{2}{*}{\makecell[l]{\textbf{Interaction with Runtime}:
    Interaction between \JIT and other components  \\ \textit{Example:
    \JDKBug{JDK-8155638}, \VBug{42209849}}}} &
    \multirow{2}{*}{\UseMacro{empirical-hotspot-rc-runtime}} &
    \multirow{2}{*}{\UseMacro{empirical-graal-rc-runtime}} &
    \multirow{2}{*}{\UseMacro{empirical-v8-rc-runtime}} &
    \multirow{2}{*}{\UseMacro{empirical-sm-rc-runtime}} &
    \multirow{2}{*}{\UseMacro{empirical-total-rc-runtime}} \\
    & & & & & \\
    \hline
    \multirow{2}{*}{\makecell[l]{\textbf{Long Compilation}: \JIT
    compilers spend a significant amount of time compiling code \\
    \textit{Example: \MozillaBug{1318926}, \GraalBug{2548}}}} &
    \multirow{2}{*}{\UseMacro{empirical-hotspot-rc-lc}} &
    \multirow{2}{*}{\UseMacro{empirical-graal-rc-lc}} &
    \multirow{2}{*}{\UseMacro{empirical-v8-rc-lc}} &
    \multirow{2}{*}{\UseMacro{empirical-sm-rc-lc}} &
    \multirow{2}{*}{\UseMacro{empirical-total-rc-lc}} \\
    & & & & & \\
    \hline
    \multirow{2}{*}{\makecell[l]{\textbf{Other}: All the bugs not
    included in the above categories \\ \textit{Example:
    \JDKBug{JDK-8302736}, \VBug{40782714}}}} &
    \multirow{2}{*}{\UseMacro{empirical-hotspot-rc-other}} &
    \multirow{2}{*}{\UseMacro{empirical-graal-rc-other}} &
    \multirow{2}{*}{\UseMacro{empirical-v8-rc-other}} &
    \multirow{2}{*}{\UseMacro{empirical-sm-rc-other}} &
    \multirow{2}{*}{\UseMacro{empirical-total-rc-other}} \\
    & & & & & \\
    \hline
    \end{tabular}
    \vspace{-10pt}
\end{table}

\vspace{3pt}
\noindent
While benchmark suites are widely used to evaluate the overall
performance of \JIT compilers, we find that they detect only about
one-fourth of all performance bugs. In contrast, nearly half of the
reported bugs are demonstrated by micro-benchmarks--small, focused
programs that test specific compiler features. These micro-benchmarks
are either explicitly designed to exercise particular language
features or distilled from real-world examples. Because benchmark
suites consist of a large fixed set of programs, they are not
well-suited for revealing previously unknown performance issues and
take a long time to run. By comparison, although micro-benchmarks
target narrower and sometimes less common scenarios, their focused
nature enables them to uncover unknown issues and subtle corner cases,
while remaining lightweight, fast to execute, and easier to diagnose.

Moreover, during our dataset construction, we observed that the
\VEight issue tracker contains numerous false positives (\ie, reports
later marked as ``Won't Fix'' or ``Obsolete'') generated by periodic
benchmark runs~\cite{v8-fp}. This observation highlights the
limitations of existing benchmark suites, which may fail to isolate
compiler-specific regressions from performance noise.

\answerrqbox{RQ1}{Micro-benchmarks play a crucial role in exposing
  \JIT performance bugs, either by testing specific compiler
  optimizations or runtime behaviors, or by isolating the minimal
  program patterns from real-world examples. For the automated
  detection of performance bugs in \JIT compilers, future work should
  focus on generating small, feature-specific programs that exercise
  distinct compiler optimizations or runtime behaviors, such as
  inlining, loop unrolling, speculative checks, etc.}

\subsection{RQ2: Symptoms}
\label{rq2}

Performance bugs are typically identified through abnormal execution
behaviors. We summarize the common symptoms observed in the collected
bug reports (the middle part of Table~\ref{table-empirical:char}):

\begin{itemize}[topsep=2pt,itemsep=3pt,partopsep=0ex,parsep=0ex,leftmargin=*]
\item \textbf{Performance regression}
      (\UseMacro{empirical-total-sym-pr-percentage}): A significant
  degradation in performance after \JIT compiler version updates.
    \item \textbf{Performance difference}
      (\UseMacro{empirical-total-sym-pd-percentage}): Noticeable
      performance gaps between equivalent executions--such as across
      platforms, vendors, or semantically identical program
      variants--often indicate compiler inefficiencies or missed
      optimizations for certain platforms or code patterns.
    \item \textbf{Log}
      (\UseMacro{empirical-total-sym-log-percentage}): Abnormal entries
      in compiler or runtime logs, or anomalies revealed by monitoring
      and profiling tools, signaling unexpected performance behaviors.
    \item \textbf{Native code}
    (\UseMacro{empirical-total-sym-nc-percentage}): (Manually)
    inspecting the generated native code reveals inefficient
    instruction sequences or missing optimizations.
    \item \textbf{Crash/Hang}
    (\UseMacro{empirical-total-sym-crash-percentage}): Severe
    performance issues may lead to crashes, hangs, or assertion
    failures triggered by incorrect compiler assumptions.
    \item \textbf{Other}
    (\UseMacro{empirical-total-sym-other-percentage}): Other irregular
    behaviors, such as performance fluctuations, that indicate
    unstable or inconsistent optimization decisions.
\end{itemize}

\vspace{3pt}
\noindent
Unlike \functional bugs, whose oracles are well-defined (\ie, program
crashes or incorrect outputs), performance bugs lack a clear ground
truth.
The diversity observed in the list above underscores the difficulty of
defining reliable oracles for performance validation.

The strong dependence on comparative signals suggests that
\emph{differential testing}~\cite{mckeeman1998differential} is
particularly suitable for uncovering performance bugs in \JIT
compilers. Effective oracle design should try to combine quantitative
metrics (\eg, execution time, compilation cost) with qualitative
indicators (\eg, anomalous log entries or inefficient native code) to
detect subtle performance degradations more systematically.

\answerrqbox{RQ2}{Performance bugs in \JIT compilers lack simple
  oracles and are often detected through various comparative
  evidence. Test oracles often employ a differential
  approach--comparing execution across compiler versions,
  configurations, or semantically equivalent programs--while
  integrating lightweight checks such as profiling counters, log
  anomalies, or code pattern analysis to pinpoint unexpected slowdowns
  and missed optimizations.}

\subsection{RQ3: Root Causes}
\label{rq3}

To understand commonly buggy parts in \JIT compilers and how to
prevent or detect issues in these parts, we inspect the root causes in
the studied bug reports. Some common root causes include (the bottom
part of Table~\ref{table-empirical:char}):

\begin{itemize}[topsep=2pt,itemsep=3pt,partopsep=0ex,parsep=0ex,leftmargin=*]
    \item \textbf{Optimization}
      (\UseMacro{empirical-total-rc-opt-percentage}): Bugs caused by
      missed or incorrect optimizations, such as redundant
      computations, unoptimized loops, or overly aggressive
      transformations that miss certain edge cases. These bugs
      typically result from flawed heuristics or incomplete cost
      models.
    \item \textbf{Speculation}
      (\UseMacro{empirical-total-rc-spec-percentage}): Bugs caused by
      invalid or misleading speculative assumptions derived from
      profiling data, or overly aggressive speculation that fails too
      often with high deoptimization and recompilation costs.
    \item \textbf{Code generation}
      (\UseMacro{empirical-total-rc-codegen-percentage}):
      Inefficiencies in instruction selection, register allocation, or
      scheduling that produce suboptimal native code. Such issues are
      similar to those in \AOT compilers but are complicated by
      additional runtime guards or stubs inserted by \JIT compilers.
    \item \textbf{Interaction with runtime}
      (\UseMacro{empirical-total-rc-runtime-percentage}): Performance
      issues caused by inefficient coordination between the \JIT
      compiler and other runtime components, such as code cache
      management, dynamic compilation triggers, or garbage collectors.
    \item \textbf{Long compilation}
      (\UseMacro{empirical-total-rc-lc-percentage}): Bugs due to
      excessive compilation time or resource usage, which delay code
      execution and reduce overall responsiveness.
    \item \textbf{Other}
      (\UseMacro{empirical-total-rc-other-percentage}): Issues related
      to internal design decisions, such as suboptimal data structures
      or algorithms, or cross-component interactions.
\end{itemize}

\noindent
While optimizations, speculations, and code generation are closely
intertwined and often influence each other, we distinguish between
them using the following criteria. We classify speculation bugs as
those related to the collection or use of profiling data, including
setting profiling data, generating intermediate representation (IR)
based on such data, and the deoptimization process. In contrast, we
classify optimization bugs as those arising from IR transformations
that are independent of profiling data, and code generation bugs as
those related to the backend of a compiler---transforming IR to native
code.

For example, in \JDKBug{JDK-8280320}, certain profiling data is
incorrectly set, causing the compiler to miss certain loop
optimizations even though the loop optimizations themselves are
correctly implemented, so we classify it as a speculation bug. On the
other hand, consider the case of \JDKBug{JDK-8279888}, there is an
implementation flaw in the loop optimization itself, so we classify it
as an optimization bug, since the problem will arise whenever the
optimization is performed.

Traditionally, the optimization and code generation phases have been
recognized as error-prone in \AOT compilers. Our study reveals that
\JIT-specific components, particularly speculation and runtime
interaction, are also significant sources of performance bugs,
especially for dynamically-typed languages like \JS.

These components are difficult to validate using conventional compiler
testing, because their correctness depends on dynamic runtime
behavior, profiling feedback, and adaptive decisions made during
execution. To improve robustness, \JIT compiler testing should focus
on systematically validating speculative and adaptive mechanisms under
diverse runtime conditions. This includes stress-testing
profiling-guided optimizations, evaluating deoptimization and
recompilation triggers, and ensuring stable performance across varying
workload characteristics.

\answerrqbox{RQ3}{Beyond traditional optimization and code generation
  testing, \JIT compilers require focused validation of dynamic
  behaviors. Testing frameworks should systematically emulate diverse
  profiling patterns, workload phases, and execution environments to
  capture the impact of dynamic feedback on compilation decisions. In
  particular, they should stress-test speculation heuristics,
  deoptimization triggers, and tiered compilation transitions to
  reveal unstable or overly aggressive adaptive behaviors that lead to
  performance degradation.}

\subsection{Fixes}

We also examined the fixes proposed in each bug report to understand
how performance issues are addressed. Unlike general software
performance bugs, where fixes often follow recognizable patterns such
as adjusting condition checks or adding early
exits~\cite{jin2012understanding}, we did not observe clear or
recurring fix patterns in \JIT compilers.

Two main factors contribute to this observation. (1)~\JIT compilers
consist of tightly coupled components and phases, so fixes frequently
involve coordinated edits across multiple modules. (2)~The performance
impact of source-level changes is difficult to predict, as even small
modifications can alter the generated machine code in complex ways.
Consequently, resolving these issues often requires deep,
domain-specific understanding of compiler internals and optimization
interactions.

These findings highlight the importance of modular design in \JIT
compilers to improve maintainability and isolate performance-critical
logic. For instance, \SpiderMonkey recently introduced
CacheIR~\cite{de2023cacheir}, a structured intermediate representation
for inline caches, an essential speculative mechanism for optimizing
dynamic property access~\cite{deutsch1984efficient}, to make this
component more maintainable and extensible.


\section{\Tool: \xName}
\label{sec:tech}

Based on our empirical study on input artifacts (Section~\ref{rq1}),
\UseMacro{empirical-total-art-tc-percentage} of performance bugs can
be demonstrated through micro-benchmarks, which are simple test cases
that target specific features of \JIT compilers. Thus, the goal of our
approach is to generate a large number of small programs that can
potentially reveal performance issues. We use various program
generators, which have been previously shown effective at finding
\functional bugs in both \AOT and \JIT
compilers~\cite{yang2011finding, ZangETAL22JAttack, li2023validating,
  li2024boosting, gross2023fuzzilli}. Their potential for
finding performance bugs has never been studied. Moreover, based on
the common symptoms observed in real-world performance bugs
(Section~\ref{rq2}) such as performance difference
(\UseMacro{empirical-total-sym-pd-percentage}) and performance
regression (\UseMacro{empirical-total-sym-pr-percentage}),
differential testing proved to be an effective approach for
detecting performance bugs.  To balance the trade-off between test
efficiency and accuracy from a large number of generated programs, we
designed and developed \Tool, a tool that implements \emph{\xname}
with test oracles for detecting performance bugs in \JIT compilers.

We applied \Tool to the \HotSpot and \Graal \JIT compilers and found
\NumOfFound previously unknown performance bugs with \NumOfBugs
confirmed or fixed by developers.

\begin{figure*}[t]
    \centering
    \includegraphics[width=\columnwidth]{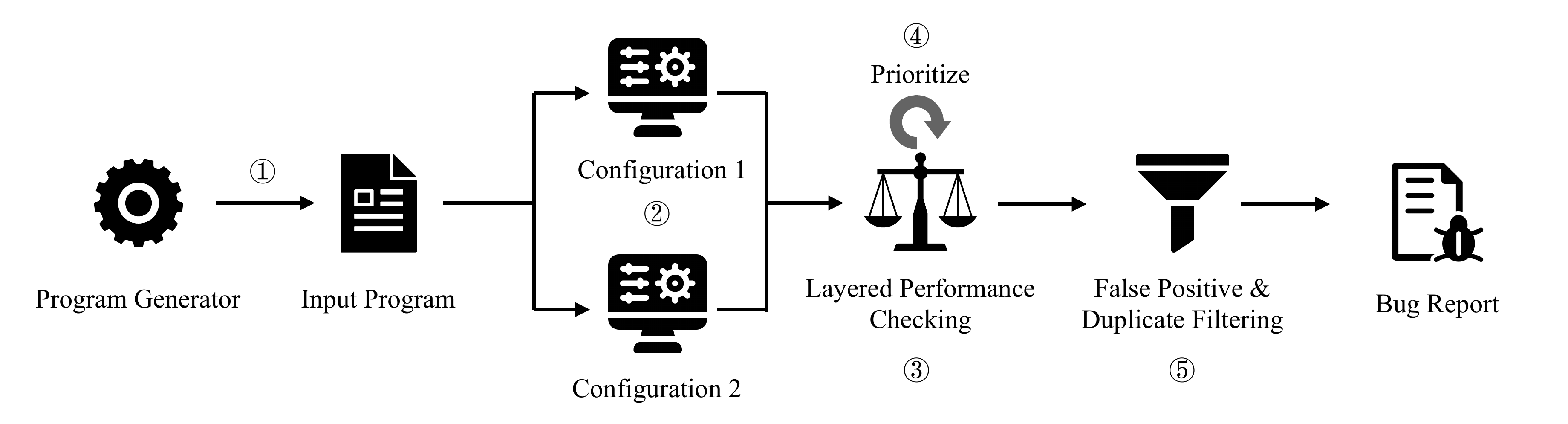}
    \vspace{-20pt}
    \caption{Overview of \xname.\label{fig:tech-ov}}
\end{figure*}

\subsection{Overview}

Figure~\ref{fig:tech-ov} shows the overview of \Tool. We next describe
the key parts of \Tool.

\MyPara{Program generators (\textcircled{1})}
The first step is to generate input micro-benchmarks that are likely
to trigger certain \JIT features. Due to the abundance of existing
program generators for finding \functional bugs, we leverage existing
tools to generate programs.  We use three generators in our
experiments: \Artemis~\cite{li2023validating},
\JavaFuzzer~\cite{javafuzzer}, and \Lejit~\cite{ZangETAL24LeJit}.

\Artemis is a mutation-based generator that mutates given programs in
a semantics-preserving way to trigger different \JIT compilation
decisions. \JavaFuzzer is a random program generator, which is also
used by \Artemis to generate seed programs.  \Lejit is a
template-based generator that generates programs by filling in holes
in templates extracted from existing \Java projects.
To ensure \JIT compilation is triggered, we invoke generated code a
large number of times (by wrapping the code in a loop).

\MyPara{Differential configurations (\textcircled{2})}
The next step is to run the generated programs with a pair of
configurations. There are many possible choices of such configurations
as described in Section~\ref{rq2} (\eg, the original program and a
mutated but semantically equivalent version of it). We explore two
kinds of configurations in this paper: (1)~Regression Pair (RP) and
(2)~Level Pair (LP). RP uses two different versions of the \JIT
compiler under test. LP leverages the tiered compilation feature of
modern \JIT compilers (Section~\ref{sec:bg-tiered}).
In this paper, we use levels L1 and L4 for \HotSpot and \Graal, which
exercises \COne and \CTwo compilers.
This way we enable different sets of optimizations to be applied, as
well as different use of profiling and speculative optimizations.

\MyPara{\Leveled performance checking (\textcircled{3})}
We gradually filter out programs that are not likely to trigger
performance bugs to balance test efficiency and accuracy, as
doing rigorous benchmarking for every program is costly. Recall that
we execute the entry method in a loop with a set number of iterations
to trigger \JIT compilation. We choose an ascending list of iteration
counts ($Ns$). We refer to each iteration count as a \emph{performance
checking \level}. We run each program against these \levels and when a
program does not show an interesting performance difference, we skip
the remaining \levels.  When we increase the iteration count for
programs of interest, the benefit is two-fold: (1)~the increased
iteration count allows the warmup overhead to be amortized over more
executions, reducing its impact on the measured performance; (2)~the
increased iteration count can augment potential performance issues
and make them more observable.
      
\MyPara{Test inputs prioritization (\textcircled{4})}
Across different performance checking \levels, we can use runtime
information from previous \level{}(s) to prioritize which test inputs
(namely programs) to run first in the next \level. This step can help
find bugs faster with limited resources.

\MyPara{False positive and duplicate filtering (\textcircled{5})}
After all the performance testing is done, although the remaining
programs are very likely to be true positives, there could potentially
exist false positives due to inherent noise in performance measurement
and duplicate bugs with the same root cause. To mitigate this problem,
we use a few simple heuristics to automatically filter out false
positives or duplicates.  We find these heuristics to be effective and
can greatly reduce manual effort.
Finally, we do more thorough and rigorous performance measurement for
remaining candidates before reporting them to compiler developers.

\begin{small}
\begin{algorithm}[t!]
    \caption{\xName with Prioritization}
    \label{algo:tdt}
    \begin{algorithmic}[1]
        \REQUIRE Programs $Ps$, \newline\hspace*{0.5cm} Configuration Pair $CP = (C_1, C_2)$, \newline
                 \hspace*{0.5cm} Iterations $Ns$, Thresholds $THs$, Top-K limits $TOP\_Ks$
        \ENSURE List of booleans $Rs$ indicating if a potential performance bug exists for each program
        \STATE $n \gets |Ns|$ \label{line:init}
        \STATE Initialize historical measurements $H \gets [\,]$ \label{line:hist-init}
        \STATE Initialize result list $Rs[P] \gets \text{true}$ for all $P \in Ps$ \label{line:init-rs}
        \FOR{$i \gets 1$ \TO $n$} \label{line:outer-loop-start}
            \STATE $Ps' \gets \text{Prioritize}(Ps, H)$ \label{line:prioritize}
            \IF{$i > 1$ \AND $TOP\_Ks[i-1] > 0$} \label{line:topk-check}
                \STATE $k \gets TOP\_Ks[i-1]$
                \STATE $T \gets \{P \in Ps' \mid Rs[P]=\text{true}\}$
                \STATE $Ps' \gets Ps'[0:\min(k,|T|)]$ \label{line:topk}

            \ENDIF
            \FORALL{$P$ in $Ps'$} \label{line:prog-loop-start}
                \IF{$Rs[P] = \text{false}$}
                    \STATE \textbf{continue} \label{line:skip} 
                \ENDIF
                \STATE $m_1 \gets \text{Execute}(P, C_1, Ns[i])$ \label{line:exec-c1}
                \STATE $m_2 \gets \text{Execute}(P, C_2, Ns[i])$ \label{line:exec-c2}
                \IF{$\neg \text{Check}(m_1, m_2, THs[i])$} \label{line:check}
                    \STATE $Rs[P] \gets \text{false}$ \label{line:mark-false}
                \ENDIF
                \STATE $H[P] \gets \text{UpdateHistory}(H[P], (m_1, m_2))$ \label{line:update-history}
            \ENDFOR \label{line:prog-loop-end}
        \ENDFOR \label{line:outer-loop-end}
        \RETURN $Rs$ \label{line:final-return}
    \end{algorithmic}
\end{algorithm}
\end{small}

\subsection{Algorithm}
\label{sec-app:algo}

We formally describe the \xname (Algorithm~\ref{algo:tdt}). The
algorithm takes as input a set of programs $Ps$, a configuration pair
$CP = (C_1, C_2)$, a list of iteration counts $Ns$ (one per \level),
thresholds $THs$ (one per \level), and a list of top-$k$ limits
$TOP\_Ks$ (with $|Ns| = |THs| = n$ and $|TOP\_Ks| = n - 1$).

Initially (line~\ref{line:init}), the algorithm determines the number
of \levels $n$, which equals the length of $Ns$.  It then
initializes an empty history structure $H$ to store past performance
measurements (line~\ref{line:hist-init}) and prepares a result list
$Rs$, where each entry corresponds to a program and is initialized to
\texttt{true} (line~\ref{line:init-rs}).

For each iteration, the algorithm reorders the program list using the
prioritization function $\text{Prioritize}(Ps, H)$
(line~\ref{line:prioritize}), which leverages previous execution
history to determine the order in which programs should be tested.
Starting from the second iteration ($i > 1$), only the top-$k$
programs after prioritization are selected, where $k = TOP\_Ks[i-1]$,
and we skip the top-$k$ selection if $TOP\_Ks[i-1] < 0$
(line~\ref{line:topk-check}).
If fewer than $k$ programs remain active (\ie, still marked as
\texttt{true} in $Rs$), all remaining programs are executed
(line~\ref{line:topk}).

Within each iteration, the algorithm evaluates every selected program
$P$ (line~\ref{line:prog-loop-start}).
If a program has already been filtered out in a previous iteration, it
is skipped (line~\ref{line:skip}). Otherwise, the program is executed
twice--once under configuration $C_1$ and once under configuration
$C_2$ (lines~\ref{line:exec-c1}--\ref{line:exec-c2})--to obtain two
performance measurements $m_1$ and $m_2$. The function
$\text{Check}(m_1, m_2, THs[i])$ (line~\ref{line:check}) compares
these measurements to determine whether the ratio between them exceeds
the current threshold $THs[i]$. If this condition is not satisfied,
the corresponding entry $Rs[P]$ is updated to \texttt{false}
(line~\ref{line:mark-false}), indicating that the program is unlikely
to reveal a performance bug and will not be tested in future
iterations. Finally, the historical record $H[P]$ is updated with the
new measurement pair $(m_1, m_2)$ (line~\ref{line:update-history}).

There are some customizable functions in the algorithm:
$\text{Execute}(P, C, N)$, $\text{Check}(m_1, m_2, TH)$,
$\text{UpdateHistory}(H[P], (m_1, m_2))$, and $\text{Prioritize}(Ps,
H)$. In this paper, we define them in a straightforward way, but
future work should explore more sophisticated approaches:

\begin{itemize}[topsep=2pt,itemsep=3pt,partopsep=0ex,parsep=0ex,leftmargin=*]
    \item \textbf{Execute:} We run entry method in the program $P$
      under the given configuration $C$ for $N$ iterations and measure
      the execution time using hardware performance
      counters~\cite{hpc}.  Other information (\eg, logs and native
      code as we show in Section~\ref{rq2}), besides the execution
      time, could also be collected and used.
    \item \textbf{Check:} We assume there is a configuration (\ie, a
      newer compiler version or a higher compilation level) which
      runs faster than another configuration (an older compiler or a
      lower compilation level).  We check if the ratio ($m_2 / m_1$)
      of the two measurements exceeds the threshold $TH$.  If so, we
      consider this program is likely to reveal a performance bug and
      return true.
    \item \textbf{UpdateHistory:} We do not consider all previous
      history in this paper, but only the latest measurement pair.
    \item \textbf{Prioritize:} We prioritize programs based on their
      previous performance measurements. We compute the ratio of the
      two measurements from the latest history and sort programs in
      descending order based on this ratio. The intuition is that a
      larger ratio indicates a more significant performance difference
      between the two configurations, making it more likely to reveal
      a performance bug. A more sophisticated prioritization method
      can also exploit features from the program under test itself and
      combine the static and dynamic information.
\end{itemize}

\subsection{Filtering False Positives and Duplicates}
We now describe how we use simple heuristics to filter out false
positives and duplicate bugs when the algorithm finishes.

For false positive filtering, consider the case where the performance
difference for executing the entry method once is denoted as $m_d =
|m_1 - m_2|$. As the iteration count increases, the accumulated
difference $m_d \times Ns[i]$ is expected to grow proportionally.  If
a program's accumulated performance difference across multiple
performance-checking \levels does not exhibit such an increasing
trend, it is likely a false positive caused by constant compilation
overheads or transient environmental noise. This heuristic effectively
eliminates cases where the observed performance gap is not a true
reflection of runtime behavior differences for some short running
programs.

We apply duplicate filtering to programs generated using \Lejit.
Different programs that are generated from the same template (\ie,
they are structurally identical and only differ in certain values or
expressions) often expose bugs that have identical root cause. To
avoid redundant reporting, we record the template in a known-bug set
and exclude subsequent findings with identical templates.
In addition, we observe that exceptions frequently correlate with
performance bugs.  Therefore, we also group programs (for each program
generator) by their thrown exception types (if any) and filter out new
bugs that share the same exception. This combined strategy efficiently
removes redundant bugs while retaining distinct performance issues for
manual inspection.

\subsection{Evaluation}

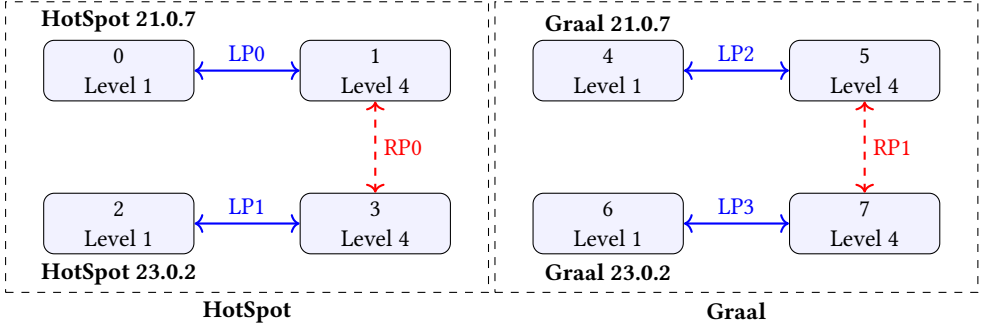
\begin{figure}[t!]
  \centering
  \begin{small}
\begin{tikzpicture}[
  node distance=1.2cm and 1.4cm,
  level/.style={rectangle, draw, rounded corners, fill=blue!5, align=center, minimum width=2cm},
  regression/.style={<->, thick, red, dashed},
  levelpair/.style={<->, thick, blue},
  every label/.style={font=\small\bfseries, align=center}
]

\node[level, label=above:{HotSpot 21.0.7}] (hs21l1) {0\\Level 1};
\node[level, right=of hs21l1] (hs21l4) {1\\Level 4};

\node[level, below=of hs21l1, label=below:{HotSpot 23.0.2}] (hs23l1) {2\\Level 1};
\node[level, right=of hs23l1] (hs23l4) {3\\Level 4};

\node[level, right=4.5cm of hs21l1, label=above:{Graal 21.0.7}] (gr21l1) {4\\Level 1};
\node[level, right=of gr21l1] (gr21l4) {5\\Level 4};

\node[level, below=of gr21l1, label=below:{Graal 23.0.2}] (gr23l1) {6\\Level 1};
\node[level, right=of gr23l1] (gr23l4) {7\\Level 4};

\draw[levelpair] (hs21l1) -- node[above]{LP0} (hs21l4);
\draw[levelpair] (hs23l1) -- node[above]{LP1} (hs23l4);
\draw[levelpair] (gr21l1) -- node[above]{LP2} (gr21l4);
\draw[levelpair] (gr23l1) -- node[above]{LP3} (gr23l4);

\draw[regression] (hs21l4) -- node[right]{RP0} (hs23l4);
\draw[regression] (gr21l4) -- node[right]{RP1} (gr23l4);

\node[draw, dashed, fit=(hs21l1)(hs23l4), inner sep=0.5cm, label=below:{\textbf{HotSpot}}] {};
\node[draw, dashed, fit=(gr21l1)(gr23l4), inner sep=0.5cm, label=below:{\textbf{Graal}}] {};

\end{tikzpicture}
\end{small}
\vspace{-10pt}
\caption{Visualization of JDK configurations setup.  Blue lines
  (LP0-LP3) connect optimization levels within each version; red
  dashed lines (RP0-RP1) connect regression pairs across
  versions.\label{fig:eval-setup}}
\vspace{-5pt}
\end{figure}

In this section, we describe the steps we took to assess the benefits
of using \Tool, and we report our results and findings.

\MyPara{Experiment setup}
We use two \JDK versions 21.0.7 and 23.0.2, with each version
corresponding to two \JIT engines---\HotSpot and \Graal. This setup gives
us four \JIT compilers, eight configurations, four pairs of level
configurations, and two pairs of regression pairs as depicted in
Figure~\ref{fig:eval-setup}. We run using an Ubuntu 24.04.2 LTS
machine with an Intel(R) Xeon(R) w5-3433 CPU and 128GB memory. We use
the option \CodeIn{-XX:TieredStopAtLevel} to limit the reachable
tiered execution levels, and the option \CodeIn{-Xbatch} to turn off
background compilation.  We report detailed results for \Lejit as we
did more extensive experiments with it.

\begin{table}[p]
\centering
\scriptsize
\caption{Evaluation results of \Tool using \Lejit. Columns show the
  project name; numbers of extracted templates and generated programs;
  programs filtered at each \level; remaining programs after each
  filtering step; and total execution time (excluding the first
  \level).
  \label{table-eval:tdt}}
\vspace{-5pt}
\begin{tabular}{lrrrrrrrrrrrr}
\toprule
\multirow{2}{*}{\centering \rotatebox{60}{\textbf{Project}}} & \multicolumn{2}{c}{\textbf{Input}} & \multirow{2}{*}{\centering \rotatebox{60}{\textbf{Pair}}} & \multicolumn{4}{c}{\textbf{\Leveled Filtering}} & \multirow{2}{*}{\centering \rotatebox{60}{\textbf{Pass}}} & \multirow{2}{*}{\textbf{\shortstack{False\\Positive\\Filtering}}} & \multirow{2}{*}{\textbf{\shortstack{Duplicate\\Filtering}}} & \multirow{3}{*}{\rotatebox{60}{\textbf{Unique}}} & \multirow{3}{*}{\rotatebox{60}{\textbf{\shortstack{Time\\{}[min]}}}} \\
\cmidrule(lr){2-3} \cmidrule(lr){5-8}
 & \#Template & \#Gen & & $TH_0$ & $TH_1$ & $TH_2$ & $TH_3$ & & & & & \\
\midrule
\multirow{6}{*}{codec} & \multirow{6}{*}{\UseMacro{table-tdt-codec-tmpl}} & \multirow{6}{*}{\UseMacro{table-tdt-codec-gen}} & LP0 & \UseMacro{table-tdt-codec-lp0-th0} & \UseMacro{table-tdt-codec-lp0-th1} & \UseMacro{table-tdt-codec-lp0-th2} & \UseMacro{table-tdt-codec-lp0-th3} & \UseMacro{table-tdt-codec-lp0-th4} & \UseMacro{table-tdt-codec-lp0-fp-filtered} & \UseMacro{table-tdt-codec-lp0-dup-filtered} & \UseMacro{table-tdt-codec-lp0-tp} & \multirow{6}{*}{\UseMacro{table-tdt-codec-time}} \\
 &  &  & LP1 & \UseMacro{table-tdt-codec-lp1-th0} & \UseMacro{table-tdt-codec-lp1-th1} & \UseMacro{table-tdt-codec-lp1-th2} & \UseMacro{table-tdt-codec-lp1-th3} & \UseMacro{table-tdt-codec-lp1-th4} & \UseMacro{table-tdt-codec-lp1-fp-filtered} & \UseMacro{table-tdt-codec-lp1-dup-filtered} & \UseMacro{table-tdt-codec-lp1-tp} & \\
 &  &  & LP2 & \UseMacro{table-tdt-codec-lp2-th0} & \UseMacro{table-tdt-codec-lp2-th1} & \UseMacro{table-tdt-codec-lp2-th2} & \UseMacro{table-tdt-codec-lp2-th3} & \UseMacro{table-tdt-codec-lp2-th4} & \UseMacro{table-tdt-codec-lp2-fp-filtered} & \UseMacro{table-tdt-codec-lp2-dup-filtered} & \UseMacro{table-tdt-codec-lp2-tp} & \\
 &  &  & LP3 & \UseMacro{table-tdt-codec-lp3-th0} & \UseMacro{table-tdt-codec-lp3-th1} & \UseMacro{table-tdt-codec-lp3-th2} & \UseMacro{table-tdt-codec-lp3-th3} & \UseMacro{table-tdt-codec-lp3-th4} & \UseMacro{table-tdt-codec-lp3-fp-filtered} & \UseMacro{table-tdt-codec-lp3-dup-filtered} & \UseMacro{table-tdt-codec-lp3-tp} & \\
 &  &  & RP0 & \UseMacro{table-tdt-codec-rp0-th0} & \UseMacro{table-tdt-codec-rp0-th1} & \UseMacro{table-tdt-codec-rp0-th2} & \UseMacro{table-tdt-codec-rp0-th3} & \UseMacro{table-tdt-codec-rp0-th4} & \UseMacro{table-tdt-codec-rp0-fp-filtered} & \UseMacro{table-tdt-codec-rp0-dup-filtered} & \UseMacro{table-tdt-codec-rp0-tp} & \\
 &  &  & RP1 & \UseMacro{table-tdt-codec-rp1-th0} & \UseMacro{table-tdt-codec-rp1-th1} & \UseMacro{table-tdt-codec-rp1-th2} & \UseMacro{table-tdt-codec-rp1-th3} & \UseMacro{table-tdt-codec-rp1-th4} & \UseMacro{table-tdt-codec-rp1-fp-filtered} & \UseMacro{table-tdt-codec-rp1-dup-filtered} & \UseMacro{table-tdt-codec-rp1-tp} & \\
\midrule
\multirow{6}{*}{compress} & \multirow{6}{*}{\UseMacro{table-tdt-compress-tmpl}} & \multirow{6}{*}{\UseMacro{table-tdt-compress-gen}} & LP0 & \UseMacro{table-tdt-compress-lp0-th0} & \UseMacro{table-tdt-compress-lp0-th1} & \UseMacro{table-tdt-compress-lp0-th2} & \UseMacro{table-tdt-compress-lp0-th3} & \UseMacro{table-tdt-compress-lp0-th4} & \UseMacro{table-tdt-compress-lp0-fp-filtered} & \UseMacro{table-tdt-compress-lp0-dup-filtered} & \UseMacro{table-tdt-compress-lp0-tp} & \multirow{6}{*}{\UseMacro{table-tdt-compress-time}} \\
 &  &  & LP1 & \UseMacro{table-tdt-compress-lp1-th0} & \UseMacro{table-tdt-compress-lp1-th1} & \UseMacro{table-tdt-compress-lp1-th2} & \UseMacro{table-tdt-compress-lp1-th3} & \UseMacro{table-tdt-compress-lp1-th4} & \UseMacro{table-tdt-compress-lp1-fp-filtered} & \UseMacro{table-tdt-compress-lp1-dup-filtered} & \UseMacro{table-tdt-compress-lp1-tp} & \\
 &  &  & LP2 & \UseMacro{table-tdt-compress-lp2-th0} & \UseMacro{table-tdt-compress-lp2-th1} & \UseMacro{table-tdt-compress-lp2-th2} & \UseMacro{table-tdt-compress-lp2-th3} & \UseMacro{table-tdt-compress-lp2-th4} & \UseMacro{table-tdt-compress-lp2-fp-filtered} & \UseMacro{table-tdt-compress-lp2-dup-filtered} & \UseMacro{table-tdt-compress-lp2-tp} & \\
 &  &  & LP3 & \UseMacro{table-tdt-compress-lp3-th0} & \UseMacro{table-tdt-compress-lp3-th1} & \UseMacro{table-tdt-compress-lp3-th2} & \UseMacro{table-tdt-compress-lp3-th3} & \UseMacro{table-tdt-compress-lp3-th4} & \UseMacro{table-tdt-compress-lp3-fp-filtered} & \UseMacro{table-tdt-compress-lp3-dup-filtered} & \UseMacro{table-tdt-compress-lp3-tp} & \\
 &  &  & RP0 & \UseMacro{table-tdt-compress-rp0-th0} & \UseMacro{table-tdt-compress-rp0-th1} & \UseMacro{table-tdt-compress-rp0-th2} & \UseMacro{table-tdt-compress-rp0-th3} & \UseMacro{table-tdt-compress-rp0-th4} & \UseMacro{table-tdt-compress-rp0-fp-filtered} & \UseMacro{table-tdt-compress-rp0-dup-filtered} & \UseMacro{table-tdt-compress-rp0-tp} & \\
 &  &  & RP1 & \UseMacro{table-tdt-compress-rp1-th0} & \UseMacro{table-tdt-compress-rp1-th1} & \UseMacro{table-tdt-compress-rp1-th2} & \UseMacro{table-tdt-compress-rp1-th3} & \UseMacro{table-tdt-compress-rp1-th4} & \UseMacro{table-tdt-compress-rp1-fp-filtered} & \UseMacro{table-tdt-compress-rp1-dup-filtered} & \UseMacro{table-tdt-compress-rp1-tp} & \\
\midrule
\multirow{6}{*}{lang} & \multirow{6}{*}{\UseMacro{table-tdt-lang-tmpl}} & \multirow{6}{*}{\UseMacro{table-tdt-lang-gen}} & LP0 & \UseMacro{table-tdt-lang-lp0-th0} & \UseMacro{table-tdt-lang-lp0-th1} & \UseMacro{table-tdt-lang-lp0-th2} & \UseMacro{table-tdt-lang-lp0-th3} & \UseMacro{table-tdt-lang-lp0-th4} & \UseMacro{table-tdt-lang-lp0-fp-filtered} & \UseMacro{table-tdt-lang-lp0-dup-filtered} & \UseMacro{table-tdt-lang-lp0-tp} & \multirow{6}{*}{\UseMacro{table-tdt-lang-time}} \\
 &  &  & LP1 & \UseMacro{table-tdt-lang-lp1-th0} & \UseMacro{table-tdt-lang-lp1-th1} & \UseMacro{table-tdt-lang-lp1-th2} & \UseMacro{table-tdt-lang-lp1-th3} & \UseMacro{table-tdt-lang-lp1-th4} & \UseMacro{table-tdt-lang-lp1-fp-filtered} & \UseMacro{table-tdt-lang-lp1-dup-filtered} & \UseMacro{table-tdt-lang-lp1-tp} & \\
 &  &  & LP2 & \UseMacro{table-tdt-lang-lp2-th0} & \UseMacro{table-tdt-lang-lp2-th1} & \UseMacro{table-tdt-lang-lp2-th2} & \UseMacro{table-tdt-lang-lp2-th3} & \UseMacro{table-tdt-lang-lp2-th4} & \UseMacro{table-tdt-lang-lp2-fp-filtered} & \UseMacro{table-tdt-lang-lp2-dup-filtered} & \UseMacro{table-tdt-lang-lp2-tp} & \\
 &  &  & LP3 & \UseMacro{table-tdt-lang-lp3-th0} & \UseMacro{table-tdt-lang-lp3-th1} & \UseMacro{table-tdt-lang-lp3-th2} & \UseMacro{table-tdt-lang-lp3-th3} & \UseMacro{table-tdt-lang-lp3-th4} & \UseMacro{table-tdt-lang-lp3-fp-filtered} & \UseMacro{table-tdt-lang-lp3-dup-filtered} & \UseMacro{table-tdt-lang-lp3-tp} & \\
 &  &  & RP0 & \UseMacro{table-tdt-lang-rp0-th0} & \UseMacro{table-tdt-lang-rp0-th1} & \UseMacro{table-tdt-lang-rp0-th2} & \UseMacro{table-tdt-lang-rp0-th3} & \UseMacro{table-tdt-lang-rp0-th4} & \UseMacro{table-tdt-lang-rp0-fp-filtered} & \UseMacro{table-tdt-lang-rp0-dup-filtered} & \UseMacro{table-tdt-lang-rp0-tp} & \\
 &  &  & RP1 & \UseMacro{table-tdt-lang-rp1-th0} & \UseMacro{table-tdt-lang-rp1-th1} & \UseMacro{table-tdt-lang-rp1-th2} & \UseMacro{table-tdt-lang-rp1-th3} & \UseMacro{table-tdt-lang-rp1-th4} & \UseMacro{table-tdt-lang-rp1-fp-filtered} & \UseMacro{table-tdt-lang-rp1-dup-filtered} & \UseMacro{table-tdt-lang-rp1-tp} & \\
\midrule
\multirow{6}{*}{math} & \multirow{6}{*}{\UseMacro{table-tdt-math-tmpl}} & \multirow{6}{*}{\UseMacro{table-tdt-math-gen}} & LP0 & \UseMacro{table-tdt-math-lp0-th0} & \UseMacro{table-tdt-math-lp0-th1} & \UseMacro{table-tdt-math-lp0-th2} & \UseMacro{table-tdt-math-lp0-th3} & \UseMacro{table-tdt-math-lp0-th4} & \UseMacro{table-tdt-math-lp0-fp-filtered} & \UseMacro{table-tdt-math-lp0-dup-filtered} & \UseMacro{table-tdt-math-lp0-tp} & \multirow{6}{*}{\UseMacro{table-tdt-math-time}} \\
 &  &  & LP1 & \UseMacro{table-tdt-math-lp1-th0} & \UseMacro{table-tdt-math-lp1-th1} & \UseMacro{table-tdt-math-lp1-th2} & \UseMacro{table-tdt-math-lp1-th3} & \UseMacro{table-tdt-math-lp1-th4} & \UseMacro{table-tdt-math-lp1-fp-filtered} & \UseMacro{table-tdt-math-lp1-dup-filtered} & \UseMacro{table-tdt-math-lp1-tp} & \\
 &  &  & LP2 & \UseMacro{table-tdt-math-lp2-th0} & \UseMacro{table-tdt-math-lp2-th1} & \UseMacro{table-tdt-math-lp2-th2} & \UseMacro{table-tdt-math-lp2-th3} & \UseMacro{table-tdt-math-lp2-th4} & \UseMacro{table-tdt-math-lp2-fp-filtered} & \UseMacro{table-tdt-math-lp2-dup-filtered} & \UseMacro{table-tdt-math-lp2-tp} & \\
 &  &  & LP3 & \UseMacro{table-tdt-math-lp3-th0} & \UseMacro{table-tdt-math-lp3-th1} & \UseMacro{table-tdt-math-lp3-th2} & \UseMacro{table-tdt-math-lp3-th3} & \UseMacro{table-tdt-math-lp3-th4} & \UseMacro{table-tdt-math-lp3-fp-filtered} & \UseMacro{table-tdt-math-lp3-dup-filtered} & \UseMacro{table-tdt-math-lp3-tp} & \\
 &  &  & RP0 & \UseMacro{table-tdt-math-rp0-th0} & \UseMacro{table-tdt-math-rp0-th1} & \UseMacro{table-tdt-math-rp0-th2} & \UseMacro{table-tdt-math-rp0-th3} & \UseMacro{table-tdt-math-rp0-th4} & \UseMacro{table-tdt-math-rp0-fp-filtered} & \UseMacro{table-tdt-math-rp0-dup-filtered} & \UseMacro{table-tdt-math-rp0-tp} & \\
 &  &  & RP1 & \UseMacro{table-tdt-math-rp1-th0} & \UseMacro{table-tdt-math-rp1-th1} & \UseMacro{table-tdt-math-rp1-th2} & \UseMacro{table-tdt-math-rp1-th3} & \UseMacro{table-tdt-math-rp1-th4} & \UseMacro{table-tdt-math-rp1-fp-filtered} & \UseMacro{table-tdt-math-rp1-dup-filtered} & \UseMacro{table-tdt-math-rp1-tp} & \\
\midrule
\multirow{6}{*}{statistics} & \multirow{6}{*}{\UseMacro{table-tdt-statistics-tmpl}} & \multirow{6}{*}{\UseMacro{table-tdt-statistics-gen}} & LP0 & \UseMacro{table-tdt-statistics-lp0-th0} & \UseMacro{table-tdt-statistics-lp0-th1} & \UseMacro{table-tdt-statistics-lp0-th2} & \UseMacro{table-tdt-statistics-lp0-th3} & \UseMacro{table-tdt-statistics-lp0-th4} & \UseMacro{table-tdt-statistics-lp0-fp-filtered} & \UseMacro{table-tdt-statistics-lp0-dup-filtered} & \UseMacro{table-tdt-statistics-lp0-tp} & \multirow{6}{*}{\UseMacro{table-tdt-statistics-time}} \\
 &  &  & LP1 & \UseMacro{table-tdt-statistics-lp1-th0} & \UseMacro{table-tdt-statistics-lp1-th1} & \UseMacro{table-tdt-statistics-lp1-th2} & \UseMacro{table-tdt-statistics-lp1-th3} & \UseMacro{table-tdt-statistics-lp1-th4} & \UseMacro{table-tdt-statistics-lp1-fp-filtered} & \UseMacro{table-tdt-statistics-lp1-dup-filtered} & \UseMacro{table-tdt-statistics-lp1-tp} & \\
 &  &  & LP2 & \UseMacro{table-tdt-statistics-lp2-th0} & \UseMacro{table-tdt-statistics-lp2-th1} & \UseMacro{table-tdt-statistics-lp2-th2} & \UseMacro{table-tdt-statistics-lp2-th3} & \UseMacro{table-tdt-statistics-lp2-th4} & \UseMacro{table-tdt-statistics-lp2-fp-filtered} & \UseMacro{table-tdt-statistics-lp2-dup-filtered} & \UseMacro{table-tdt-statistics-lp2-tp} & \\
 &  &  & LP3 & \UseMacro{table-tdt-statistics-lp3-th0} & \UseMacro{table-tdt-statistics-lp3-th1} & \UseMacro{table-tdt-statistics-lp3-th2} & \UseMacro{table-tdt-statistics-lp3-th3} & \UseMacro{table-tdt-statistics-lp3-th4} & \UseMacro{table-tdt-statistics-lp3-fp-filtered} & \UseMacro{table-tdt-statistics-lp3-dup-filtered} & \UseMacro{table-tdt-statistics-lp3-tp} & \\
 &  &  & RP0 & \UseMacro{table-tdt-statistics-rp0-th0} & \UseMacro{table-tdt-statistics-rp0-th1} & \UseMacro{table-tdt-statistics-rp0-th2} & \UseMacro{table-tdt-statistics-rp0-th3} & \UseMacro{table-tdt-statistics-rp0-th4} & \UseMacro{table-tdt-statistics-rp0-fp-filtered} & \UseMacro{table-tdt-statistics-rp0-dup-filtered} & \UseMacro{table-tdt-statistics-rp0-tp} & \\
 &  &  & RP1 & \UseMacro{table-tdt-statistics-rp1-th0} & \UseMacro{table-tdt-statistics-rp1-th1} & \UseMacro{table-tdt-statistics-rp1-th2} & \UseMacro{table-tdt-statistics-rp1-th3} & \UseMacro{table-tdt-statistics-rp1-th4} & \UseMacro{table-tdt-statistics-rp1-fp-filtered} & \UseMacro{table-tdt-statistics-rp1-dup-filtered} & \UseMacro{table-tdt-statistics-rp1-tp} & \\
\midrule
\multirow{6}{*}{text} & \multirow{6}{*}{\UseMacro{table-tdt-text-tmpl}} & \multirow{6}{*}{\UseMacro{table-tdt-text-gen}} & LP0 & \UseMacro{table-tdt-text-lp0-th0} & \UseMacro{table-tdt-text-lp0-th1} & \UseMacro{table-tdt-text-lp0-th2} & \UseMacro{table-tdt-text-lp0-th3} & \UseMacro{table-tdt-text-lp0-th4} & \UseMacro{table-tdt-text-lp0-fp-filtered} & \UseMacro{table-tdt-text-lp0-dup-filtered} & \UseMacro{table-tdt-text-lp0-tp} & \multirow{6}{*}{\UseMacro{table-tdt-text-time}} \\
 &  &  & LP1 & \UseMacro{table-tdt-text-lp1-th0} & \UseMacro{table-tdt-text-lp1-th1} & \UseMacro{table-tdt-text-lp1-th2} & \UseMacro{table-tdt-text-lp1-th3} & \UseMacro{table-tdt-text-lp1-th4} & \UseMacro{table-tdt-text-lp1-fp-filtered} & \UseMacro{table-tdt-text-lp1-dup-filtered} & \UseMacro{table-tdt-text-lp1-tp} & \\
 &  &  & LP2 & \UseMacro{table-tdt-text-lp2-th0} & \UseMacro{table-tdt-text-lp2-th1} & \UseMacro{table-tdt-text-lp2-th2} & \UseMacro{table-tdt-text-lp2-th3} & \UseMacro{table-tdt-text-lp2-th4} & \UseMacro{table-tdt-text-lp2-fp-filtered} & \UseMacro{table-tdt-text-lp2-dup-filtered} & \UseMacro{table-tdt-text-lp2-tp} & \\
 &  &  & LP3 & \UseMacro{table-tdt-text-lp3-th0} & \UseMacro{table-tdt-text-lp3-th1} & \UseMacro{table-tdt-text-lp3-th2} & \UseMacro{table-tdt-text-lp3-th3} & \UseMacro{table-tdt-text-lp3-th4} & \UseMacro{table-tdt-text-lp3-fp-filtered} & \UseMacro{table-tdt-text-lp3-dup-filtered} & \UseMacro{table-tdt-text-lp3-tp} & \\
 &  &  & RP0 & \UseMacro{table-tdt-text-rp0-th0} & \UseMacro{table-tdt-text-rp0-th1} & \UseMacro{table-tdt-text-rp0-th2} & \UseMacro{table-tdt-text-rp0-th3} & \UseMacro{table-tdt-text-rp0-th4} & \UseMacro{table-tdt-text-rp0-fp-filtered} & \UseMacro{table-tdt-text-rp0-dup-filtered} & \UseMacro{table-tdt-text-rp0-tp} & \\
 &  &  & RP1 & \UseMacro{table-tdt-text-rp1-th0} & \UseMacro{table-tdt-text-rp1-th1} & \UseMacro{table-tdt-text-rp1-th2} & \UseMacro{table-tdt-text-rp1-th3} & \UseMacro{table-tdt-text-rp1-th4} & \UseMacro{table-tdt-text-rp1-fp-filtered} & \UseMacro{table-tdt-text-rp1-dup-filtered} & \UseMacro{table-tdt-text-rp1-tp} & \\
\midrule
\multirow{6}{*}{vectorz} & \multirow{6}{*}{\UseMacro{table-tdt-vectorz-tmpl}} & \multirow{6}{*}{\UseMacro{table-tdt-vectorz-gen}} & LP0 & \UseMacro{table-tdt-vectorz-lp0-th0} & \UseMacro{table-tdt-vectorz-lp0-th1} & \UseMacro{table-tdt-vectorz-lp0-th2} & \UseMacro{table-tdt-vectorz-lp0-th3} & \UseMacro{table-tdt-vectorz-lp0-th4} & \UseMacro{table-tdt-vectorz-lp0-fp-filtered} & \UseMacro{table-tdt-vectorz-lp0-dup-filtered} & \UseMacro{table-tdt-vectorz-lp0-tp} & \multirow{6}{*}{\UseMacro{table-tdt-vectorz-time}} \\
 &  &  & LP1 & \UseMacro{table-tdt-vectorz-lp1-th0} & \UseMacro{table-tdt-vectorz-lp1-th1} & \UseMacro{table-tdt-vectorz-lp1-th2} & \UseMacro{table-tdt-vectorz-lp1-th3} & \UseMacro{table-tdt-vectorz-lp1-th4} & \UseMacro{table-tdt-vectorz-lp1-fp-filtered} & \UseMacro{table-tdt-vectorz-lp1-dup-filtered} & \UseMacro{table-tdt-vectorz-lp1-tp} & \\
 &  &  & LP2 & \UseMacro{table-tdt-vectorz-lp2-th0} & \UseMacro{table-tdt-vectorz-lp2-th1} & \UseMacro{table-tdt-vectorz-lp2-th2} & \UseMacro{table-tdt-vectorz-lp2-th3} & \UseMacro{table-tdt-vectorz-lp2-th4} & \UseMacro{table-tdt-vectorz-lp2-fp-filtered} & \UseMacro{table-tdt-vectorz-lp2-dup-filtered} & \UseMacro{table-tdt-vectorz-lp2-tp} & \\
 &  &  & LP3 & \UseMacro{table-tdt-vectorz-lp3-th0} & \UseMacro{table-tdt-vectorz-lp3-th1} & \UseMacro{table-tdt-vectorz-lp3-th2} & \UseMacro{table-tdt-vectorz-lp3-th3} & \UseMacro{table-tdt-vectorz-lp3-th4} & \UseMacro{table-tdt-vectorz-lp3-fp-filtered} & \UseMacro{table-tdt-vectorz-lp3-dup-filtered} & \UseMacro{table-tdt-vectorz-lp3-tp} & \\
 &  &  & RP0 & \UseMacro{table-tdt-vectorz-rp0-th0} & \UseMacro{table-tdt-vectorz-rp0-th1} & \UseMacro{table-tdt-vectorz-rp0-th2} & \UseMacro{table-tdt-vectorz-rp0-th3} & \UseMacro{table-tdt-vectorz-rp0-th4} & \UseMacro{table-tdt-vectorz-rp0-fp-filtered} & \UseMacro{table-tdt-vectorz-rp0-dup-filtered} & \UseMacro{table-tdt-vectorz-rp0-tp} & \\
 &  &  & RP1 & \UseMacro{table-tdt-vectorz-rp1-th0} & \UseMacro{table-tdt-vectorz-rp1-th1} & \UseMacro{table-tdt-vectorz-rp1-th2} & \UseMacro{table-tdt-vectorz-rp1-th3} & \UseMacro{table-tdt-vectorz-rp1-th4} & \UseMacro{table-tdt-vectorz-rp1-fp-filtered} & \UseMacro{table-tdt-vectorz-rp1-dup-filtered} & \UseMacro{table-tdt-vectorz-rp1-tp} & \\
\midrule
\multirow{6}{*}{zxing} & \multirow{6}{*}{\UseMacro{table-tdt-zxing-tmpl}} & \multirow{6}{*}{\UseMacro{table-tdt-zxing-gen}} & LP0 & \UseMacro{table-tdt-zxing-lp0-th0} & \UseMacro{table-tdt-zxing-lp0-th1} & \UseMacro{table-tdt-zxing-lp0-th2} & \UseMacro{table-tdt-zxing-lp0-th3} & \UseMacro{table-tdt-zxing-lp0-th4} & \UseMacro{table-tdt-zxing-lp0-fp-filtered} & \UseMacro{table-tdt-zxing-lp0-dup-filtered} & \UseMacro{table-tdt-zxing-lp0-tp} & \multirow{6}{*}{\UseMacro{table-tdt-zxing-time}} \\
 &  &  & LP1 & \UseMacro{table-tdt-zxing-lp1-th0} & \UseMacro{table-tdt-zxing-lp1-th1} & \UseMacro{table-tdt-zxing-lp1-th2} & \UseMacro{table-tdt-zxing-lp1-th3} & \UseMacro{table-tdt-zxing-lp1-th4} & \UseMacro{table-tdt-zxing-lp1-fp-filtered} & \UseMacro{table-tdt-zxing-lp1-dup-filtered} & \UseMacro{table-tdt-zxing-lp1-tp} & \\
 &  &  & LP2 & \UseMacro{table-tdt-zxing-lp2-th0} & \UseMacro{table-tdt-zxing-lp2-th1} & \UseMacro{table-tdt-zxing-lp2-th2} & \UseMacro{table-tdt-zxing-lp2-th3} & \UseMacro{table-tdt-zxing-lp2-th4} & \UseMacro{table-tdt-zxing-lp2-fp-filtered} & \UseMacro{table-tdt-zxing-lp2-dup-filtered} & \UseMacro{table-tdt-zxing-lp2-tp} & \\
 &  &  & LP3 & \UseMacro{table-tdt-zxing-lp3-th0} & \UseMacro{table-tdt-zxing-lp3-th1} & \UseMacro{table-tdt-zxing-lp3-th2} & \UseMacro{table-tdt-zxing-lp3-th3} & \UseMacro{table-tdt-zxing-lp3-th4} & \UseMacro{table-tdt-zxing-lp3-fp-filtered} & \UseMacro{table-tdt-zxing-lp3-dup-filtered} & \UseMacro{table-tdt-zxing-lp3-tp} & \\
 &  &  & RP0 & \UseMacro{table-tdt-zxing-rp0-th0} & \UseMacro{table-tdt-zxing-rp0-th1} & \UseMacro{table-tdt-zxing-rp0-th2} & \UseMacro{table-tdt-zxing-rp0-th3} & \UseMacro{table-tdt-zxing-rp0-th4} & \UseMacro{table-tdt-zxing-rp0-fp-filtered} & \UseMacro{table-tdt-zxing-rp0-dup-filtered} & \UseMacro{table-tdt-zxing-rp0-tp} & \\
 &  &  & RP1 & \UseMacro{table-tdt-zxing-rp1-th0} & \UseMacro{table-tdt-zxing-rp1-th1} & \UseMacro{table-tdt-zxing-rp1-th2} & \UseMacro{table-tdt-zxing-rp1-th3} & \UseMacro{table-tdt-zxing-rp1-th4} & \UseMacro{table-tdt-zxing-rp1-fp-filtered} & \UseMacro{table-tdt-zxing-rp1-dup-filtered} & \UseMacro{table-tdt-zxing-rp1-tp} & \\
\bottomrule
\end{tabular}
\end{table}

\MyPara{Evaluation results}
We show the evaluation results when using \Lejit as the input
generator for 8 \Java projects, which were used to evaluate \Lejit
itself~\cite{ZangETAL24LeJit}.
We choose (empirically) the following parameters for our algorithm
(Algorithm~\ref{algo:tdt}): $n=4$, $Ns=[100000, 500000, 1000000,
  3000000]$, $THs=[1.2, 1.2, 1.3, 1.4]$ for LP and $THs=[1.1, 1.1,
  1.2, 1.3]$ for RP. We disable the $TOP\_Ks$ limit in this experiment,
and report results when using $TOP\_Ks$ later in this section.

Table~\ref{table-eval:tdt}\footnote{Because there are duplicate bugs
across projects and \JDK versions, and some bugs are not reproduced in
this set of experiments, the counts in the second-to-last column do
not sum up to the total number of bugs reported in this paper.} shows
the evaluation results. The first column lists the project names,
followed by two columns showing the number of templates extracted and
the number of programs generated by \Lejit (each template may
correspond to multiple programs).
The next column shows the used configuration pair.
The next four columns present the number of programs filtered at each
performance-checking \level.
The following three columns show, respectively, the number of programs
that remain after all \levels (``Pass''), after false-positive
filtering (``False Positive Filtering''), and after duplicate
filtering (``Duplicate Filtering'').
The last two columns report the number of unique true positive
performance bugs (``Unique'') and the total time spent on the entire
process (``Time''), excluding the first performance \level in which
all programs are executed.  We exclude time for the first \level
because \Tool is designed to \emph{augment} existing program
generators, which typically already execute all generated programs to
check for \functional bugs.
The first \level can therefore be reused both to detect \functional
bugs and collect runtime information, upon which our \xname proceeds
to identify performance bugs.

We can observe from Table~\ref{table-eval:tdt} that the first two
\levels ($TH_0$ and $TH_1$) filter out the majority of programs,
significantly improving the efficiency of the testing process since
the later \levels involve more iterations and are more computationally
expensive. The false positive and duplicate filtering stages further
reduce the number of programs requiring manual inspection, leaving
only a few cases to be examined manually.  Although these two
automated filtering steps can eliminate a large number of programs,
they are not perfect, which explains the discrepancy between the
counts in the second-to-last and third-to-last columns.

\begin{table*}[t!]
\centering
\small
\caption{Comparison of execution time with and without test
  prioritization across evaluation projects.\label{table-eval:time}}
\vspace{-8pt}
\begin{tabular}{lccc}
\toprule

\textbf{Project} & \shortstack{Time [min]\\w/o prioritization}  & \shortstack{Time [min]\\w/ prioritization} & \shortstack{Time \\reduction (\%)} \\
\midrule
codec & \UseMacro{table-tdt-codec-time} & \UseMacro{table-tdt-topk500_100_50-codec-time} &\timediff{\UseMacro{table-tdt-codec-time}}{\UseMacro{table-tdt-topk500_100_50-codec-time}} \\
compress & \UseMacro{table-tdt-compress-time} & \UseMacro{table-tdt-topk500_100_50-compress-time} & \timediff{\UseMacro{table-tdt-compress-time}}{\UseMacro{table-tdt-topk500_100_50-compress-time}} \\
lang & \UseMacro{table-tdt-lang-time} & \UseMacro{table-tdt-topk500_100_50-lang-time} & \timediff{\UseMacro{table-tdt-lang-time}}{\UseMacro{table-tdt-topk500_100_50-lang-time}} \\
math & \UseMacro{table-tdt-math-time} & \UseMacro{table-tdt-topk500_100_50-math-time} & \timediff{\UseMacro{table-tdt-math-time}}{\UseMacro{table-tdt-topk500_100_50-math-time}} \\
statistics & \UseMacro{table-tdt-statistics-time} & \UseMacro{table-tdt-topk500_100_50-statistics-time} & \timediff{\UseMacro{table-tdt-statistics-time}}{\UseMacro{table-tdt-topk500_100_50-statistics-time}} \\
text & \UseMacro{table-tdt-text-time} & \UseMacro{table-tdt-topk500_100_50-text-time} & \timediff{\UseMacro{table-tdt-text-time}}{\UseMacro{table-tdt-topk500_100_50-text-time}} \\
vectorz & \UseMacro{table-tdt-vectorz-time} & \UseMacro{table-tdt-topk500_100_50-vectorz-time} & \timediff{\UseMacro{table-tdt-vectorz-time}}{\UseMacro{table-tdt-topk500_100_50-vectorz-time}} \\
zxing & \UseMacro{table-tdt-zxing-time} & \UseMacro{table-tdt-topk500_100_50-zxing-time} & \timediff{\UseMacro{table-tdt-zxing-time}}{\UseMacro{table-tdt-topk500_100_50-zxing-time}} \\
\midrule
\textbf{Total} & \UseMacro{table-tdt-total-time} & \UseMacro{table-tdt-topk500_100_50-total-time} & \UseMacro{table-tdt-total-time-reduction} \\

\bottomrule
\end{tabular}
\end{table*}

\begin{table*}[t!]
    \centering
    \small
    \caption{Comparison of execution time (in minutes) between \Tool
      using a single \level and \Tool using multiple
      \levels.\label{table-eval:single-tier}}
    \vspace{-8pt}
    \begin{tabular}{lcccc}
    \toprule
    
    \textbf{Project} & \shortstack{Multiple \Levels\\w/o prioritization}  & \shortstack{Single \Level\\w/o prioritization} & \shortstack{Multiple \Levels\\w/ prioritization} & \shortstack{Single \Level\\w/ prioritization} \\
    \midrule
    codec & \UseMacro{table-tdt-codec-time} & \UseMacro{table-single-tier-codec-time} (+\timediffr{\UseMacro{table-tdt-codec-time}}{\UseMacro{table-single-tier-codec-time}}\%) & \UseMacro{table-tdt-topk500_100_50-codec-time} & \UseMacro{table-single-tier-codec-topk500-time} (+\timediffr{\UseMacro{table-tdt-topk500_100_50-codec-time}}{\UseMacro{table-single-tier-codec-topk500-time}}\%) \\
    compress & \UseMacro{table-tdt-compress-time} & \UseMacro{table-single-tier-compress-time} (+\timediffr{\UseMacro{table-tdt-compress-time}}{\UseMacro{table-single-tier-compress-time}}\%) & \UseMacro{table-tdt-topk500_100_50-compress-time} & \UseMacro{table-single-tier-compress-topk500-time} (+\timediffr{\UseMacro{table-tdt-topk500_100_50-compress-time}}{\UseMacro{table-single-tier-compress-topk500-time}}\%) \\
    lang & \UseMacro{table-tdt-lang-time} & \UseMacro{table-single-tier-lang-time} (+\timediffr{\UseMacro{table-tdt-lang-time}}{\UseMacro{table-single-tier-lang-time}}\%) & \UseMacro{table-tdt-topk500_100_50-lang-time} & \UseMacro{table-single-tier-lang-topk500-time} (+\timediffr{\UseMacro{table-tdt-topk500_100_50-lang-time}}{\UseMacro{table-single-tier-lang-topk500-time}}\%) \\
    math & \UseMacro{table-tdt-math-time} & \UseMacro{table-single-tier-math-time} (+\timediffr{\UseMacro{table-tdt-math-time}}{\UseMacro{table-single-tier-math-time}}\%) & \UseMacro{table-tdt-topk500_100_50-math-time} & \UseMacro{table-single-tier-math-topk500-time} (+\timediffr{\UseMacro{table-tdt-topk500_100_50-math-time}}{\UseMacro{table-single-tier-math-topk500-time}}\%) \\
    statistics & \UseMacro{table-tdt-statistics-time} & \UseMacro{table-single-tier-statistics-time} (+\timediffr{\UseMacro{table-tdt-statistics-time}}{\UseMacro{table-single-tier-statistics-time}}\%) & \UseMacro{table-tdt-topk500_100_50-statistics-time} & \UseMacro{table-single-tier-statistics-topk500-time} (+\timediffr{\UseMacro{table-tdt-topk500_100_50-statistics-time}}{\UseMacro{table-single-tier-statistics-topk500-time}}\%) \\
    text & \UseMacro{table-tdt-text-time} & \UseMacro{table-single-tier-text-time} (+\timediffr{\UseMacro{table-tdt-text-time}}{\UseMacro{table-single-tier-text-time}}\%) & \UseMacro{table-tdt-topk500_100_50-text-time} & \UseMacro{table-single-tier-text-topk500-time} (+\timediffr{\UseMacro{table-tdt-topk500_100_50-text-time}}{\UseMacro{table-single-tier-text-topk500-time}}\%) \\
    vectorz & \UseMacro{table-tdt-vectorz-time} & \UseMacro{table-single-tier-vectorz-time} (+\timediffr{\UseMacro{table-tdt-vectorz-time}}{\UseMacro{table-single-tier-vectorz-time}}\%) & \UseMacro{table-tdt-topk500_100_50-vectorz-time} & \UseMacro{table-single-tier-vectorz-topk500-time} (+\timediffr{\UseMacro{table-tdt-topk500_100_50-vectorz-time}}{\UseMacro{table-single-tier-vectorz-topk500-time}}\%) \\
    zxing & \UseMacro{table-tdt-zxing-time} & \UseMacro{table-single-tier-zxing-time} (+\timediffr{\UseMacro{table-tdt-zxing-time}}{\UseMacro{table-single-tier-zxing-time}}\%) & \UseMacro{table-tdt-topk500_100_50-zxing-time} & \UseMacro{table-single-tier-zxing-topk500-time} (+\timediffr{\UseMacro{table-tdt-topk500_100_50-zxing-time}}{\UseMacro{table-single-tier-zxing-topk500-time}}\%) \\

    \midrule
    \textbf{Total} & 19484.92 & 30387.39 (+55.95\%) & 1481.69 & 2832.93 (+\timediffr{1481.69}{2832.93}\%) \\

    
    \bottomrule
    \end{tabular}
    \end{table*}

\MyPara{Effects of prioritization}
Table~\ref{table-eval:time} presents the execution time of \Tool with
and without top-$k$ prioritization (configured with $TOP\_Ks = [500,
  100, 50]$). As shown in the table, enabling prioritization
substantially reduces the overall testing time, achieving an average
time reduction of \UseMacro{table-tdt-total-time-reduction}\%. This
improvement stems from the fact that the prioritization strategy
guides the testing process toward programs that are more likely to
expose performance anomalies, thereby reducing the number of less
promising candidates that require costly evaluation in later \levels.
Importantly, we observe that applying top-$k$ prioritization does not
lead to any loss of true positive bugs in our experiments.  This
finding indicates that our prioritization effectively accelerates the
testing process while preserving its effectiveness in identifying
genuine performance issues.  These results demonstrate that
prioritization can serve as a practical and scalable enhancement to
\Tool.

\MyPara{Effects of multiple \levels}
Using multiple \levels enables gradual filtering of programs: with
only a single \level, there is limited flexibility in selecting
iteration counts ($Ns$) and thresholds ($THs$).  For example, if the
iteration count is set too low, many false positives may remain (as
shown with $TH_0$ in Table~\ref{table-eval:tdt}), and if it is set too
high, the approach becomes expensive, since many programs must be
executed with a large number of iterations.  Thus, we opt for a multi-\level approach.

Table~\ref{table-eval:single-tier} compares the execution time of
single-\level and multi-\level settings.  The settings for
multi-\level remain the same as in Table~\ref{table-eval:tdt}.  For
the single-\level setting, we only keep the last \level besides the
first \level, and thus we set the following parameters: $Ns=[100000,
  3000000]$, $THs=[1.2, 1.4]$ for LP and $THs=[1.1, 1.3]$ for RP, and
$TOP\_Ks=[500]$ (columns ``w/ prioritization'').
As shown in the table, the single-\level configuration incurs
substantially higher execution time than the multi-\level approach,
both with and without prioritization, since many more programs are
executed in the final \level.  Note that the single-\level setting can
avoid missing true bugs as it skips the intermediate filtering
\levels.  We did find a new regression bug between the two versions of
\Graal compilers used in our experiments (but the issue is not present
in the latest version and is therefore not reported).  Besides, we
also observe that the execution-time gap varies across projects, \eg,
in codec we observe
+\timediffr{\UseMacro{table-tdt-codec-time}}{\UseMacro{table-single-tier-codec-time}}\%
difference and in lang we observe
+\timediffr{\UseMacro{table-tdt-lang-time}}{\UseMacro{table-single-tier-lang-time}}\%
difference.  This is because the runtime of some simple programs is
relatively insensitive to high iteration counts, while the
multi-\level approach introduces additional overhead from repeated
instrumentation of different iteration counts and compilation of
programs into bytecode.

\subsection{Discovered Bugs}

\begin{table}[t]
    \centering
    \caption{Summarization of bugs found by \Tool.\label{table-eval:bugs}}
    \footnotesize
    \vspace{-8pt}
    \begin{tabular}{l|ccc|cc|c|c|c}
        \toprule
        \multirow{2}{*}{\centering \textbf{ID}} & \multicolumn{3}{c|}{\textbf{Generator}} & \multicolumn{2}{c|}{\textbf{Oracle}} & \multirow{2}{*}{\textbf{Project(s)}} & \multirow{2}{*}{\textbf{Root Cause}} & \multirow{2}{*}{\textbf{Status}}\\
        \cmidrule(lr){2-4} \cmidrule(lr){5-6}
        & \Lejit & \Artemis & \JavaFuzzer & Level & Regression & & & \\
        \midrule
        \JDKBug{JDK-8345766} & \Select & - & - & \Select & - & math (2) & \RcOpt & \StFixed \\
        \JDKBug{JDK-8346989} & \Select & \Select & & \Select & - & math (1) & \RcSpec & \StFixed \\
        \JDKBug{JDK-8349401} & \Select & - & - & - & \Select & math (2) & \RcCodegen & \StFixed \\
        \JDKBug{JDK-8349452} & \Select & - & - & \Select & \Select & zxing (1) & \RcCodegen & \StConfirmed \\
        \JDKBug{JDK-8349364} & \Select & - & - & \Select & - & vectorz (2) & \RcOpt & \StConfirmed \\
        \JDKBug{JDK-8350494} & - & \Select & - & \Select & - & N/A & \RcOpt & \StDup \\
        \JDKBug{JDK-8350720} & - & \Select & - & \Select & - & N/A & \RcOpt & \StConfirmed \\
        \JDKBug{JDK-8353638} & \Select & \Select & \Select & \Select & - & compress (6)& \RcSpec & \StFixed \\
        \GraalBug{10565} & \Select & \Select  & \Select & \Select & - & lang (6)& \RcSpec & \StFixed \\
        \GraalBug{10609} & \Select & - & \Select & \Select & - & math (2) & \RcUnknown & \StConfirmed \\
        \GraalBug{10776} & \Select & - & - & - & \Select & codec (1) & \RcSpec & \StFixed \\
        \bottomrule
    \end{tabular}
    \vspace{-5pt}
\end{table}

\Tool discovered \NumOfFound previously unknown performance bugs in
\HotSpot and \Graal \JIT compilers, with \NumOfBugs of them confirmed
or fixed by developers.

Table~\ref{table-eval:bugs} summarizes the bugs we found. For each
bug, we show the corresponding generators and pairs used to find the
bugs (if multiple choices are selected, it means the bug was found by
several generators or pairs)\footnote{The goal of our work is
\emph{not} to compare generators, but to showcase that \Tool is able
to detect performance bugs with programs obtained from various
sources.}. We also include corresponding projects if the bug is found
by \Lejit (using the first source project of the bug-reporting
template, with the number of projects shown in parentheses when
multiple projects find the bug), as well as the root causes when such
information has been confirmed by developers for our reports.

As shown in the table, \Tool is capable of uncovering performance bugs
spanning multiple compiler phases--not only those arising from
optimization and code generation, but also speculation-related issues
that are unique to \JIT compilers.  Besides, two of the bugs are
severe performance regressions in standard libraries, and several are
related to basic arithmetic calculations, which could be widely used
by many programs.  These findings demonstrate the generality and
practical relevance of \Tool in detecting diverse and impactful
performance issues.
We briefly describe four representative bugs detected by \Tool, which
spread across different phases of compilers.

\begin{figure}[t!]
    \subcaptionbox{
        \JDKBug{JDK-8346989}
        \label{figure:eval-bug1}
    }
    {
\begin{lstlisting}[language=java-pretty]
public static int square(int a) {
  return Math.multiplyExact(a, a);
}
public static void main(String[] a) {
  for (int i = 0; i < 5000000; i++) {
    try { square(i);}
    catch (Throwable e) {}
  }
}
\end{lstlisting}
    }
    \hspace{0.8cm}
    \subcaptionbox{
        \JDKBug{JDK-8349452}
        \label{figure:eval-bug2}
    }
    {
\begin{lstlisting}[language=java-pretty]
public static void main(String[] a) {
  byte[][] bytes = new byte[90][1];

  for (int i = 0; i < 100000; ++i) {
    for (byte[] aByte : bytes) {
      Arrays.fill(aByte, i);
    }
  }
}
\end{lstlisting}
    }
    \subcaptionbox{
        \JDKBug{JDK-8345766}
        \label{figure:eval-bug3}
    }
    {
\begin{lstlisting}[language=java-pretty]
public static double f(final double x) {
  double w = 0.1;
  double p = 0;
  p = (3.10E307 % (w % Z));
  p = (7.61E307 / (x % x));
  return (x * p);
}
public static void main(String[] a) {
  for (int i = 0; i < 100000; i++) {
    f(1.0);
  }
}
\end{lstlisting}
    }
    \hspace{0.4cm}
    \subcaptionbox{
        \GraalBug{10776}
        \label{figure:eval-bug4}
    }
    {
\begin{lstlisting}[language=java-pretty]
public static void write() {
  final ByteArrayOutputStream buffer =  \
        new ByteArrayOutputStream();
  for (int i = 0; i < 100; i++) {
    buffer.write(1);
  }
}
public static void main(String[] a) {
  for (int i = 0; i < 100000; i++) {
      write();
  }
}
\end{lstlisting}
    }
    \caption{Examples of performance bugs found by \Tool in \HotSpot
      and \Graal.\label{fig:eval-bugs}}
\end{figure}

Figure~\ref{figure:eval-bug1} shows a speculation bug where the
\HotSpot compiler continuously hits a deoptimization point and fails
to self-correct the speculation.  The \CodeIn{multiplyExact} method
will throw an exception when the result overflows.  When problematic
arguments are continuously passed to the method, it will always hit
the deoptimization point and move to the slow path. Instead, it should
change its speculative assumption with updated profiling data.

Figure~\ref{figure:eval-bug2} shows a code generation bug where newly
added optimizations may cause performance regression for other cases
where the optimization may not be beneficial.  A previous commit
optimizes the \CodeIn{Arrays.fill()} stub for AVX512 targets,
providing 2-5 times gains for large fill sizes. However, for small
arrays, the call overhead to the optimized stub dominates the runtime,
making the new path substantially slower.

Figure~\ref{figure:eval-bug3} shows a missed optimization bug in
\HotSpot \CTwo compiler. \CTwo currently lacks native lowering rules
for floating-point remainder on most platforms. As a result, it always
emits a runtime call instead of generating machine code, except on the
obsolete x86-32 backend. This design prevents constant folding and
other optimizations in the \CTwo pipeline--operations that \COne can
perform successfully.  The proposed fix converts \CodeIn{ModFNode} and
\CodeIn{ModDNode} into macro nodes that are created during parsing,
enabling standard \CTwo optimizations (\eg, constant folding and
idealization) before being lowered to runtime calls during macro
expansion.

Figure~\ref{figure:eval-bug4} shows a performance regression in
\CodeIn{ByteArrayOutputStream.write()} under \Graal, which is caused
by conservative speculations. Calling \CodeIn{write()} in a loop
repeatedly enters a small synchronized block on a newly allocated
object. In older \Graal versions, escape analysis correctly identified
that the object did not escape the thread, replaced the monitor with a
virtual lock, and completely eliminated synchronization. After the
regression, however, \Graal conservatively treated these locks as
having side effects and emitted real \CodeIn{monitorenter} and
\CodeIn{monitorexit} calls, forcing every iteration down the slow
runtime locking path. The fix restores the original performance by
reintroducing speculative virtual-lock support, allowing the compiler
to assume locks are side-effect-free and safely virtualize them unless
a deoptimization proves otherwise.


\section{Limitations and Future Work}
\label{sec:discussion}

We discuss the limitations of \Tool and outline directions for future
work. Key challenges in detecting \JIT compiler performance bugs lie
in test input generation and test oracle design.

\MyPara{Test input generation}
Detecting performance bugs effectively requires generating inputs that
can trigger diverse \JIT behaviors.  Existing generators designed for
finding \functional bugs already explore many compiler paths, but how
to tailor them toward performance-related behaviors remains an open
question.  For example, optimizations such as vectorization or
inlining depend on specific program patterns (\eg, certain loop
structures).  Moreover, generating programs that systematically expose
speculative optimizations and deoptimizations--behaviors unique to
\JIT compilers--presents an intriguing avenue for future research.

\MyPara{Test oracle}
Designing test oracles for performance bugs is harder than for
\functional testing.  The first challenge is choosing appropriate
metrics.  In addition to timing, logs and program states can also
serve as performance indicators.  Another challenge is measurement
accuracy, which can be influenced by hardware, OS, and runtime
factors.  Accurately measuring the performance of \JIT compilers is
particularly challenging~\cite{barrett2017virtual}. Although existing
methods improve accuracy~\cite{georges2007statistically, traini2024ai,
  zheng2015accurate}, they often reduce throughput. \Tool therefore
uses coarse-grained measurements for scalability, and future work could
explore the use of fine-grained measurements.


\section{Related Work}
\label{sec:related}

We discuss several related works in this section: (a)~detecting
performance bugs, (b)~compiler testing, (c)~finding missed
optimizations in compilers, (d)~\JIT compiler benchmarking, and
(e)~formal verification of compilers.

\MyPara{Detecting performance bugs}
Jin et al.~\cite{jin2012understanding} conducted one of the
earliest empirical studies on performance bugs in real-world
applications and proposed a rule-based approach for their detection.
Subsequent empirical studies have also investigated performance bugs
from various perspectives~\cite{selakovic2016performance,
azad2023empirical, zaman2012qualitative, han2016empirical}, but all
primarily focus on application-level software rather than compiler
infrastructures.

In addition to empirical analyses, several automated techniques have
been proposed to detect performance bugs. Fuzzing-based approaches
such as JITProf~\cite{gong2015jitprof},
PerfFuzz~\cite{lemieux2018perffuzz}, and
others~\cite{toffola2018synthesizing, li2025spider,
han2018perflearner} generate diverse test inputs to expose
pathological performance behaviors. Li et
al.~\cite{li2021understanding} studied \textit{long compilation} bugs
in markdown compilers, emphasizing inefficiencies during compilation
rather than runtime execution. Other domain-specific tools target
specialized systems, such as for database
engines~\cite{jung2019apollo,ba2024cert} and for machine learning
libraries~\cite{tizpaz2020detecting}. Our work differs from these in
that we focus on detecting performance bugs in \JIT compilers
themselves, which are not caused by the application source code but by
the compiler's optimization and code generation phases.

\MyPara{Compiler testing}
Extensive research has focused on testing \functional correctness of both \AOT
and \JIT compilers~\cite{chen2020survey}, spanning a wide range of
languages such as C/C++~\cite{yang2011finding, le2014compiler,
  le2015finding, livinskii2023fuzzing, fan2024high, li2024boosting},
\Java~\cite{javafuzzer, chen2016coverage, zhao2022history,
  jia2023detecting, li2023validating, ZangETAL22JAttack,
  ZangETAL24LeJit}, \JS~\cite{gross2023fuzzilli, wang2023fuzzjit,
  wachterdumpling, wong2025extraction}, and Smalltalk~\cite{polito2022interpreter}.

Zhao et al.~\cite{zhao2022history} synthesized realistic test programs
by mining code snippets from real-world repositories to better expose
\functional bugs.  Zang et al.~\cite{ZangETAL22JAttack,
  ZangETAL24LeJit} proposed a template-based generation approach that
automatically extracts program templates from real programs. Li et
al.~\cite{li2023validating} developed a systematic method named
compilation space exploration to trigger different compilation
decisions in \JIT compilers, improving coverage of dynamic behaviors.

Typing-related bugs in compilers have also been
explored~\cite{dewey2015fuzzing, chaliasos2021well,
chaliasos2022finding}, alongside complementary efforts on test case
prioritization~\cite{chen2016test, chen2017learning} and
reduction~\cite{sun2018perses, regehr2012test}.  Several empirical
studies have further analyzed compiler bug characteristics across
ecosystems, including GCC/LLVM~\cite{sun2016toward} and
Rustc~\cite{liu2025empirical}. Marcozzi et
al.~\cite{marcozzi2019compiler} examined how compiler bugs found by
fuzzers affect real-world applications, bridging the gap between
research results and practical software reliability. Related compiler
testing approaches have also been recently extended to other
semantics-preserving code transformation tools such as \JS obfuscators
and deobfuscators~\cite{jiang2026obsmith, jiang2026cascade}.

Unlike this prior work, we are the first to focus on understanding and
detecting performance bugs in \JIT compilers.

\MyPara{Finding missed optimizations in compilers}
Moseley et al.~\cite{moseley2009optiscope} investigated performance
accountability in optimizing compilers, aiming to better attribute
performance effects to specific optimization decisions.
Barany~\cite{barany2018finding} applied differential testing to
identify missed optimizations by analyzing discrepancies in the
generated binaries. Theodoridis et al.~\cite{theodoridis2022finding,
  theodoridis2024refined} introduced techniques using dead-code
markers and refined information to assess how effectively compilers
eliminate redundant computations or exploit precise data-flow
facts. Liu et al.~\cite{liu2023exploring} analyzed missed
optimizations in WebAssembly compilers, while Gao et
al.~\cite{gao2024shoot} proposed a source-level transformation
framework to detect compiler-induced performance differences. The
techniques used in prior works to derive semantically equivalent
programs could potentially be extended to testing of \JIT compilers.
\Tool can naturally incorporate such similar program pairs by treating
them as two distinct configurations.

Pe\v{c}im\'{u}th et al.~\cite{pevcimuth2023diagnosing} studied \Java
\JIT compilers by representing optimizations as trees and comparing
them via tree edit distance to identify behavioral differences across
compiler versions. However, their work mainly focuses on optimization
phases, leaving other \JIT-specific components--such as speculation,
tiered compilation, and runtime interaction--largely unexplored.
Profiling-based approaches have also been used to support
compiler-level performance debugging~\cite{basso2023optimization,
  burchell2024towards} by analyzing internal compiler events to
identify potential missed optimizations, though care is needed to avoid
instrumentation interfering with compiler optimizations.

\MyPara{\JIT compiler benchmarking}
Benchmarking \JIT compilers is more challenging than benchmarking \AOT
compilers since it involves warm-up stage and other runtime
components~\cite{barrett2017virtual, traini2024ai}. Several benchmarks
have been proposed to represent real-world
workloads~\cite{blackburn2006dacapo, prokopec2019renaissance}. Some
methods are proposed to more rigorously evaluate \JIT
compilers~\cite{georges2007statistically, zheng2015accurate}. Zheng et
al.~\cite{zheng2015accurate} studied deoptimizations in \Graal \JIT
compiler and compared alternative deoptimization strategies. Southern
et al. and Parravicini et al.~\cite{southern2016overhead,
  parravicini2021cost} studied deoptimization and speculation overhead
in \VEight \JIT compiler. These benchmarks and methodologies help
characterize \JIT performance and can inform test oracle design.  We
focus on understanding performance testing of \JIT compilers and
designing the first approach for finding such bugs.

\MyPara{Formal verification of compilers}
Formal verification is a promising approach to ensure the correctness
of compilers. CompCert is the pioneering work in this
area~\cite{leroy2009formal}. Recently, several other works have also
focused on the formal verification of \JIT compilers, including formal
verification of speculative optimizations with dynamic
deoptimization~\cite{fluckiger2017correctness, barriere2021formally},
formal verification of native code
generation~\cite{barriere2023formally}. Icarus~\cite{smith2024icarus}
uses symbolic meta-execution to formally verify parts of real-world
\JIT compilers. All prior works focus on \functional correctness of (\JIT) compilers;
verifying absence of performance bugs in generated code remains an open
direction.


\section{Conclusion}
\label{sec:conclusion}

This paper provides the first systematic investigation into
performance bugs in just-in-time (\JIT) compilers. Through an
empirical study of real-world issues across four widely used \JIT
compilers for Java and JavaScript, we identify recurring patterns in
how these bugs arise and the compiler components most often
affected. Guided by these observations, we develop \Tool, a
lightweight tool based on \xname that effectively uncovers \JIT
compiler performance anomalies while minimizing testing overhead and
manual effort through prioritization and filtering. \Tool not only
revealed multiple previously unknown performance bugs, several of
which have been confirmed and fixed by developers, but also
demonstrated the feasibility of automated approaches for diagnosing
efficiency regressions in complex runtime systems. We hope our
findings, \Tool, and dataset will inspire future research toward
ensuring more robust, high-performance \JIT compilers.



\section*{Acknowledgements}

We thank Ivan Grigorik, Tong-Nong Lin, Aditya Thimmaiah, Zhiqiang
Zang, Linghan Zhong, Jiyang Zhang, and anonymous reviewers for their
feedback on this work.
We are also grateful to compiler developers for investigating and
addressing our bug reports.
This work is partially supported by the US National Science Foundation
under Grant Nos.~CCF-2107291, CCF-2217696, CCF-2313027, and
CCF-2403036.

\bibliography{bib}

\end{document}